\documentclass[galaxies,review,submit,moreauthors,pdftex,10pt,a4paper]{mdpi}

\pdfoutput=1
\preto{\abstractkeywords}{\nolinenumbers}

\usepackage{aas_macros}
\usepackage{amssymb}
\usepackage{amsmath}
\usepackage{subfigure}
\usepackage{csquotes}

\makeatletter
\def\blfootnote{\xdef\@thefnmark{}\@footnotetext}
\makeatother

\firstpage{1}
\articlenumber{x}
\doinum{10.3390/------}
\pubvolume{xx}
\pubyear{2018}
\copyrightyear{2018}
\history{Received: date; Accepted: date; Published: date}

\Title{Dark matter haloes and subhaloes}

\Author{Jes\'us Zavala$^{1}$\orcidA{}, Carlos S. Frenk$^{2}$\orcidB{}}

\AuthorNames{Jes\'us Zavala, Carlos S. Frenk}

\address{
$^1$\quad Center for Astrophysics and Cosmology, Science Institute, University of Iceland, Dunhagi 5, 107 Reykjav\'ik, Iceland; jzavala@hi.is\\
$^2$\quad Institute of Computational Cosmology, Department of Physics, Durham University, South Road, Durham, DH1 3LE, UK; c.s.frenk@durham.ac.uk}

\corres{Correspondence: jzavala@hi.is}

\abstract{The development of methods and algorithms to solve the
  $N$-body problem for classical, collisionless, non-relativistic
  particles has made it possible to follow the growth and evolution of
  cosmic dark matter structures over most of the Universe's
  history. In the best studied case -- the cold dark matter or CDM
  model -- the dark matter is assumed to consist of elementary
  particles that had negligible thermal velocities at early times.
  Progress over the past three decades has led to a nearly complete
  description of the assembly, structure and spatial distribution of
  dark matter haloes, and their substructure in this model, over
  almost the entire mass range of astronomical objects.  On scales of
  galaxies and above, predictions from this standard CDM model have
  been shown to provide a remarkably good match to a wide variety of
  astronomical data over a large range of epochs, %and masses, 
  from the temperature structure of the cosmic background radiation to the
  large-scale distribution of galaxies. The frontier in this field has
  shifted to the relatively unexplored subgalactic scales, the domain
  of the central regions of massive haloes, and that 
  of low-mass haloes and subhaloes, where potentially fundamental
  questions remain. Answering them may require: %new or additional physics, particularly: 
  (i)~ the effect of known but uncertain {\it
    baryonic} processes (involving gas and stars), and/or
  (ii)~alternative models with new dark matter physics. Here we
  present a review of the field, focusing on our current understanding
  of dark matter structure from $N$-body simulations and on the challenges ahead.}

\begin{document}
\tableofcontents

\section{Introduction} \label{sec_intro}

\noindent The current theory of the formation and evolution of cosmic
structure in the Universe is based on the dark matter hypothesis in
which $\sim84\%$ of the mass-energy density of the Universe
\citep{Planck2018} is in the form of a new type of particle, or
particles, with negligible electromagnetic interactions. The evidence
for the existence of dark matter is varied and compelling. It comes
from cosmic structures on all scales and across all epochs: from the
smallest, dark-matter-dominated dwarf galaxies
(e.g. \citep{Walker2009}), through the largest clusters of galaxies
(e.g. \cite{Lokas2003}), to the large-scale structure of the Universe
(e.g. \cite{Springel2005}) and back to the very seeds of cosmic
structure reflected in the temperature of the cosmic background
radiation (CMB; e.g. \citep{Planck2018}). This body of evidence,
accumulated over the past three decades, can be accounted for within a
coherent theory of structure formation in which the gravity of the
dark matter amplifies primordial density perturbations imprinted
during an early period of cosmic inflation \cite{Guth1981,
  Linde1982}. Empirical evidence for the existence of dark matter
comes purely from its gravitational effect: despite significant
efforts, experimental searches for dark matter particles in
accelerators (e.g. for a review in LHC searches see \cite{Kahlhoefer2017}), and
dedicated detectors on Earth (e.g. \cite{Xenon2018,ADMX2018}) and in
space (e.g. \cite{Fermi2017,Horiuchi2014}) so far remain unsuccessful.
Until the particles are discovered, dark matter will remain a
hypothesis, albeit one with strong empirical support.

In addition to the dark matter hypothesis, the standard theory of
structure formation makes a specific assumption about the nature of
dark matter, which is only partially supported by observations. This
is that the dark matter consists of classical, non-relativistic,
collisionless particles which had negligible thermal velocities at
early times.  This ``cold dark matter'' (CDM) is assumed to behave as
a fluid throughout most of the Universe's history, except at very
early times when this assumption breaks down in different ways
depending on the specific mechanism of dark matter production. The
most common hypothesis is that the dark matter particles are thermal
relics from the Big Bang (e.g. \citep{Gondolo1991}). In this case,
dark matter was symmetric\footnote{Equal amounts of dark matter and
  anti-dark matter.}  and in thermal equilibrium with the
photon-baryon plasma through interactions with standard model
particles. As the Universe cools down, dark matter decouples from the
standard model particles, its creation and annihilation stops and its
co-moving density {\it freezes out}. If the strength of the
interactions is assumed to be on the scale of the weak force, then the
thermal relic abundance of these weakly interacting massive particles
(WIMPs) is quite close to the observed abundance of dark matter. This
remarkable coincidence, haplessly known as the {\it WIMP miracle}, has
enshrined WIMPs as the most popular dark matter candidates, especially
since new physics at the weak scale (and with them the emergence of
WIMP-like particles) was anticipated by Supersymmetric theories in
order to solve the hierarchy problem
(e.g. \citep{Jungman1996}). Moreover, WIMPs are the quintessential CDM
candidate because once they decouple, they are nearly collisionless
and, since they are massive ($\sim$10~GeV -- 1~TeV), they behave as a
classical (non-quantum) fluid that becomes non-relativistic very early
on.

The combination of the {\it WIMP miracle} with the success of the CDM
model in explaining the observed large-scale structure of the Universe
in the mid 1980s \citep{Davis1985} established the current paradigm of
structure formation in which gravity is the only dark matter
interaction. This model has been widely adopted by the community
working on galaxy formation and evolution and, as a result, most of
our understanding on how cosmic structure emerges comes from studies
that assume the CDM model. This is a relevant remark in the context of
this review because the properties of dark matter haloes and their
substructure depend on the nature of dark matter (see
Section~\ref{sec_halo_formation}). In reality, the range of allowed
dark matter models, motivated to varying degrees by particle physics
considerations, is vast.  In this landscape of models, only a fraction
fall in the CDM category alongside WIMPs, e.g.  the QCD axion
(motivated by a proposed solution to the strong CP problem in particle
physics \cite{Preskill1983}).

Dark matter could become non-relativistic at sufficiently late times
so as to suppress, by free streaming, the formation of low-mass
galactic-scale haloes. This case is, in fact, one of the best studied
alternatives to CDM, known as warm dark matter (WDM). In contrast to
WIMPs, these particles have masses of $\mathcal{O}$(1~keV). A sterile
neutrino, included as a part of a model that accounts for neutrino
masses and for the baryon asymmetry of the Universe, is the favourite
WDM candidate (for a recent review see \cite{Boyarsky2018}).  Another
possibility is that dark matter is made of extremely light bosons with
a $\mathcal{O}$(1~kpc) de Broglie wavelength, in which case quantum
effects would be relevant on galactic scales (such possibility falls in
the category of ``fuzzy dark matter''; for a review see
\cite{Hui2017}.)  

Although the interactions between dark matter and standard model
particles are severely constrained, the interactions among the dark
matter particles themselves are not. It is possible that dark matter
may have its own rich phenomenology hidden from the ordinary
matter. This {\it hidden dark matter sector} might possess new forces
and particles, some of which could be viable dark matter particles
that are strongly self-interacting\footnote{By strong, we mean that
  the cross-section for self-interaction is of the order of the
  nuclear cross-section for visible matter (set by the strong
  force).}. These collisional particles fall under the category of
self-interacting dark matter (SIDM; for a review see
\cite{Tulin2018}\footnote{Some SIDM models are motivated by the baryon
  asymmetry; in these models, dark matter, unlike traditional WIMPs,
  shares this asymmetry (for a review of asymmetric dark matter see
  \cite{Zurek2014}).}). Some of the hidden particles might be light
enough that they effectively act as {\it dark radiation} that prevents
the gravitational collapse of dark matter on subgalactic
scales\footnote{In contrast to WDM, the damping of small structures is
  not due to free-streaming, but to a collisional, Silk-like,
  damping.}  (e.g. \cite{Buckley2014,CyrRacine2016}). As mentioned
earlier, the CDM hypothesis is only supported to some extent:
astronomical data allow a variety of models in which dark matter
behaves significantly differently from CDM.

The goal of this paper is to provide a review of the formation,
evolution and dynamics of dark matter haloes and subhaloes, as
revealed primarily by $N$-body simulations.  Although no
account of the properties of haloes based purely on gravitational
dynamics can be complete since baryonic processes play a
significant role in galaxy formation, and new dark matter physics could
also do so, we focus on the standard CDM paradigm of
structure formation in part because the subfield of cosmological
$N$-body simulations has historically been developed in this context,
and also because the emergence and properties of dark matter
structures are most simply understood in the context of
CDM. Alternative dark matter models with additional physical
ingredients to gravity, albeit appealing, are more complicated. In various parts of this
review, we will explore how different assumptions for the nature of
dark matter can lead to different predictions from CDM.

\section{Formation of dark matter haloes} \label{sec_halo_formation}

\subsection{Initial conditions: the primordial power spectrum in the linear regime}\label{subsec_linear}

\noindent A theory of structure formation aims to explain the
evolution of the Universe from a nearly homogeneous initial state,
with tiny matter density perturbations, $\delta\rho/\rho$, seeded by
inflation, which grow to leave an imprint on the CMB (emitted at the
time of recombination, $z\sim1100$, when
$\delta\rho/\rho\sim10^{-3}$), through the emergence of the
self-gravitating dark matter haloes where galaxies form
($\delta\rho/\rho\gg1$), to the Universe we observe today
characterised by a web of filamentary large-scale structure
($\delta\rho/\rho\sim1$).

The starting point is the end of cosmic inflation when dark matter
perturbations are predicted to have a nearly scale-invariant power
spectrum, $\Delta^2\propto k^{3+n_s}$, where
$\Delta^2(k)=k^3P(k)/2\pi^2$ is the dimensionless power spectrum, and
the spectral index $n_s=0.965$ \cite{Planck2018}. The growth of
dark matter perturbations in the expanding Universe is driven by
self-gravity. As long as the perturbations are small,
$\delta\rho/\rho\ll1$ (the {\it linear regime}), this growth can be
described by linear perturbation theory in which each perturbation
evolves independently of all others.

Two important processes occur in the linear regime, which modify the
primordial power spectrum. The first (known as the M\'esz\'aros effect
\cite{Meszaros1974}) operates during the period when the energy
density in the Universe is dominated by radiation: the growth of dark
matter perturbations on scales smaller than the horizon stagnates,
while super-horizon scales continue to grow. This situation pertains
until matter overcomes radiation as the dominant component of the
energy density, after which all perturbations grow at the same rate.
The transition introduces a characteristic scale in the power
spectrum, the size of the horizon at the time of matter-radiation
equality. On scales smaller than this, the power spectrum
flattens. The second important scale, a cutoff in the power spectrum,
is of non-gravitational origin and reflects the particle nature of
dark matter. The physical mechanism that imposes this cutoff is model
dependent. For thermal relics (like many WIMP models and certain types
of WDM), the mechanism is free-streaming, a form of collisionless
(Landau) damping, whose scale is given by the horizon size at the
epoch when the dark matter particles become non-relativistic; the more
massive the particle, the earlier this epoch, and thus the smaller the
(co-moving) free-streaming scale is\footnote{The (co-moving)
  free-streaming scale is given by:
  $l_{\rm fs}=2ct_{\rm nr}/a_{\rm nr}\left[1+{\rm ln}(a_{\rm
      eq}/a_{\rm nr})\right]$, where $t_{\rm nr}$ is the age of the
  Universe at the time when the dark matter particles become
  non-relativistic (at a temperature $3k_BT_{\rm nr}\sim m_\chi c^2$);
  $a_{\rm nr}=1/(1+z_{\rm nr})$ is the scale factor at $t_{\rm nr}$
  ($a\propto t^{1/2}$ in the radiation-dominated era); and
  $a_{\rm eq}$ is the scale factor at the time of matter-radiation
  equality.}, $k_{\rm fs}=2\pi/l_{\rm fs}$.  This is the best-known
cutoff mechanism, which has been traditionally used to classify dark
matter into three categories (where $m_\chi$ denotes the mass of the
particle): cold\footnote{For cold particles, we have assumed CDM
  WIMPs, which requires taking into account the kinetic decoupling
  temperature and epoch; specifically we took Eq.~43 of
  \cite{Green2005}. } ($m_\chi\sim100$~GeV,
$k_{\rm fs}\sim2.5\times10^6$~h/Mpc); warm ($m_\chi\sim1$~keV,
$k_{\rm fs}\sim3.8$~h/Mpc); and hot ($m_\chi\sim$~30 eV,
$k_{\rm fs}\sim0.3$~h/Mpc).

A different type of damping is collisional damping,
which %can either result from the interactions of dark matter particles and standard model particles in the early Universe; in either case, the effect is an opposition to
prevents the gravitational collapse of small structures, resulting in
an effective cutoff in the power spectrum. An example is kinetic
coupling of WIMPs, which effectively keeps dark matter coupled to the
photon-baryon plasma until the Universe cools enough that the
interactions become inefficient, damping perturbations beyond a scale
in the range ($2.6\times10^5 - 1.2\times 10^8$)~h/Mpc
\cite{Bringmann2009}. Another example is collisional damping due to
interactions between dark matter and relativistic particles in the
early universe (either photons or neutrinos,
e.g. \cite{Boehm2002,Boehm2014}, or, in non-standard models, dark
radiation in {\it hidden dark sector} models,
e.g. \cite{Buckley2014,CyrRacine2016}). The relativistic particles
create an effective radiation pressure that counteracts the
gravitational collapse, driving oscillations in the density
perturbations, akin to the well-known baryon acoustic oscillations
(BAOs), but on much smaller scales; by analogy they are called dark
acoustic oscillations, DAOs \footnote{Note that acoustic oscillations are also present in WIMP-CDM models (e.g. \cite{Loeb2005}), but they occur at much smaller scales than in relevant hidden dark sector models where they can be of galactic scale.}. 
%{\color{blue} Jesus, is this meant to happen for CDM as well?} 
Once the Universe cools down, the dark
matter decouples from the relativistic particles, imprinting a
characteristic scale (the size of the sound horizon at the time of
decoupling) in the power spectrum, followed by a Silk-like damping
cutoff.

\begin{figure}[H]
\centering
\includegraphics[width=10 cm, height=8cm]{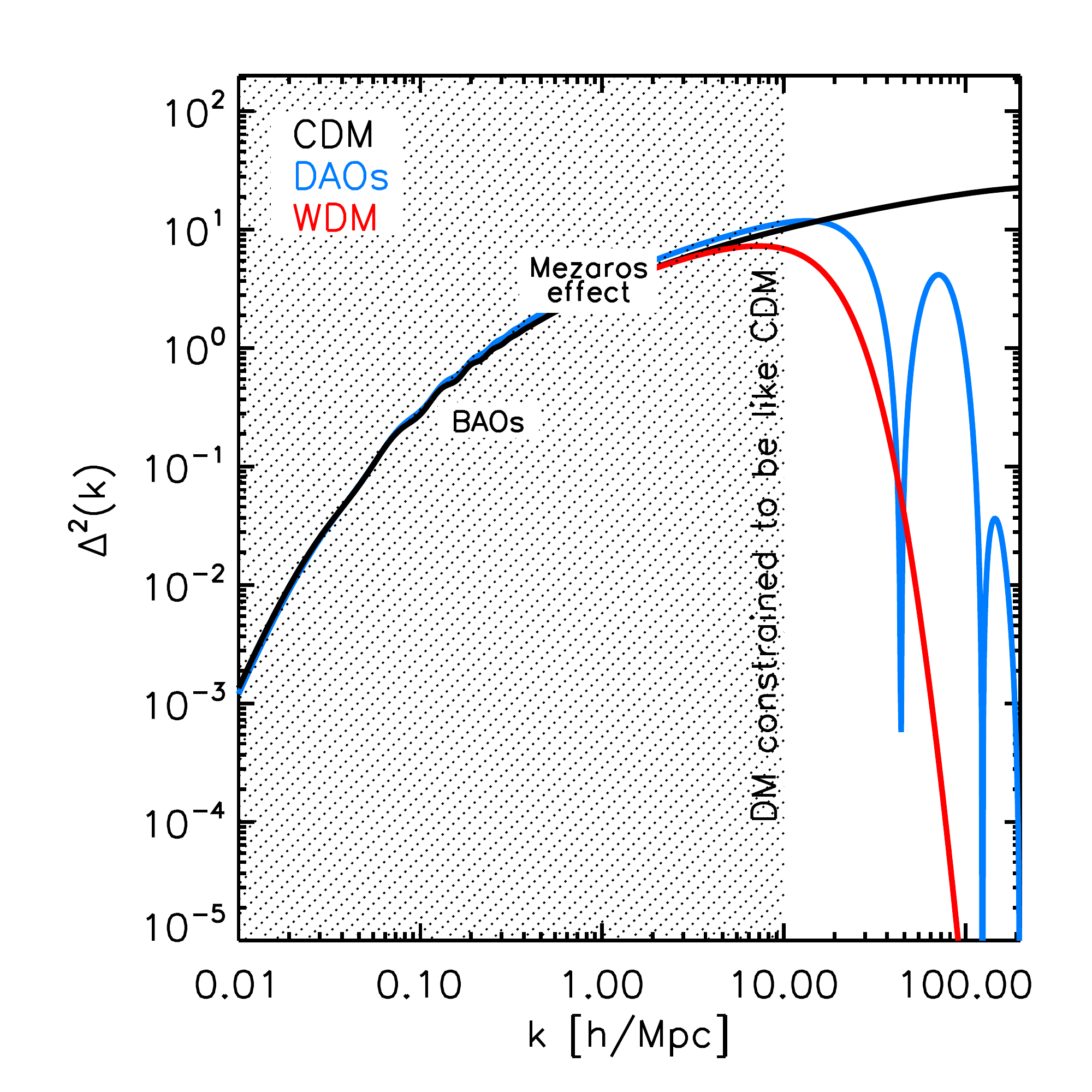}
\caption{Dimensionless linear dark matter power spectrum in different
  dark matter models. In the current paradigm, cold dark matter (CDM),
  the power spectrum keeps on rising to well below subgalactic
  scales. Alternative models such as warm dark matter (WDM) or
  interacting dark matter (DAOs) have a cutoff at or sligthly below galactic scales,
  which determines the abundance and structure of small-mass dark
  matter haloes and subhaloes and the galaxies within. In the black hashed area, the dark
  matter is constrained by the observed large-scale distribution of
  galaxies \cite[e.g.][]{Cole2005,Percival2007} and the Ly-$\alpha$
  forest constraints on WDM \cite{Viel2013} to behave as CDM. Figure adapted from \cite{Zavala2019}.}
\label{linear_pk}
\end{figure}

The main features of the clustered dark matter distribution during the
linear regime are illustrated in Fig.~\ref{linear_pk}. On the largest
scales, not affected by the M\'ez\'aros effect, the power spectrum is
nearly scale-invariant, $\Delta^2\propto k^{3+n_s}$; on smaller scales
it bends to increasingly shallower slopes. For CDM (black line), the
power spectrum remains featureless well below galactic scales. For
reference, a dark matter halo today hosting a typical dwarf galaxy
would have a mass $\sim10^{10}$~M$_\odot$, roughly corresponding to a
(co-moving) wavenumber $\sim12$~h/Mpc\footnote{We use
  $M=4\pi/3{\overline\rho}(\pi/k)^3$, where ${\overline\rho}$ is the mean dark
  matter density today.}. Measurements of galaxy clustering on scales
larger than individual galaxies, together with constraints from the
flux spectrum of the Ly-$\alpha$ forest (e.g. \cite{Viel2013})
constrain the power spectrum to be like CDM to the left of the hashed
area in Fig.~\ref{linear_pk}. On smaller scales the power spectrum
could have a damping cutoff due to either collisionless (as in WDM
models; red line) or to collisional (as in models with DAOs; blue
line) processes. We have not included the cutoff characteristic of
{\it fuzzy dark matter} models, but we note that it is also
oscillatory like the DAOs models (but due to quantum rather than
collisional effects; see e.g. Fig. 2 of \cite{Schive2016}).

As long as the dark matter perturbations remain linear
($\delta\rho/\rho\ll1$), they grow at a rate that does not depend on
their co-moving scale, $\Delta^2(k;t)\propto D^2(t)$, where $D(t)$ is
the growth factor, which depends only on the mean density of matter
and dark energy (see e.g. \cite{Carroll1992}). Once the density
contrast is no longer small ($\delta\rho/\rho\sim0.1$), perturbation
theory breaks down since gravity couples perturbations on different
scales and their evolution can no longer be calculated as independent
modes. 

\subsection{The non-linear regime: $N$-body simulation methods}\label{Nbody}

\noindent To follow the evolution of dark matter density perturbations
beyond the linear regime, a number of approaches are possible
depending on the problem of interest. (i)~High order perturbation
theory which can be used to study the quasi-linear regime
($\delta\rho/\rho\lesssim1$), particularly in a modern reformulation
such as the {\em Effective field theory of large-scale structure}
\citep{Baumann2012,Carrasco2012}.  (ii)~Analytical models with
simplified assumptions for the growth, turnaround (i.e., decoupling
from the expansion of the Universe), collapse and virialization (i.e.,
the formation of a gravitationally self-bound structure) of individual
perturbations. The best known examples are the {\em Spherical
  collapse} \cite{Gunn1972} and {\em Ellipsoidal collapse}
\cite{Sheth2001} models which link a primordial perturbation to the
final equilibrium configuration: the dark matter halo.  (iii)~The {\it
  halo model} (for a review see \cite{Cooray2002}), which combines the
analytical models in (ii) with the assumption of a Gaussian density
field and can be used to compute the abundance of virialised haloes as
a function of their mass (the halo mass function); together with a
model for the dark matter distribution within haloes, it can be used
to model the non-linear dark matter power spectrum on all
scales. (iv)~Models based on the {\it Stable clustering hypothesis}
(\cite{Davis1977} see also \cite{Smith2003}), which assumes that the
number of neighbouring dark matter particles within a fixed physical
separation remains constant, and can be used to study the deeply
non-linear regime; a recent re-formulation in phase space has been
shown to be a promising alternative to the halo model
\cite{Afshordi2010,Zavala2014,Zavala2016}.  (v) Numerical {\em
  $N$-body simulations}, which solve {\em ab initio} the gravitational
evolution in phase space of a distribution of $N$ particles sampled
from an initial power spectrum. This is the most general and powerful
approach to study the clustering evolution on all scales and is the
focus of this review.  (vi) Techniques that avoid the particle
discretisation inherent in $N$-body simulations by following the
phase-space distribution function directly \cite{Hahn2013}. These are
particularly useful to study evolution from truncated power spectra
such as for hot or warm dark matter for which standard $N$-body
techniques suffer from artifical fragmentation \cite{Angulo2013}. 

%As we mentioned earlier, the development of the theory of structure
%formation, and particularly in the non-linear regime, has been done
%within the context of CDM, and as such, the approaches and references
%described in the previous paragraph have CDM as an underlying
%assumption, i.e. they deal exclusively with gravitational dynamics in
%an expanding Universe. In some cases, some of these approaches have
%been extended to include a non-CDM nature (most notable for
%WDM). Since the focus of this review is based on $N$-body simulations,
%we will restrict our discussion to this technique and how it is
%modified to include a different dark matter nature.

In the case of classical, non-relativistic, collisionless particles, i.e., CDM, 
$N$-body simulations follow the evolution of the dark matter
phase-space distribution function, $f(\vec{x}, \vec{v}; t)$, which in principle is given by the
collisionless Boltzmann equation coupled with the Poisson equation
for the gravitational field, $\Phi(\vec{x})$ (a combination known as
the Vlasov-Poisson equation):
\begin{eqnarray}\label{CBE}
\frac{df}{dt}&=& \frac{\partial f}{\partial t}+\sum_iv_i\frac{\partial f}{\partial x_i}-\sum_i\frac{\partial \Phi}{\partial x_i}\frac{\partial f}{\partial v_i}=0\\
\rho_\chi(\vec{x};t)&=&\int f(\vec{x},\vec{v}; t)d^3\vec{v}\\
\nabla^2\Phi(\vec{x})&=&4\pi G\rho_\chi(\vec{x})
\end{eqnarray}
where $d/dt$ is the Lagrangian derivative. Cosmological $N$-body
simulations\footnote{For a review see e.g. Section 3 of
  \cite{Dehnen2011}.} solve this equation in an expanding universe
using a co-moving reference frame (with the expansion included
explicitly through the solution of the Friedmann equations for the
scale factor), discretizing the distribution function as an ensemble
of $N$ phase-space elements or ``particles'',
$\{\vec{x_i},\vec{v_i}\}$, with $i=1,...,N$. Since the collisionless
Boltzmann equation implies that the phase-space distribution remains
constant in time along any trajectory $\{\vec{x}(t),\vec{v}(t)\}$, the
distribution obtained by following the $N$ particles from initial
conditions sampled from the phase-space distribution at $t=0$,
constitute a representative Monte-Carlo sampling of the distribution
function at any subsequent time, $t$. The $N$ particles are thus a
statistical representation of the coarse-grained\footnote{By this we
  mean an average of the fine-grained distribution function in the
  collisionless Boltzmann equation over the scales resolved in the
  simulation, typically several times the interparticle separation.}
distribution function:
\begin{eqnarray}
\tilde{f}(\vec{x},\vec{v})&\sim&\sum_im_iW(\vert\vec{x}-\vec{x_i}\vert;\varepsilon)\delta^3(\vec{v}-\vec{v_i});\ \ \ \ \ \ \ \dfrac{d\tilde{f}}{dt}=0\\
\tilde{\rho}(\vec{x})&=&\int \tilde{f}(\vec{x},\vec{v})d^3\vec{v}\sim\sum_im_iW(\vert\vec{x}-\vec{x_i}\vert;\varepsilon)\\
%\tilde{\Phi}(\vec{x})&\sim&-G\sum_im_iW_\Phi(\vert\vec{x}-\vec{x_i}\vert;\varepsilon)
\tilde{\Phi}(\vec{x})&=&\int g(\vec{x}-\vec{x}')\tilde{\rho}(\vec{x}')d^3\vec{x}'
\label{CBE_discrete}
\end{eqnarray}
where $m_i$ is the mass of the simulation particle,
$\delta^3(\vec{v}-\vec{v_i})$ is the Dirac delta funcion in 3D, $W$ is
a {\it kernel density} with a softening length
$\varepsilon$\footnote{In principle each particle can have an
  individual softening, see e.g. Section 4 of \cite{Dehnen2001}.},
introduced to obtain a smooth density field from the set of $N$
discrete particles; i.e., the kernel effectively models each
simulation particle as an extended mass distribution\footnote{The
  introduction of a softening scale in the density (or potential)
  suppresses gravitational two-body large-angle scatterings which are
  artificial for an approximately continuous dark matter density
  distribution.}; finally, the last equation for the potential is the
general solution to Poisson's equation as a convolution of the density
field with a suitable Green's function\footnote{In Fourier
    space, Eq.~\ref{CBE_discrete} is simply a multiplication
    $\hat{\tilde{\Phi}}(\vec{k})=\hat{g}(\vec{k})\hat{\tilde{\rho}}(\vec{k})$. }.
%  $W_\Phi$ is the {\it softening kernel} for
% the potential (connected to $W$ through Poisson's equation). 
Since each simulation particle represents a region of phase space containing
a very large number, $m_i/m_\chi$, of real dark matter particles, the
information in an $N$-body simulation is always incomplete, limited by
the phase space resolution and the softening length.

With the discretization method employed in an $N$-body simulation,
calculating the evolution of the phase-space distribution is reduced
to following self-consistently the dynamics of a system of $N$
particles (usually in terms of the Hamiltonian of the system in the
co-moving frame) according to the potential derived from the particle
distribution.  Modern codes used to solve this problem employ
efficient methods for computing the gravitational potential and
integrating the Hamiltonian system forward in time. Early cosmological
simulation codes used the particle-mesh (PM) technique in Fourier
space (e.g. \cite{Klypin1983,Melott1983}) or direct integration of the
$N^2$ interactions (e.g. \cite{Frenk1983}). The former is limited in
resolution by the size of the mesh while the latter is limited by
speed. These two shortcomings can be overcome by combining both
techniques in the P$^3$M method
(e.g. \cite{Hockney1988,Efstathiou1981}), in which the long-range
forces acting on a particle are calculated on a PM grid and the
short-range forces by direct $N^2$ summation. An alternative approach
is the hierarchical tree method \cite{Barnes1986} in which an octree
is used to divide the volume recursively into cubic cells and
increasingly coarse cells are used to compute the forces on a particle
at increasingly large distances. The most widely used cosmological
simulation code is GADGET-2 \cite{Gadget2005}, which uses the treePM
algorithm, whereby short-range forces are computed with the tree
method and long-range forces with Fourier techniques\footnote{For a
  review of the force computation methods see Section~3.5 of
  \cite{Dehnen2011}.}.

If dark matter cannot be treated as CDM, then the fundamental
equations may need to be modified. For models that only deviate from
CDM because of a cutoff in the initial power spectrum (such as hot or
warm dark matter and certain DAO models), the $N$-body equations
(\ref{CBE}-\ref{CBE_discrete}) and methods used for CDM are still
valid as long as the dark matter behaves as a collisionless, classical
system, and the simulation starts well after the dark matter particles
have become non-relativistic ; all that is needed is a modification of
the initial conditions (see \ref{ICs} below).  On the other hand, if
dark matter is non-relativistic but no longer collisionless, like in
SIDM, then the collisionless Boltzmann equation needs to be replaced
by the full collisional Boltzmann equation, which has an extra term
(the collisional operator) in the right-hand-side of eq.~\ref{CBE}, to
account for the effect of dark matter collisions according to a
self-scattering cross-section. It is possible to incorporate this new
term within the Monte Carlo approach of traditional $N$-body
simulations by adding ``collisions'' between each simulation particle
and its immediate neighbours in a probabilistic way that reflects the
effective scattering rate given by the cross-section
(e.g. \cite{Kochanek2000,Dave2001,Vogelsberger2012,Rocha2013,Robertson2017}; see
Appendix~A of \cite{Rocha2013} for a detailed derivation). An
alternative to the $N$-body approach is the ``gravothermal fluid''
approximation \cite{LyndenBell1980}, which considers an SIDM dark
matter halo as a self-gravitating, spherically symmetric, ideal gas
with an effective thermal conductivity (related to the self-scattering
cross-section, see e.g. \cite{Koda2011}). Although this approach is
restricted, it provides physical insight into the evolution of SIDM
haloes, and a degree of validation of SIDM $N$-body
simulations. Finally, if quantum effects are important for the dark
matter fluid, then the Vlasov-Poisson equation needs to be replaced by
the Schr\"{o}dinger-Poisson equation, whose solution requires
numerical methods quite distinct from the $N$-body approach
(e.g. \cite{Schive2014,Mocz2017}).

\subsection{The non-linear regime: initial conditions and the
  emergence of the cosmic web}\label{ICs}

\noindent The techniques of Section~\ref{Nbody} can be used to
integrate forward in time a particle distribution starting from an
initial state, the initial conditions, usually taken to be in the
linear regime described by perturbation theory. The basic techniques
for generating general initial conditions were laid out in
\cite{Davis1985} and \cite{Efstathiou1985} and have been refined over
the years (e.g. \cite{Hahn2011,Jenkins2013}; for a review see
\cite{Sirko2005} or Appendix C1.1.4 of \cite{Mo2010}). They provide a
particle realization with the statistical properties of the linear
dark matter density field described by the power spectrum. In general
the procedure can be divided into two steps:
\begin{itemize}
\item[i)] create a realization of an unperturbed cube of side $L$ by
  distributing $N$ particles homogeneously in a lattice or in a
  glass-like configuration\footnote{The particles are initially placed
    at random in the simulation cube and then left to evolve under a
    repulsive force by reversing the sign of the gravitational force
    until they reach an equilibrium configuration that has no
    discernible grid pattern \cite{White1996}.} to avoid imprinting a
  grid-like pattern in the simulation.
\item[ii)] perturbations of wavelength $\lambda$ down to the Nyquist
  frequency of the particle distribution are represented by plane
  waves of spatial frequency in Fourier space, $k=2\pi/\lambda$, whose
  amplitudes and phases are drawn at random from a Gaussian
  distribution with variance proportional to the desired linear power
  spectrum. The density field and its gravitational potential in real
  space are then obtained by an inverse Fourier transform. Using the
  Zel'dovich approximation \citep{Zeldovich1970}, or the more accurate
  second-order Lagrangian perturbation theory
  (e.g. \cite{Jenkins2010}), these fields are used to compute the
  displacements needed to transform the uniform $N$-particle
  distribution in part~i) into a distribution that has the desired
  power spectrum.
\end{itemize}

\begin{figure}[H]
\centering
\includegraphics[width=15 cm, height=8cm,trim=1.05cm 0.5cm 0.5cm 0.5cm, clip=true]{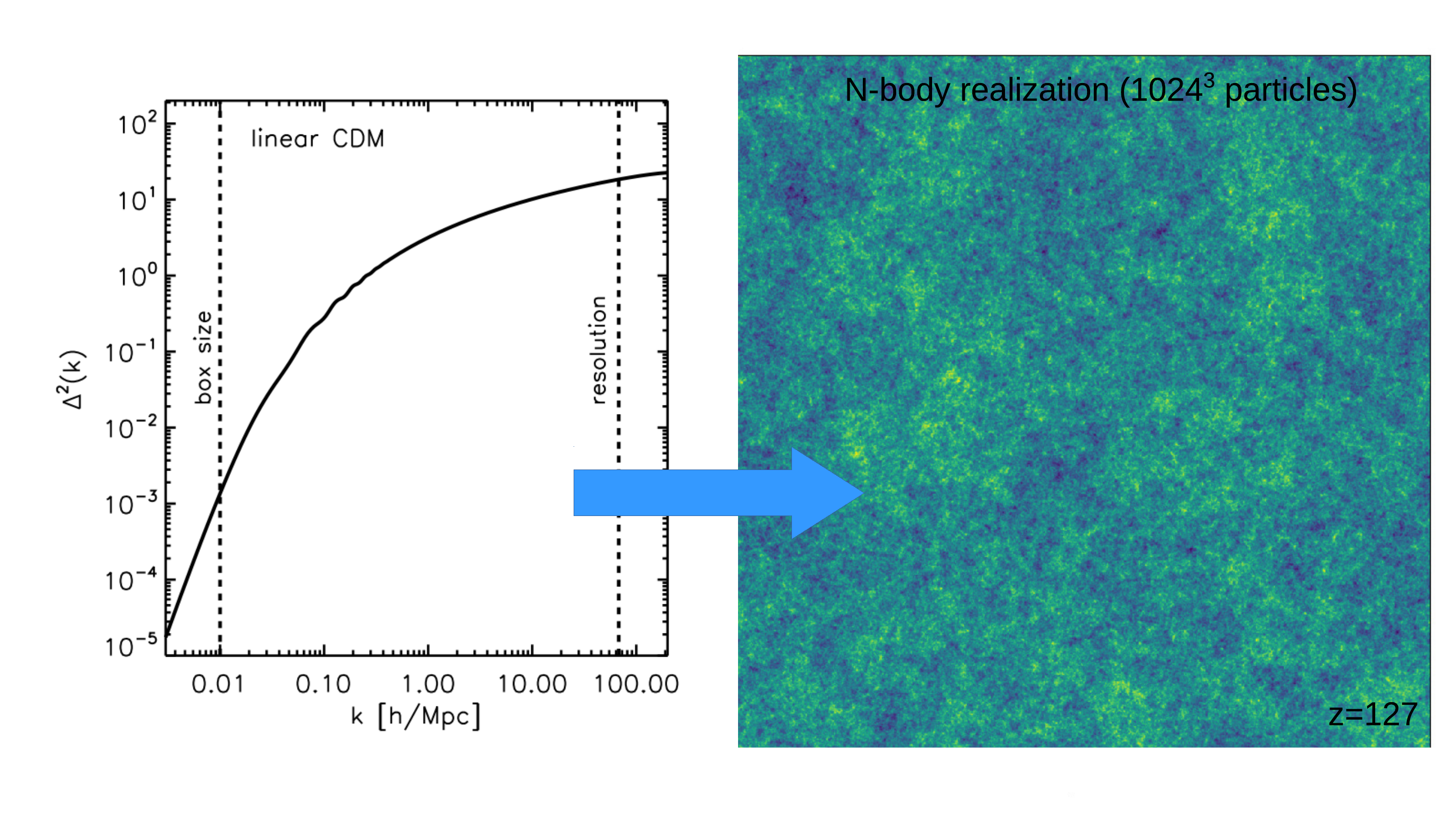}
\caption{Illustration of the initial conditions for an $N$-body
  simulation.  {\it Left:} the dimensionless linear CDM power
  spectrum. The vertical dashed lines mark the modes corresponding to
  the maximum and minimum scales that can be represented in the
  initial conditions: the fundamental mode, $2\pi/L$, and the Nyquist
  mode, $\pi/d$, where $L$ and $d$ are the cube length and
  interparticle separation, respectively. {\it Right}: a realisation
  of the dark matter density field generated from the power spectrum
  on the left at redshift $z=127$ using $N=1024^3$ particles in a
  cosmological cube of co-moving side, $L=40$~Mpc/h. The code MUSIC
  \citep{Hahn2011} was used to generate the particle distribution and
  the Pynbody package \cite{pynbody} to create the image.}
\label{initial_conditions}
\end{figure}

An illustration of the end result of this procedure is shown in
Fig.~\ref{initial_conditions}. The initial conditions generator, MUSIC
\citep{Hahn2011}, was used to construct the particle distribution on
the right, which is a statistical realisation of the CDM linear power
spectrum shown on the left. The main limitations for a cosmological
simulation are already set in the initial conditions: the maximum
lengthscale that can be simulated is determined by the (co-moving)
side of the computational cube\footnote{ A sufficiently large volume
  is needed to sample large-scale modes that remain approximately
  linear during the simulation where power is transferred from large
  to small scales; without appropriate large-scale sampling, the
  clustering is no longer accurate once perturbations on the scale of
  the cube become non-linear.}, and the minimum lengthscale that can
represented in the initial conditions is set by the Nyquist frequency
of the particle distribution\footnote{In practice, power below the
  Nyquist frequency is generated non-linearly so the resolution of the
  simulation is not limited by the Nyquist frequency but rather by the
  gravitational softening scale, $\varepsilon$.}. The choice of cube
length and particle number depends on the science goal of the
simulation and on the computing resources available. We will come back
to this point below.

The procedure illustrated in Fig.~\ref{initial_conditions} for CDM can
be readily applied to other dark matter models with different initial
power spectra. In fact, in models where dark matter only behaves
differently from CDM at very early times, e.g., in thermal-relic WDM
models, it is the different initial conditions (the lack of power on
small scales in WDM relative to CDM in the linear regime) that gives
rise to the main differences between these models (since the residual
thermal motions in WDM models of interest are negligible (see
e.g. \cite{Smith2011}). In models with a truncated initial power
spectrum, the subsequent evolution is affected by particle
discreteness in the reconstruction of the density field which
introduces an irreducible (shot-noise) power. This results in spurious
clustering on scales close to the cutoff length \citep{Wang2007} that
requires careful treatment to either remove small-scale artificial
clumps \cite{Lovell2014} or avoid their formation altogether by using
non-standard simulation techniques \cite{Angulo2013,Hobbs2016}.

\begin{figure}[H]
\centering
\includegraphics[width=15 cm, height=8cm,trim=1.05cm 0.25cm 0.5cm 0.25cm, clip=true]{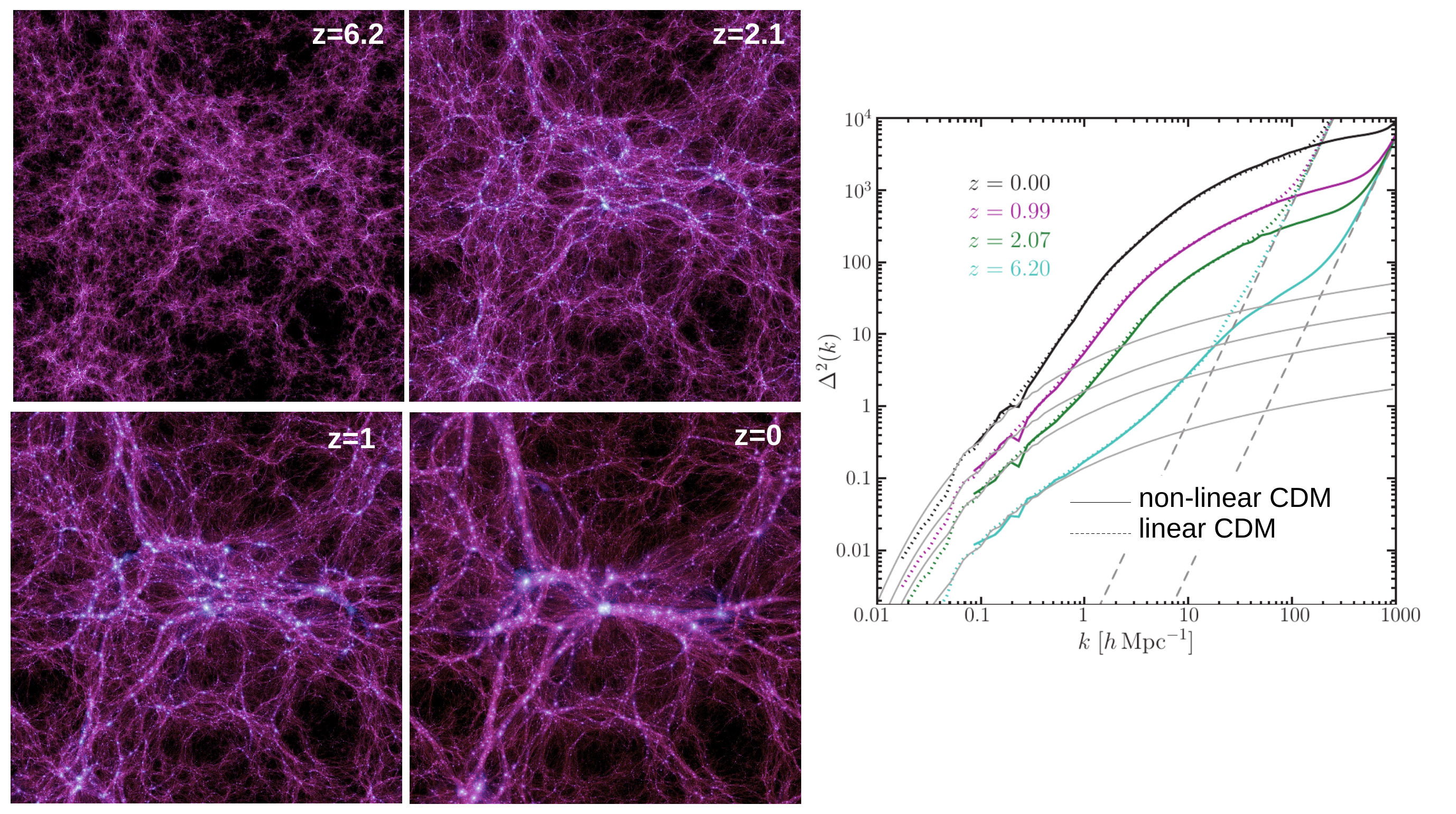}
\caption[Caption for LOF]{Emergence of the cosmic web. {\it Left:} evolution of the
  (projected) dark matter density field in a slab of length
  $L=100$~Mpc/h and thickness 15~Mpc/h  from the Millennium II simulation
  \citep{Boylan2009}. The redshift corresponding to each snapshot is
  shown on the top right.  {\it Right:} The dimensionless dark matter
  power spectrum (solid lines) at the redshifts shown on the left. For
  comparison, also shown are: the linear power spectrum (thin grey
  lines) and the non-linear power spectrum for the lower resolution
  but larger scale ($500$~Mpc/h) Millennium I simulation (in dotted
  lines; \cite{Springel2005}). The dashed lines show the Poisson 
  noise limit for the Millennium I (left) and Millennium II (right)
  simulations. Figure adapted from \cite{Boylan2009}\footnotemark.}
\label{sims_evolution}
\end{figure}

\footnotetext{\scriptsize{Reproduced from Michael Boylan-Kolchin et al. Resolving cosmic structure formation with the Millennium-II Simulation. MNRAS (2009) 398 (3): 1150-1164, doi: 10.1111/j.1365-2966.2009.15191.x. By permission of Oxford University Press on behalf of the Royal Astronomical Society. For the original article, please visit the following \href{https://academic.oup.com/mnras/article/398/3/1150/1261296}{link}. This figure is not included under the CC-BY license of this publication. For permissions, please email: journals.permissions@oup.com}}

Once the initial conditions are generated, an $N$-body simulation is
performed, most commonly in a computational cube with periodic
boundary conditions, to follow the evolution of the density and
velocity fields into the non-linear regime across all resolved scales.
An example, the Millennium~II simulation \citep{Boylan2009}, is
illustrated in Fig.~\ref{sims_evolution}. The left set of panels shows the
projected dark matter density distribution at various {\it snapshots}
corresponding to the redshifts shown at the top right of each panel.  The emergence of the cosmic
web, the result of gravitational clustering, is apparent, with its
now familiar pattern of filaments over a range of scales surrounding
voids. The right panel shows the evolution of the power spectrum at
the same snapshots (solid lines). The hierarchical onset of non-linear
structure, from small to large scales is clearly apparent by
reference to the linear power spectrum (grey lines).

The first CDM cosmological $N-$body simulations in the 1980s
\cite{Davis1985,Frenk1988} already contained all the relevant physical
processes of gravitational clustering for collisionless dark matter,
but were computationally limited; they could follow the evolution of
only $\mathcal{O}(10^4)$ particles. In the decades since then, the
tremendous improvement in computational capabilities has been such
that cosmological ($L\gtrsim100$~Mpc/h) simulations with
$\mathcal{O}(10^9)$ particles are routinely performed\footnote{For a
  review on the state of cosmological simulations circa 2012 see
  \cite{Kuhlen2012}.}, and the most expensive simulations to date have
reached the one trillion particle milestone \cite{Potter2017}.

\begin{figure}[H]
\centering
\includegraphics[width=10 cm, height=10cm]{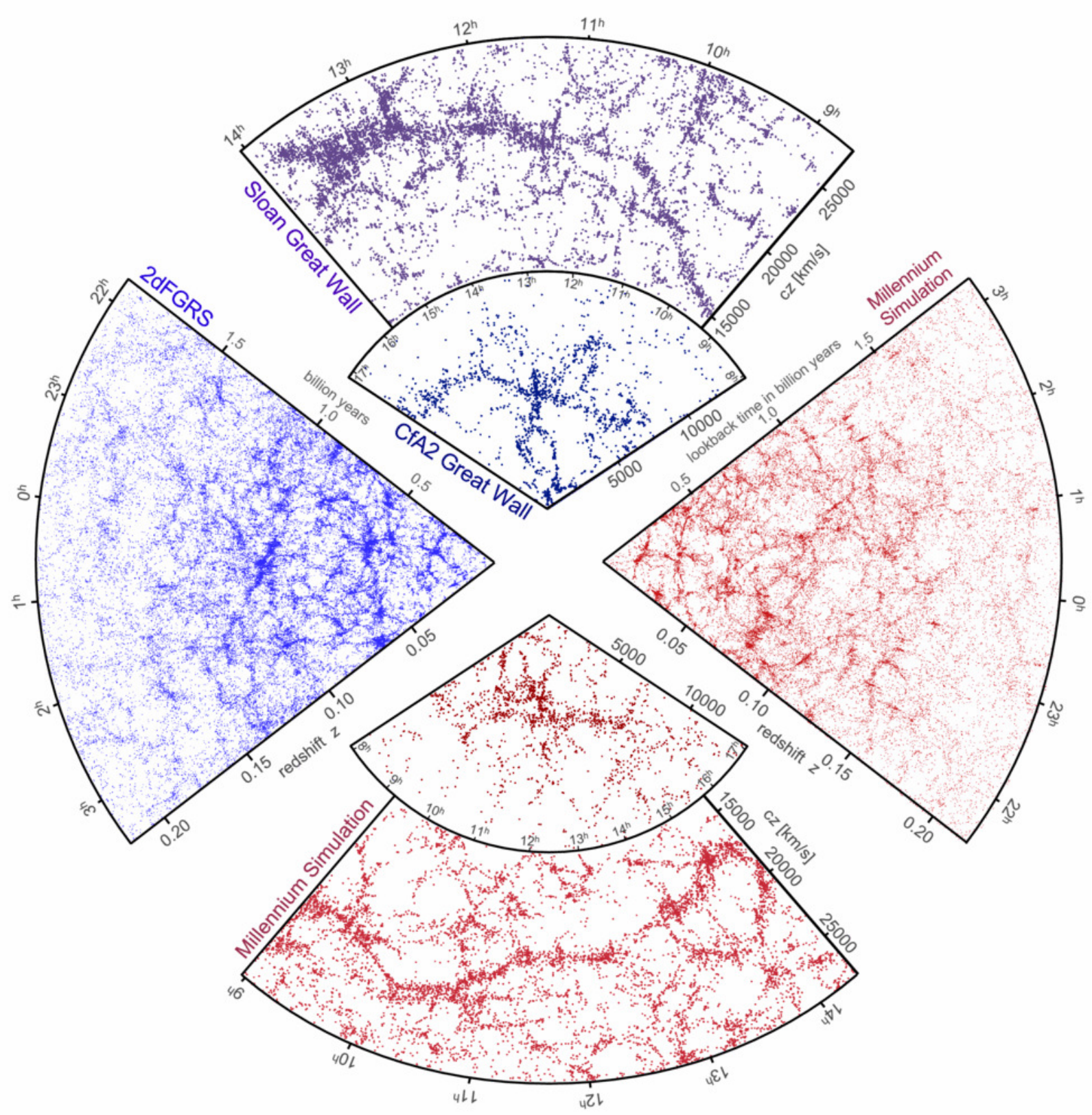}
\caption{The galaxy distribution in various redshift surveys and in
  mock catalogues constructed from the Millennium simulation
  \cite{Springel2005}. The small slice at the top shows the CfA2
  ``Great Wall'' \cite{Geller1989}, with the Coma cluster at the
  centre. Just above is a section of the Sloan Sigital Sky Survey in
  which the ``Sloan Great Wall'' \cite{Gott2005} is visible. The wedge
  on the left shows one-half of the 2-degree-field galaxy redshift
  survey \cite{Colless2001}.  At the bottom and on the right, mock
  galaxy surveys constructed using a semi-analytic model applied to
  the simulation \cite{Croton2006} are shown, selected to have geometry
  and magnitude limits matching the corresponding real
  surveys. Adapted from \cite{Springel2006}.}
\label{millennium}
\end{figure}

To compare the simulations to astronomical data it is necessary to
make a correspondence between dark matter haloes and the galaxies that
would form within them. In the earliest simulations, galaxies were
identified with high peaks of the suitably filtered density field, an
assumption known as ``biased galaxy formation''
\cite{Davis1985,Bardeen1986}. Physically based models of galaxy
formation that could be grafted onto $N$-body simulations were
developed in the early 1990s \citep{White1991}. These are known as
``semi-analytic models'' because they encapsulate the relevant
physical models in a set of coupled differential equations that are
solved numerically. These equations assume spherical symmetry and
describe the cooling of gas, the formation of stars, chemical
evolution, the growth and merging of central supermassive black holes
and feedback effects arising from energy injected into the gas during
the course of stellar evolution and by active galactic nuclei
triggered by accretion of gas onto the central black hole. The model
is applied at every stage of the gravitational evolution of the
merging hierarchy of haloes, described by a merger
tree (see Section \ref{sec_smooth_mergers}). Semi-analytic models have been extremely successful in linking
the distribution of dark matter computed in an $N$-body simulation to
the observed Universe \cite{Kauffmann1997,Kauffmann1999,Benson2000,Springel2001} %cite{Kauffmann1993,Cole2000,Somerville2008} 
and have become very sophisticated in predicting visible galaxy properties
over a large range of wavelengths (e.g. \cite{Lacey2016}).
%(e.g. \cite{Monaco2007,Lacey2016}). 
An example based on the Millennium simulation is shown in Fig.~\ref{millennium}.

Dark matter $N-$body simulations are the cornerstone of the current
understanding of how galaxies form and evolve and, as illustrated in
Fig.~\ref{millennium}, have been very successful in explaining the
large-scale structure of the Universe \cite{Springel2006}. The latter
accomplishment is non-trivial and demands certain conditions about the
nature of dark matter. For instance, already in the 1980s, light
neutrinos were ruled out as the dominant component of dark matter by
their incompatibility with the observed large-scale structure
\citep{White1983}, thus demonstrating the potential of $N$-body
simulations to test models for the nature of the dark matter.  By
contrast, the fact that CDM matched the observations available at the
time remarkably well contributed greatly to its establishment as the
standard model of cosmogony \citep{Davis1985}. By now, it is firmly
established that whatever the dark matter is, it must behave as CDM on
large scales (see Figs.~\ref{linear_pk} and~\ref{millennium}). It is
important to recognize, however, that a wide range of dark matter
candidates behave just as CDM on large scales and thus are also
allowed by the large-scale structure data, as we discussed in
Sections~\ref{sec_intro} and~\ref{subsec_linear}. In this sense, the
success of the CDM model in explaining the large-scale structure of
the Universe is shared by allowed WDM, SIDM and {\it fuzzy dark
  matter} models.

\subsection{The structural properties of dark matter haloes}\label{sec_dm_nature}

As a consequence of gravitational clustering, the tiny density
perturbations present when the CMB was emitted grow in time,
eventually separating from the expansion of the Universe and becoming
self-gravitating bound structures known as dark matter haloes. This
process of forming {\it virialised} haloes can be understood from the
simple spherical collapse model \cite{Gunn1972}. Haloes become
increasingly more massive with time, {\it smoothly} by accreting mass
from their surroundings or {\it merging} with other, smaller
haloes. The latter thus become subhaloes, which is the topic of
Section~\ref{sec_subhaloes}. Although the large-scale environment and
spatial clustering of dark matter haloes are clearly relevant, here we
focus on the abundance (halo mass function) and internal structure of
dark matter haloes. These properties are the most useful when
attempting to differentiate dark matter models. Currently, however,
the halo mass function and halo structure on the key subgalactic
scales are only weakly constrained observationally. We should also
note that not all dark matter is contained within haloes. The fraction
of {\it unclustered} dark matter is naturally a strong function of
time reflecting the growth of collapsed objects by hierarchical
clustering. Even today at most $\sim20\%$ is expected to be
unclustered in the CDM model \cite{Angulo2010}. A recent update of
this work (applied also to WDM) puts the fraction even lower, at the
percent level \cite{Stuecker2018}.

{\it Definition of a halo.---} Because of the dynamic nature of haloes
and their lack of spherical symmetry, precisely defining the boundary
of a halo, and thus its mass, is, to some extent, arbitrary
\cite{Davis1985,White2001,Cuesta2008}. A variety of definitions exist
in the literature with the most common ones being: (i)~the FOF
(friends-of-friends) mass, defined as the mass of the set of particles
that are linked together by a percolation scale, defined by a linking
length, $b\sim0.2$ in units of the mean interparticle separation
\cite{Davis1985}; (ii)~a spherical overdensity mass, $M_\Delta$,
contained within a sphere centred on the halo (with the centre placed
at the minimum of the gravitational potential of the halo), with a
radius given by the spherical collapse model, whereby the
collapsed region that defines a halo contains an average density
$\Delta(z)$ times the critical density for closure
\cite{Cole1996}. The overdensity, $\Delta(z)$, is a redshift-dependent
function of cosmology \cite{Eke1996,Bryan1998}, but for the
Einstein-de Sitter cosmology, $\Delta\sim178$ at all times; (iii)~the
viral mass, defined with $\Delta=200$, which early simulations
identified as the radius that separates the region of the halo that is
in dynamical equilibrium from the surrounding region that is still
collapsing \cite{Cole1996}.  Given the simplicity of the latter, its
relation to dynamical equilibrium, and its connection with the
Einstein-de Sitter spherical collapse overdensity, the radius,
$r_{200}$, and the enclosed mass, $M_{200}$, are widely used in the
field as the boundary and mass of dark matter haloes, respectively.

{\it The halo mass function.---} The mass function of dark matter
haloes, i.e., the number density of haloes of different mass, has been
characterised quite precisely in the last couple of decades by
$N$-body simulations
\citep[e.g.][]{Jenkins2001,Warren2006,Lukic2007,Tinker2008,Boylan2009,Trujillo2011,Hellwing2016},
and is now well determined over the full range of epochs and masses
relevant to galaxy formation, from $\mathcal{O}(10^{8}~{\rm M}_\odot)$
dwarf-size haloes to $\mathcal{O}(10^{15}~{\rm M}_\odot)$ cluster-size
haloes.  The number density of haloes per unit mass scales as:

 \begin{equation}\label{eq_halo_mf}
 \frac{dn}{dM}\propto M^{\alpha}, \ \  {\rm where} \ \ \alpha\sim-1.9 \ \ {\rm for~low~masses,} 
 \end{equation}
  with an overall normalisation that correlates with the large
 scale environment, with denser environments having a larger halo
 abundance \citep{Frenk1988,Crain2009}.

\begin{figure}[H]
\centering
\includegraphics[width=15 cm, height=8cm,trim=1.05cm 0.25cm 0.5cm 0.25cm, clip=true]{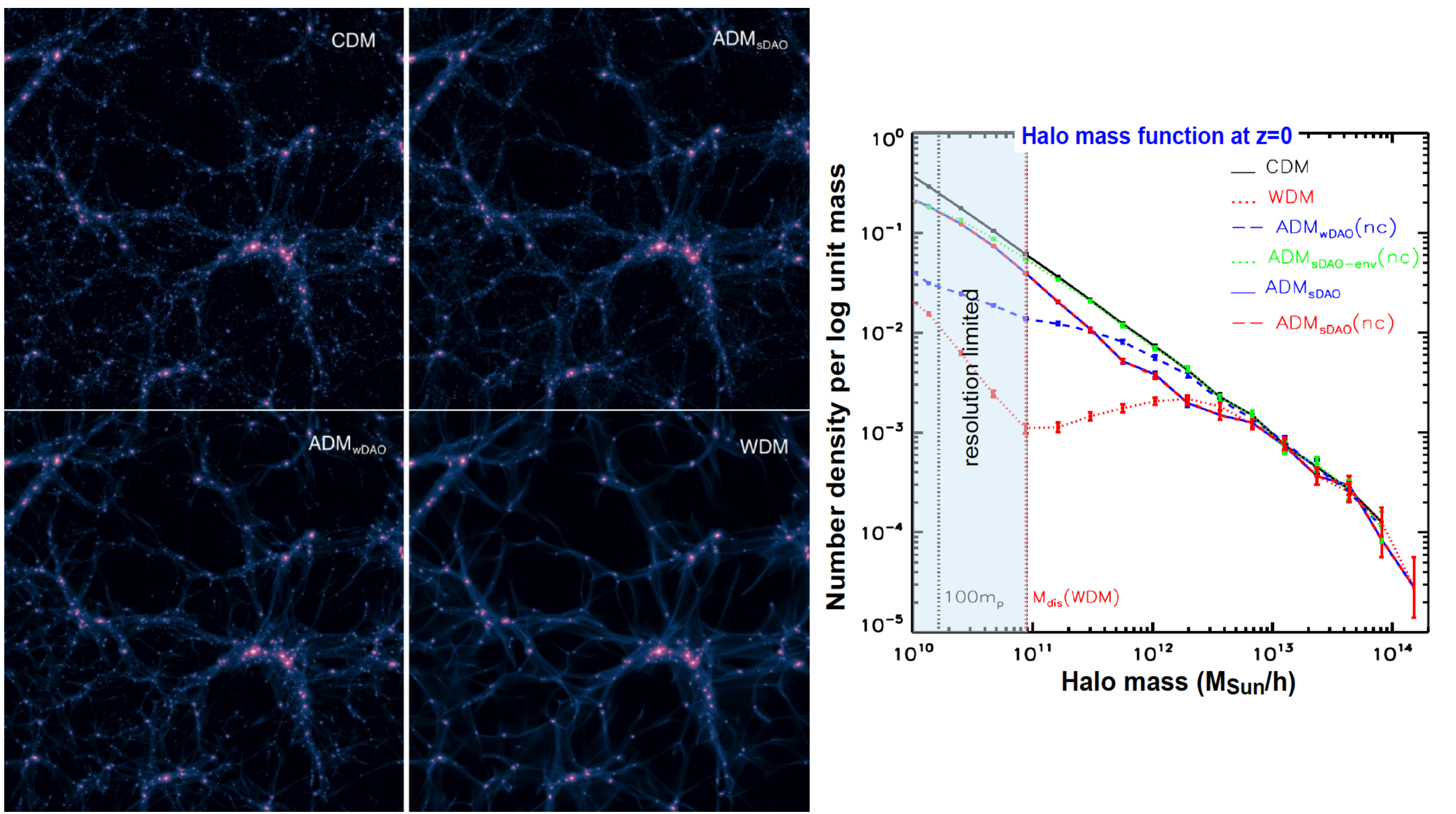}
\caption{Halo mass function for different dark matter models (adapted
  from \cite{Buckley2014}). {\it Left:} The large-scale dark matter
  distribution in a slab of a 64 Mpc/h cube in different cosmologies:
  CDM and WDM in the top left and bottom right, respectively; two
  interacting dark matter models in the other two panels.  {\it
    Right:} The halo mass function at $z=0$ for the models on the
  left. The transparent light blue region marks the resolution limit of the
  simulations.  The cutoff in the primordial linear power spectrum of
  the non-CDM models results in a lower abundance of low-mass haloes,
  visible in the panels on the left and quantified in the halo mass
  function on the right.}
\label{fig_halo_mass_function}
\end{figure}

 The shape of the halo mass function is reasonably well understood
 from statistical arguments based on the properties of the initial
 Gaussian density field (described by the power spectrum) and the
 gravitational collapse of density perturbations into virialised
 haloes as modelled by the spherical collapse model. These arguments
 are the basis of the Press-Schechter \citep{Press1974} and extended
 Press-Schechter (EPS) formalisms \cite{Bond1991,Bower1991}, which
 provide a good fit to the simulation results, particularly if the
 assumption of spherical symmetry for the collapse of overdensities is
 replaced by the assumption of ellipsoidal collapse \cite{Sheth2001}.
 At the small-mass end, the power-law form of the mass function is
 broken at a mass that depends on the nature of the dark matter. For
 example, a cutoff in the power spectrum, whether due to relativistic,
 collisional, or quantum effects, introduces a corresponding cut-off
 in the halo mass function. The mass function for WDM
 \citep[e.g.][]{Schneider2013,Angulo2013,Lovell2014,Bose2016} and
 interacting dark matter
 \citep[e.g.][]{Buckley2014,Schew2015,Vogelsberger2016} models have
 now been fairly well characterized with $N$-body simulations (with
 appropriate corrections for spurious fragmentation due to particle
 discreteness near the cutoff \cite{Wang2007,Lovell2014}). The
 Press-Schechter approach can be readily extended to provide a
 reasonable approximation to the halo mass function in these models as
 well \citep[e.g.][]{Benson2013,Leo2018,Sameie2018}.

Fig.~\ref{fig_halo_mass_function} provides an example of the effect of
a cutoff in the primordial power spectrum on the halo mass function
relative to CDM. The ``atomic dark matter model'', ADM$_{\rm sDAO}$
\cite{CyrRacine2013}, is an example of a model with dark acoustic
oscillations, while WDM is an well-known example of the free streaming
effect (see Fig.~\ref{linear_pk}). These two models give rise to
qualitatively different types of suppression in the abundance of
low-mass haloes. The halo mass function thus contains a signature of
the type of primordial power spectrum cutoff.

\begin{figure}[H]
\centering
\includegraphics[width=15 cm, height=8cm,trim=0.15cm 0.15cm 0.15cm 0.15cm, clip=true]{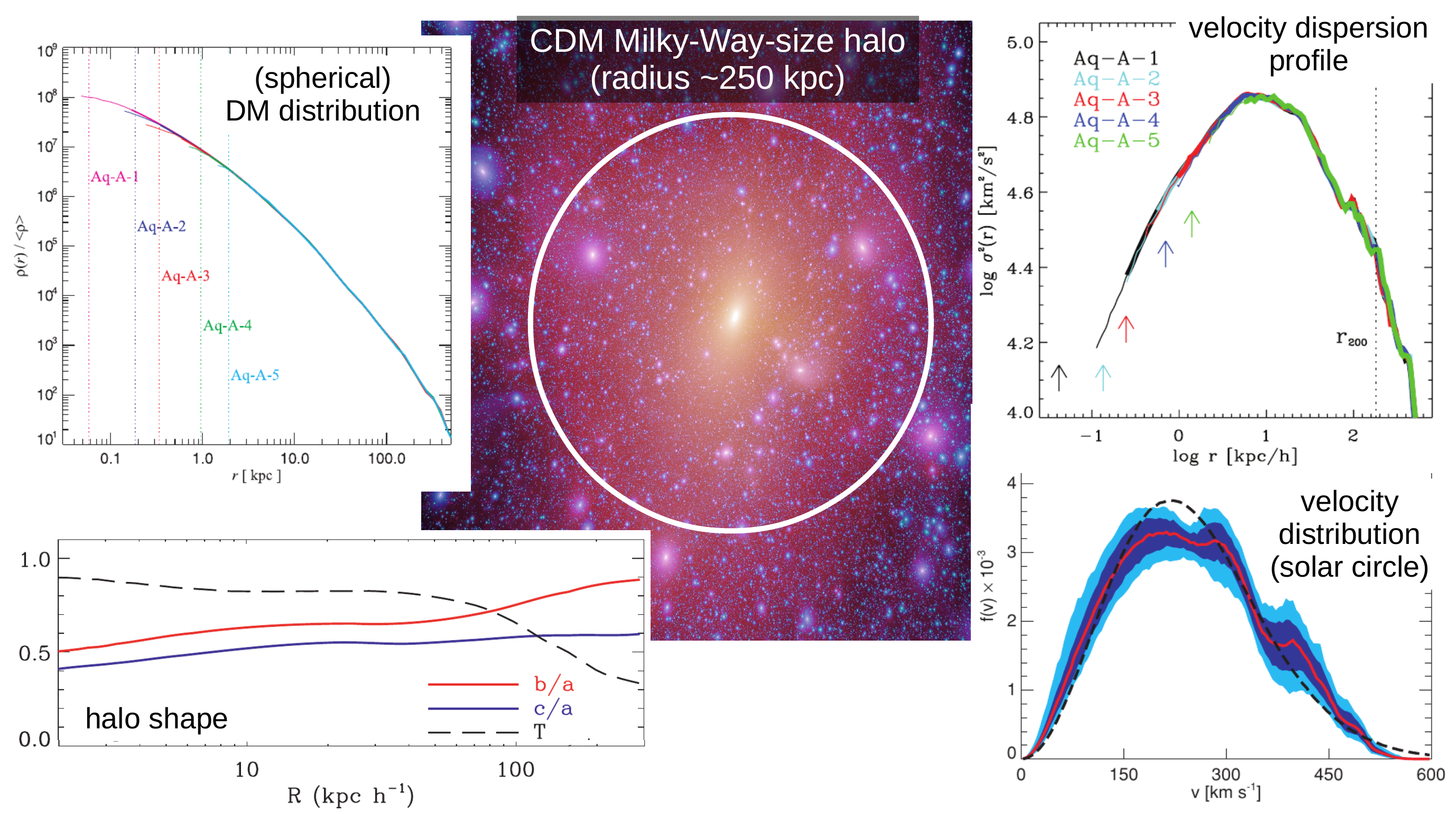}
\caption[Caption without FN]{The structure of CDM haloes. The different panels show
  several characteristics of the spatial (left) and dynamical (right)
  structure of a Milky Way-size CDM halo
  ($M_{200}\sim1.8\times10^{12}$~M$_\odot$; $r_{200}\sim250$~kpc) from
  the Aquarius project \cite{Springel2008}. The top panels show the
  spherically averaged radial density (left; \cite{Springel2008}\footnotemark) and
  velocity dispersion (right; \cite{Navarro2010}\footnotemark) profiles, which are
  nearly universal for haloes in dynamical equilibrium. The bottom
  panels show the halo shape (left: moment of inertia axis ratios, and
  triaxiality: $T=(a^2-b^2)/(a^2-c^2)$; \cite{VeraCiro2011}\footnotemark) and local
  dark matter velocity distribution near the solar circle:
  $2~{\rm kpc}<r<9~{\rm}~{\rm kpc}$ (right; \cite{Vogelsberger2009}\footnotemark).}
\label{fig_MW_structure}
\end{figure}

{\it The inner structure of dark matter haloes.---} One of the
remarkable findings of the past few decades is that the spherically
averaged mass density profiles of dark matter haloes in dynamical
equilibrium have a nearly universal form which is independent 
of halo mass, initial conditions\footnote{This is only true
  if, on the scales of interest, the primordial power spectrum grows
  monotonically towards large $k$.}  and cosmological
parameters. These profiles are quite well described by a very simple
functional form with just two parameters, the so-called
Navarro-Frenk-White (NFW) profile \cite{NFW1996,NFW1997}:
 \begin{equation}\label{eq_NFW}
 \rho_{\rm NFW}(x)=\frac{\rho_s(c)}{cx(1+cx)^2,}
 \end{equation}
 where $x=r/r_{200}$, and $c=r_{200}/r_s$ is the concentration of the
 halo; $r_s$ is the scale length, which, for the NFW profile, coincides
 with the radius, $r_{-2}$, at which the logarithmic slope of the
 profile is equal to $-2$; finally
 $\rho_s(c)=\delta_c\rho_{\rm crit}$, where
 $\rho_{\rm crit}=3H^2/8\pi G$ is the critical density of the
 Universe, and:
\begin{equation}\label{eq_delta_c}
\delta_c=\frac{\Delta c^3}{3K_c(c)}
\end{equation}  
where $K_c(c)={\rm ln}(1+c)-c/(1+c)$. Although recent simulations have
shown that a different profile, the so-called Einasto profile, which
has three parameters, is a slightly better fit to simulations
\cite{Navarro2010}, the asymptotic behaviour of the NFW profile for
$\rho({r\rightarrow0})\sim r^{-1}$ remains a remarkably good
approximation to the inner structure of CDM haloes (see top left panel
of Fig.~\ref{fig_MW_structure}). The physical origin of this divergent
{\it cusp} and the remarkably universal profile shape are not fully
understood.  It has been argued from $N$-body simulations of the early
stages of structure formation that the first CDM haloes to form, i.e.,
those near the free-streaming scale of CDM have a steeper cusp than
NFW, $\sim r^{-1.5}$, which is subsequently flattened after a few
mergers to $\sim r^{-1}$
\cite{Anderhalden2013,Ishiyama2014,Angulo2017,Delos2018}. More recent
simulations which follow the growth of the first mini-haloes all the
way to the present, seem to confirm this, suggesting that the
ubiquitous $\sim r^{-1}$ slope develops at some point after the
formation of the halo and remains until $z=0$.
%is present already as soon as the halo forms (Wang et al., in preparation).}

{\it Halo concentration.---}The remarkable simplicity of the NFW halo
density profile goes beyond Eq.~\ref{eq_NFW}: the profile is, in fact,
fully specified by a single parameter, halo mass, because the
concentration (or scale radius) correlates with mass, with lower-mass
haloes generally being more concentrated than higher-mass haloes
\cite{NFW1997,Bullock2001,Eke2001,Wechsler2002,Neto2007,Zhao2009,Prada2012,Ludlow2014,SC2014,Diemer2015,Klypin2016,Pilipenko2017}. This
correlation is ultimately a consequence of the hierarchical nature of
structure formation by gravitational instability from a primordial
power spectrum that, as in CDM, grows monotonically towards small
scales (see Fig.~\ref{linear_pk}). Lower-mass haloes form earlier,
when the mean density of the Universe is larger, and larger-mass
haloes form later when the mean density of the Universe is lower. The
inner regions of haloes collapse first \cite{Wang2011} and their
density reflects the mean density of the Universe at that time. Hence,
smaller-mass haloes are more concentrated than larger-mass
haloes. Furthermore, since (at least for CDM), the power spectrum,
$\Delta^2(k)$, becomes flatter at larger $k$ (due to the M\'esz\'aros
effect), haloes with a wide range of masses collapse in a short time
interval and this flattens the concentration-mass relation at low
masses. Models based on these simple arguments explain, at some level,
the concentration-mass relation measured in simulations
\cite{NFW1997,Prada2012,Klypin2016}, and also provide a natural
connection between the mass assembly of haloes in time and their
structure as a function of radius: each radial shell has the
characteristic density of the cosmic background density at the time
when it collapses \cite{Ludlow2014}.  Random deviations around the
mean collapse time expected for haloes of a fixed mass and a
stochastic merger history introduce scatter in the concentration-mass
relation.

\addtocounter{footnote}{-4}
 \stepcounter{footnote}\footnotetext{\scriptsize{Reproduced from Volker Springel et al. The Aquarius Project: the subhaloes of galactic haloes. MNRAS (2008) 391 (4): 1685–1711, doi: 10.1111/j.1365-2966.2008.14066.x. By permission of Oxford University Press on behalf of the Royal Astronomical Society. For the original article, please visit the following \href{https://academic.oup.com/mnras/article/391/4/1685/1747035}{link}.}} %This figure is not included under the CC-BY license of this publication. For permissions, please email: journals.permissions@oup.com}

 \stepcounter{footnote}\footnotetext{\scriptsize{Reproduced from Julio Navarro et al. The diversity and similarity of simulated cold dark matter haloes. MNRAS (2010) 402 (1): 21–34, doi: 10.1111/j.1365-2966.2009.15878.x. By permission of Oxford University Press on behalf of the Royal Astronomical Society. For the original article, please visit the following \href{https://academic.oup.com/mnras/article/402/1/21/1028856}{link}.}} %This figure is not included under the CC-BY license of this publication. For permissions, please email: journals.permissions@oup.com}

 \stepcounter{footnote}\footnotetext{\scriptsize{Reproduced from Carlos Vera-Ciro et al. The shape of dark matter haloes in the Aquarius simulations: evolution and memory. MNRAS (2011) 416 (2): 1377–1391, doi: 10.1111/j.1365-2966.2011.19134.x. By permission of Oxford University Press on behalf of the Royal Astronomical Society. For the original article, please visit the following \href{https://academic.oup.com/mnras/article/416/2/1377/1061105}{link}.}} %This figure is not included under the CC-BY license of this publication. For permissions, please email: journals.permissions@oup.com}
 
  \stepcounter{footnote}\footnotetext{\scriptsize{Reproduced from Mark Vogelsberger et al. Phase-space structure in the local dark matter distribution and its signature in direct detection experiments. MNRAS (2009) 395 (2): 797–811, doi: 10.1111/j.1365-2966.2009.14630.x. By permission of Oxford University Press on behalf of the Royal Astronomical Society. For the original article, please visit the following \href{https://academic.oup.com/mnras/article/395/2/797/1747020}{link}.}} %This figure is not included under the CC-BY license of this publication. For permissions, please email: journals.permissions@oup.com}

\blfootnote{}
\blfootnote{\scriptsize{The figures mentioned in footnotes 21$-$24 are not included under the CC-BY license of this publication. For permissions, please email: journals.permissions@oup.com}}

{\it Halo velocity distribution.---} For a spherical,
self-gravitating, collisionless system in dynamical equilibrium, with
radial density profile, $\rho(r)$, the Boltzmann equation reduces to
the well-known Jeans equation \cite{BT2008}:
\begin{equation}
\dfrac{d(\rho\sigma_r^2)}{dr}+2\frac{\beta}{r}\rho\sigma_r^2=-\rho\dfrac{d\Phi}{dr},
\end{equation}
where $\Phi(r)$ is the gravitational potential related to the density
by Poisson's equation, $\sigma_r(r)$ is the radial velocity dispersion
profile, and $\beta(r)=1-(\sigma_\theta^2+\sigma_\phi^2)/\sigma_r^2$
is the velocity anisotropy profile, which quantifies the degree of
anisotropy of the particle orbits in the dark matter halo. Haloes tend
to be isotropic only in their centres and are radially anisotropic in
their outskirts \cite{Navarro2010,Ludlow2011}, with a velocity
anisotropy that is related to the logarithmic slope of the density
profile \cite{Hansen2006}. The velocity structure of dark matter
haloes in equilibrium is thus intimately linked to their spatial
distribution (see top right panel of Fig.~\ref{fig_MW_structure}),
which is strikingly evident in the so-called {\it pseudo-phase-space
  density}, $Q\equiv\rho/\sigma^3$, where
$\sigma^2=\sigma^2_r+\sigma_\theta^2+\sigma^2_\phi$ is the square of
the 3D velocity dispersion. This quantity has been found to be an
almost perfect power law, $Q\propto r^{-1.875}$, over several orders
of magnitude \cite{Taylor2001,Navarro2010}, and is in remarkable
agreement with the self-similar solution for infall onto a point mass
in an Einstein-de Sitter universe \cite{Bert1985}. The radial
behaviour of $Q$ is a manifestation of the nearly universal structure
of dark matter haloes, which connects both their spatial and
kinematical distributions.

Besides having anisotropic particle orbits, CDM haloes have a 
non-Maxwellian velocity distribution. This may be seen in a simple way
by noting that for a purely isotropic spherical system ($\beta=0$), the
full velocity distribution function of the halo depends only on the
specific energy, $f(\mathcal{E})$, and is fully given
by the Eddington formula \cite{Eddington1916}:
\begin{equation}\label{Eddington}
f(\mathcal{E})=\frac{1}{\sqrt{8}\pi^2}\int_{0}^{\sqrt{\mathcal{E}}}du\frac{d^2\rho}{d\Psi^2}(r(\Psi(u))),
\end{equation}
where $u = \sqrt{\mathcal{E}-\Psi}$ and $\mathcal{E}$ and $\Psi(r)$
are the (negative) specific energy and gravitational potential,
respectively. If haloes were spherical and isotropic, their velocity
distribution would be given purely by the NFW density profile through
Eq.~\ref{Eddington}. Even at this level of approximation we can see
that haloes would not be described by a Maxwellian distribution
function (only the singular isothermal sphere, $\rho\propto r^{-2}$,
results in a Maxwellian distribution in Eq.~\ref{Eddington}). In
other words, haloes are non-Maxwellian simply by virtue of their NFW
density profiles. In fact, CDM haloes in simulations have {\it
  local}\footnote{Given the limited resolution of simulations, local
  in this sense refers to regions of at least $\mathcal{O}$(10
  kpc$^3$) as in \cite{Vogelsberger2009}.}  velocity distributions
that show significant departures from Maxwellian, related to the
dynamical assembly of the halo.  The features that appear in the local
velocity distribution are unique for a particular halo, and retain the
memory of its assembly history (\cite{Vogelsberger2009}; see bottom
right panel of Fig.~\ref{fig_MW_structure}).

{\it Halo shapes.---} Although to first order, CDM haloes are well
described by the spherical NFW profile, they are, in fact, triaxial
\cite{Frenk1988}. In general, CDM haloes are prolate in the inner
parts and oblate in the outskirts
(\cite{Frenk1988,Jing2002,Hayashi2007,VeraCiro2011}; see bottom left
panel of Fig.~\ref{fig_MW_structure}). This radial dependence of halo
shape seems to be related to the assembly history of the halo within
the cosmic web: the central regions, being assembled at earlier times
through accretion along narrow filaments, end up being more prolate,
while the outskirts, more recently assembled by less anisotropic
accretion end up more oblate \cite{VeraCiro2011,Punya2018}. Thus, the
halo shape profile at $z=0$ carries some memory of its assembly
history. Overall, more massive haloes are more aspherical than lower
mass haloes \cite{Bonamigo2015,Vega2017} because in the hierarchical
CDM model, the more massive haloes form more recently and thus their
shapes retain memory of the most recent accretion event
\cite{Despali2014}.

{\it Dependence on the nature of the dark matter.---} There are
significant changes in the structure of dark matter haloes if the dark
matter particles do not behave as CDM. In currently allowed models, these
deviations are mostly confined to the central regions, that is within
the scale radius, $r_s$. By introducing a new scale in the process of
structure formation, be it in the initial conditions (e.g. a cutoff in
the linear power spectrum), or during the non-linear evolution phase
(e.g. a subgalactic scale mean free path due to self-interactions),
these models break the near universality of CDM haloes. 

In models with a {\it galactic-scale cutoff in the primordial power
  spectrum}, such as WDM and interacting DM, the main changes can be
understood from the later collapse of the first generation of haloes
in these models compared to CDM. In contrast to CDM, these galactic-size haloes are
not formed hierarchically from the assembly of smaller haloes but,
instead, by {\it monolithic} collapse.  Their characteristic density
therefore reflects the mean background density at the time of the
monolithic collapse. By contrast, 
\pagebreak
%\clearpage

\begin{figure}[!htb]
\vspace{-1.07\baselineskip}
	\subfigure{\includegraphics[height=7.5cm,width=15cm,trim=-1.0cm 0cm -0.3cm 0cm, clip=true]{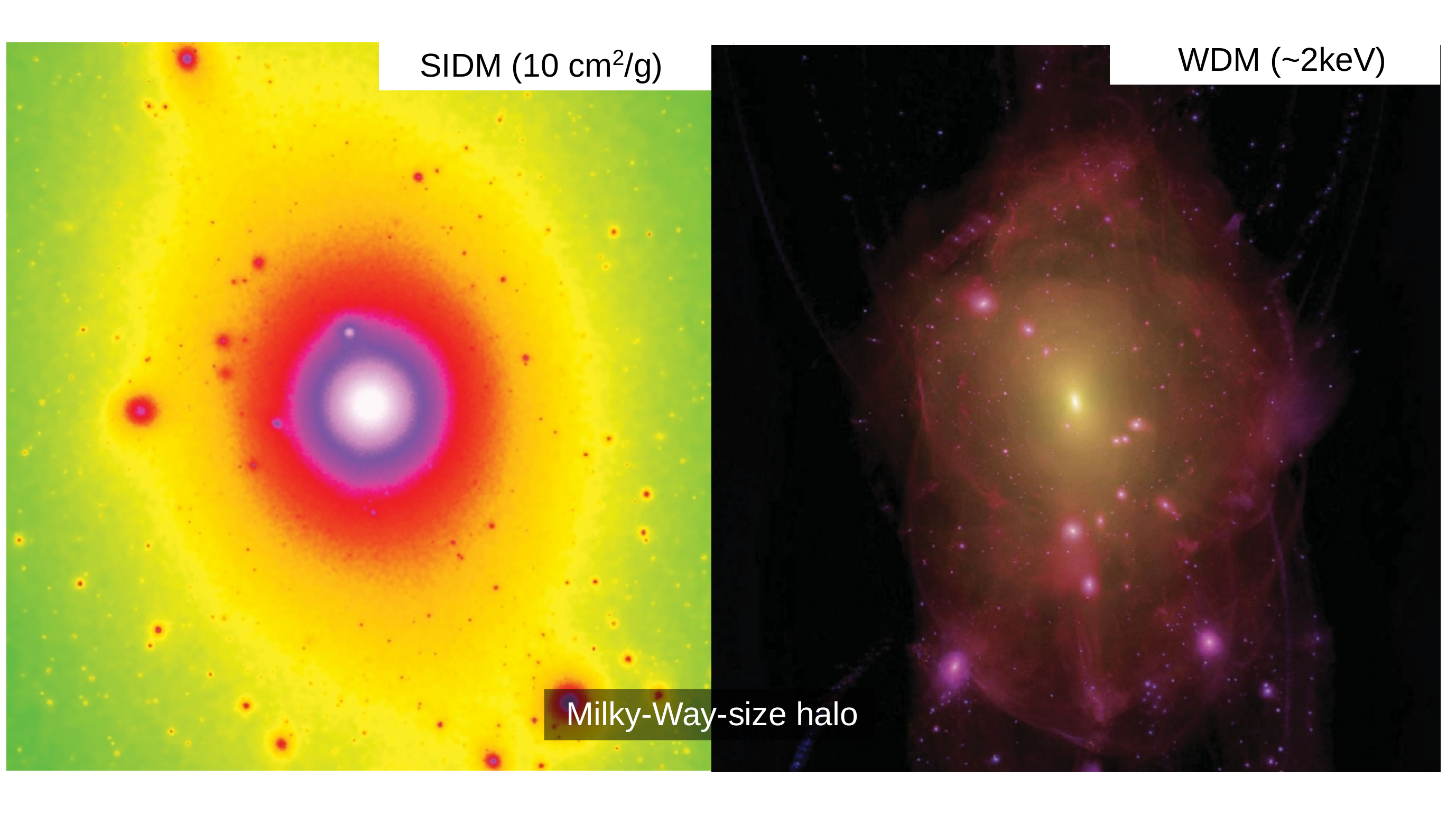}}
 	\subfigure{\includegraphics[height=7.0cm,width=7.5cm,trim=1.95cm 0.5cm 0.35cm 1.0cm, clip=true]{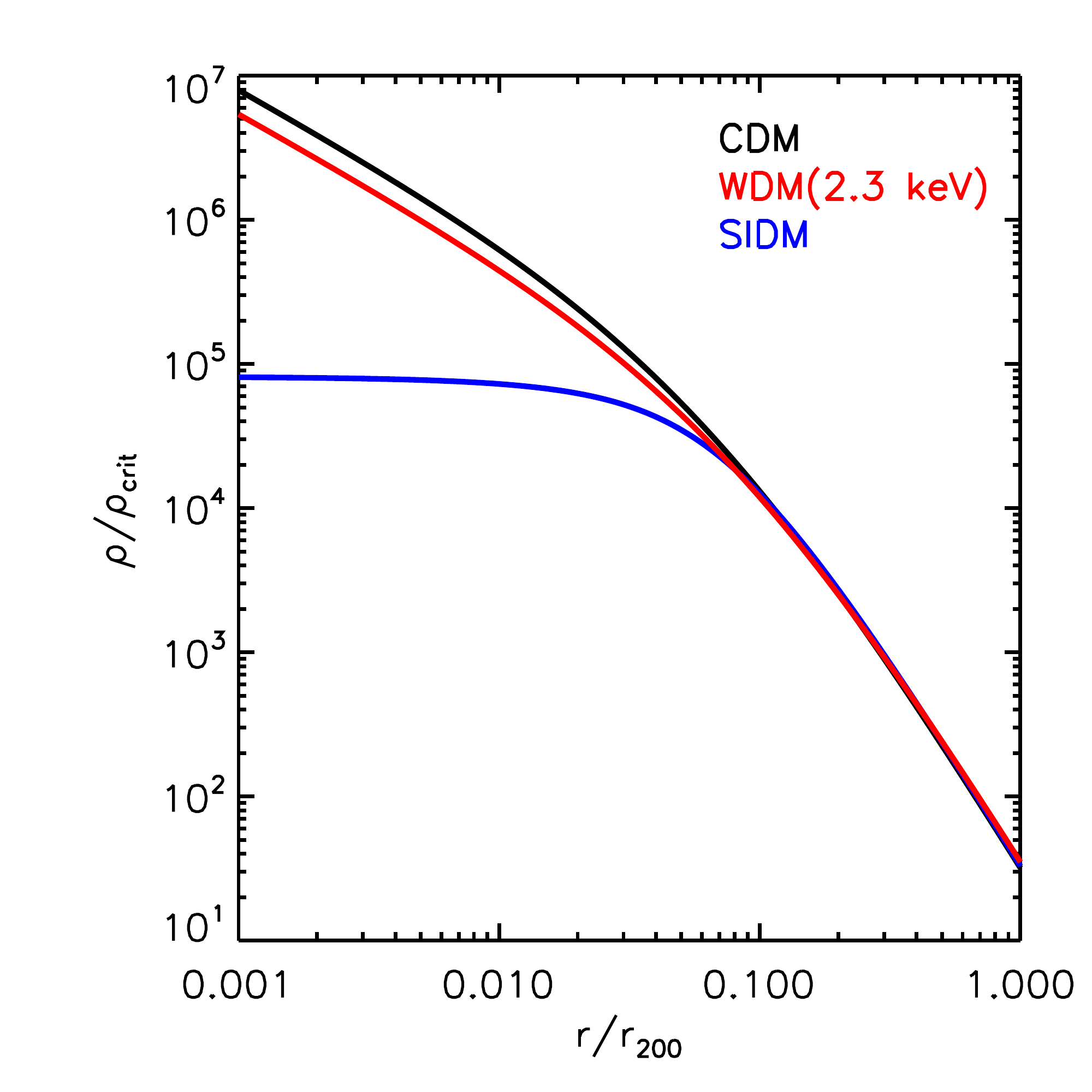}} 
    	\subfigure{\includegraphics[height=7.0cm,width=7.5cm,trim=1.75cm 0.5cm 0.35cm 1.0cm, clip=true]{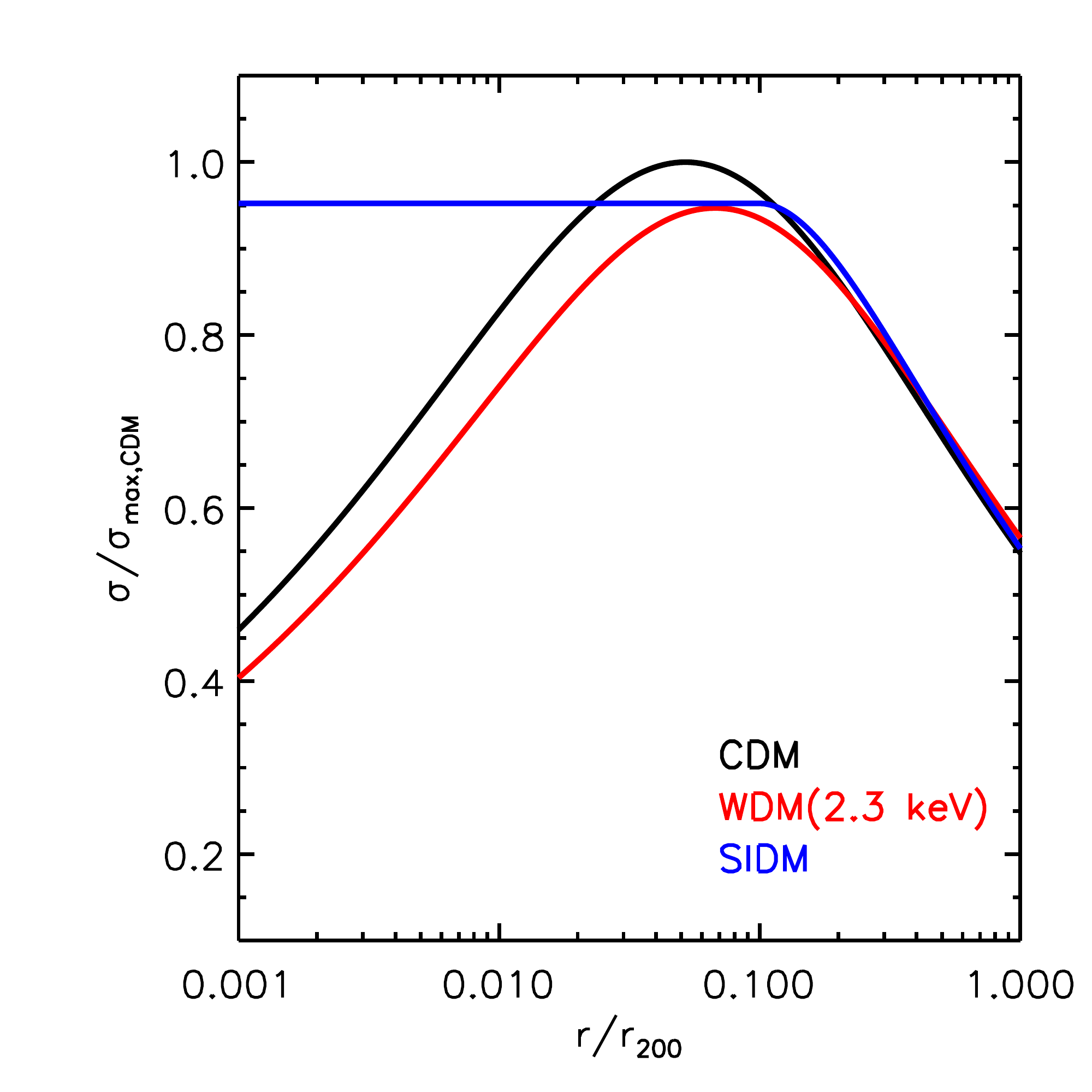}}
        \caption[Caption without FN]{Structure of haloes in models with different types of
          dark matter: collisional (SIDM;
          $\sigma_T/m_\chi\gtrsim1$~cm$^2$/g) and with a
          galactic-scale free-streaming cutoff (WDM;
          $m_\chi\sim2.3$~keV).  {\it Upper panels:} projected dark
          matter distribution of a Milky Way-size halo in the SIDM
          model (left panel; \cite {Vogelsberger2012}\protect\footnotemark) and in the WDM
          model (right panel; \cite{Lovell2012}\protect\footnotemark). {\it Bottom left:}
          spherically averaged density profiles. WDM haloes are well
          described by an NFW profile, but have lower concentrations
          than their CDM counterparts of the same mass; SIDM haloes
          develop flat density cores during a transient stage that
          inevitably ends with the collapse of the core once the
          gravothermal catastrophe is triggered. {\it Bottom right:}
          spherically averaged velocity dispersion profiles. WDM
          haloes still obey the universal scaling for the
          pseudo-phase-space density,
          $\rho/\sigma^3\propto r^{-1.875}$, at most radii, except in
          the very centre, which results from a
          similar velocity dispersion profile to that in CDM but
          shifted downwards and to the right as a result of the lower
          concentration. SIDM haloes develop isothermal density cores
          of size of the order of the scale radius.}
          \vspace{-0.5\baselineskip}
\label{fig_structure_nonCDM}
\end{figure}

\addtocounter{footnote}{-2}
 \stepcounter{footnote}\footnotetext{\scriptsize{Reproduced from Mark Vogelsberger et al. Subhaloes in self-interacting galactic dark matter haloes. MNRAS (2012) 423 (4): 3740–3752, doi: 10.1111/j.1365-2966.2012.21182.x. By permission of Oxford University Press on behalf of the Royal Astronomical Society. For the original article, please visit the following \href{https://academic.oup.com/mnras/article/423/4/3740/1749150}{link}. This figure is not included under the CC-BY license of this publication. For permissions, please email: journals.permissions@oup.com}}

 \stepcounter{footnote}\footnotetext{\scriptsize{Reproduced from Mark Lovell et al. The haloes of bright satellite galaxies in a warm dark matter universe. MNRAS (2012) 420 (3): 2318–2324, doi: 10.1111/j.1365-2966.2011.20200.x. By permission of Oxford University Press on behalf of the Royal Astronomical Society. For the original article, please visit the following \href{https://academic.oup.com/mnras/article/420/3/2318/979379}{link}. This figure is not included under the CC-BY license of this publication. For permissions, please email: journals.permissions@oup.com}}
 
%\clearpage 
\noindent a CDM halo of the same mass forms
from the assembly of smaller fragments that typically formed earlier
and are therefore denser. Simulations of WDM models have characterised
the spatial structure of WDM haloes quite accurately, showing that the
density profiles of allowed models are, in fact, well described by the
NFW profile but with a lower concentration at a given mass
\cite{Colin2000,Avila-Reese2001,Colin2008,Schneider2012,Lovell2014,Bose2016}.
The concentration-mass relation for WDM haloes can then be modeled in
an analogous way to CDM, but taking into account the cutoff in the
power spectrum \cite{Ludlow2016} which is therefore reflected in the
concentration of the haloes.  An example of this is shown in the
bottom left panel of Fig.~\ref{fig_structure_nonCDM}, where the
density profile of a CDM halo is mapped into that of a 2.3 keV thermal
relic WDM halo by simply scaling down the concentration using Eq.~39
of Ref.~\cite{Schneider2012}, which connects the concentration to the
cutoff scale in the power spectrum\footnote{This functional form has
  been corroborated by \cite{Bose2016}, but the parameters in the two
  studies are different.  The formula is nevertheless a good
  approximation to the general behaviour.}. This lower concentration
is also reflected in the velocity dispersion profile (see bottom-right
panel of Fig.~\ref{fig_structure_nonCDM}).

It is interesting to note that the pseudo-phase-space density,
$Q\equiv\rho/\sigma^3$, in WDM haloes scales with radius in the same
way as in CDM, $Q\propto r^{-1.9}$. In principle, $Q$ can never exceed
its primordial value, $Q=Q_{\rm max}$, determined by the {\it thermal}
velocities of the unclustered dark matter particles
\cite{Dalcanton2001}. This is because for a collisionless system,
Liouville's theorem requires conservation of the fine-grained
phase-space density and the coarse-grained density, approximated by
$Q$, can never exceed this value. Thus, the central regions of WDM
haloes cannot exceed a maximum density, i.e., they form a central
density core. However, the value of $Q_{\rm max}$ is so large in
allowed WDM models that the core size is tiny, $\mathcal{O}$(10~pc)
for $\sim$~keV thermal WDM relics in dwarf-size haloes
\cite{Maccio2012,Shao2013}; WDM cores are thus irrelevant in
practice. Finally, WDM haloes are slightly less triaxial than CDM
haloes as a whole for a fixed mass, at masses near the cutoff scale
\cite{Bose2016}.

In SIDM, collisions between the dark matter particles have an on 
impact in the inner structure of haloes once the timescale for
collisional relaxation at the characteristic radius of the halo $r_s$,
$t_{\rm rel, s}\sim(\rho(r_s) \langle v_{\rm rel,
  s}\rangle\sigma_T/m_\chi)^{-1}$, where
$\langle v_{\rm rel, s}\rangle$ is the characteristic local relative
velocity, becomes comparable to the age of the inner halo. The
original CDM density cusp turns into a core within the region where
this condition is satisfied. The interaction cross-section thus
introduces a new scale in structure formation -- the mean free-path
for particle collisions -- which breaks the near universality of CDM
haloes. The transformation of the cusp into a core due to elastic
collisions at the halo centre is a transitory phase that leads to a
quasi-equilibrium configuration once the core has acquired its maximum
size, which is approximately the radius at which the velocity
dispersion profile peaks (see bottom panels in
Fig.~\ref{fig_structure_nonCDM}). Prior to this, the transfer of
energy during elastic collisions proceeds from the outside in since
the velocity dispersion profile has a positive gradient in the inner
regions and so there is a net ``heat flux'' from the regions close to
the maximum of the velocity dispersion to the centre
(e.g. \cite{Colin2002}). Once the core reaches its maximum size,
subsequent collisions can only result in a net heat flux from the
inside out since the velocity dispersion profile has a negative slope
in the outer regions. This condition triggers the gravothermal
collapse of the central parts of the SIDM halo, which results in the
contraction of the core to form a new cusp, ultimately collapsing into
a black hole \cite{Balberg2002,Koda2011}\footnote{The gravothermal
  collapse \cite{LyndenBell1968} is a familiar process in globular
  clusters, where the inner regions have negative specific heat that
  is smaller than the positive specific heat in the outer regions. In the
  case of globular clusters, the collapse can be prevented by the
  formation of binary stars at the centre. In the case of a SIDM
  halo, since the interactions are purely elastic, the process is expected
  to continue until a black hole forms. The black hole efficiently accretes the
  inner core of the SIDM halo (e.g. \cite{Pollack2015}). This
  discussion refers strictly to elastic self-scattering. If collisions
  are inelastic, then the energy released needs to be taken into
  account and, in fact, it could prevent the gravothermal collapse;
  see \cite{Vogelsberger2019}.}.

For $0.1\lesssim\sigma_T/m_\chi\lesssim10~$cm$^2$/g, cosmological
$N$-body simulations have shown that SIDM haloes today should have
cores of size of the order of the scale radius
\cite{Yoshida2000,Dave2001,Colin2002,Vogelsberger2012,Rocha2013}. At
the lower end of this range of cross-sections, SIDM cores are small,
$\sim0.2r_s$ \cite{Rocha2013}, making SIDM haloes only slightly
different from their CDM counterparts. This is why below this
cross-section, SIDM is essentially indistinguishable from CDM as a
theory of structure formation \cite{Zavala2013}. At the higher end of
the cross-section range, the core sizes are slightly larger than the
scale radius and approach the full thermalization of the core, with a
maximum size bounded by the radius at which the velocity dispersion
peaks (a case like this is shown in the bottom panels of
Fig.~\ref{fig_structure_nonCDM}). Within the thermalised region, the
orbits of dark matter particles are isotropised by collisions, erasing
most of the memory of the assembly of the central regions
\cite{Brinckmann2018}. This makes haloes centrally rounder than
their CDM counterparts \cite{Peter2013} and causes them to have local
velocity distributions that are close to Maxwellian
\cite{Vogelsberger2013}. Since the onset of the gravothermal collapse
phase is expected to be $\sim(250-400)t_{\rm rel,s}$
\cite{Pollack2015}, the core phase of SIDM haloes in this range of
cross-sections is relatively long-lived.

\section{Halo mergers and the emergence of subhaloes}\label{sec_subhaloes}

\noindent In the previous section we reviewed the structural properties of dark
matter haloes in CDM and in well-known alternatives such as WDM and
SIDM. In this section we focus on dark matter subhaloes. Since these
exhibit similar structural properties to haloes, modified by a few
relevant physical processes, we draw extensively on the results of the
previous section. Our goal now is to describe these processes and how
they affect the abundance and structure of subhaloes.

\begin{figure}[H]
\centering
\vspace{-0.2\baselineskip}
\includegraphics[width=15 cm, height=8cm,trim=1.5cm 0.15cm 0.15cm 0.15cm, clip=true]{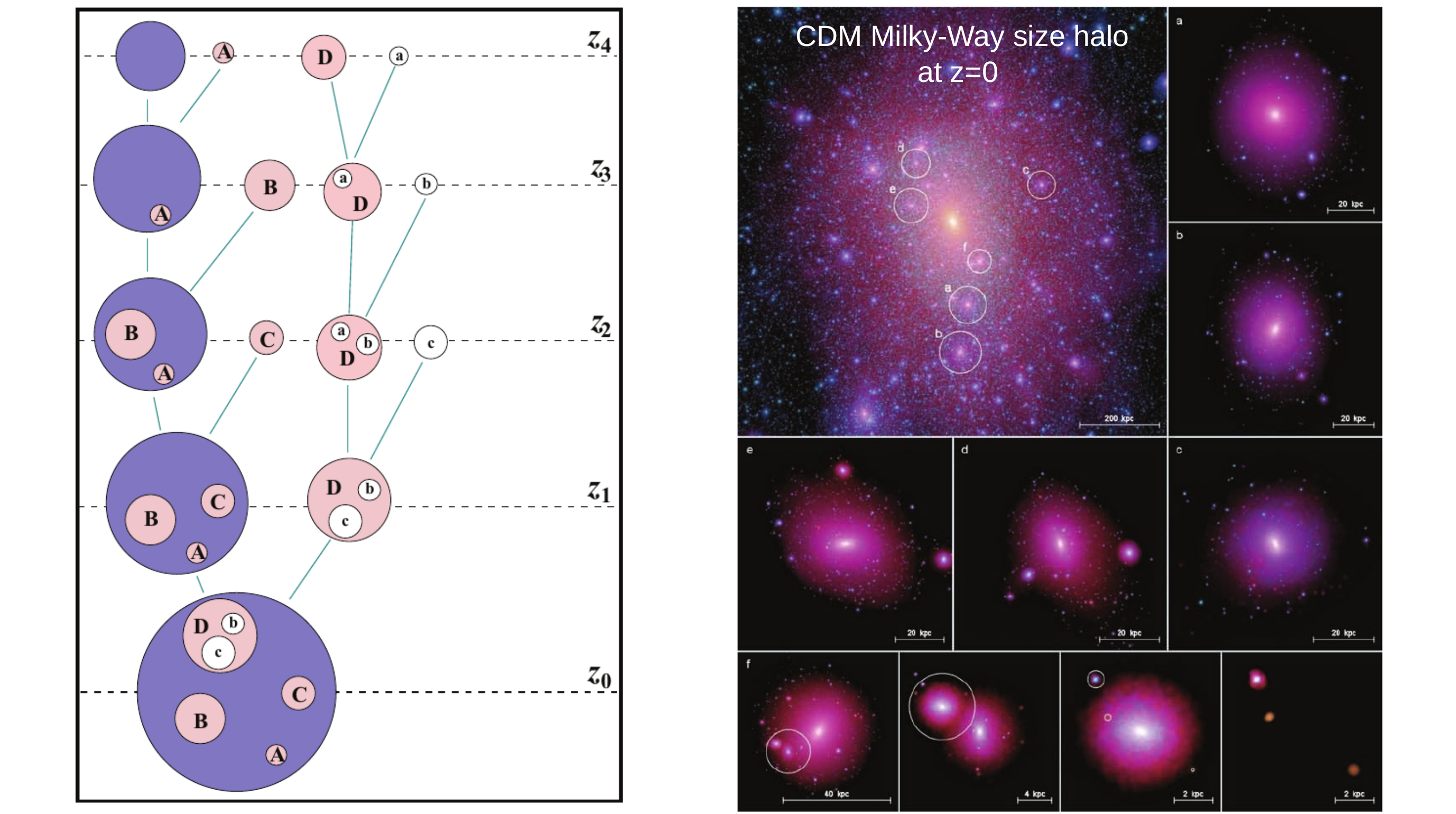}
\caption[Caption for LOF]{Dark matter subhaloes. {\it Left:} schematic representation
  of a dark matter halo {\it merger tree} (taken from
  \cite{Giocoli2010}\footnotemark) at discrete redshifts. In a hierarchical model,
  haloes grow by the accretion of smaller neighbouring haloes
  (A,B,C,D), which become subhaloes at the time when they first cross
  the virial radius of the host halo. The main branch of the tree
  represents the evolution of the main progenitor (shown in
  blue). Since this process occurs across the entire hierarchy of
  structures, there are subhaloes within subhaloes (sub-subhaloes;
  like a, b, c in system D) and so on. {\it Right:} a simulated Milky Way-size
  CDM halo from the Aquarius project (figure taken from \cite{Springel2008}\footnotemark; this is the
  same halo illustrated in Fig.~\ref{fig_MW_structure}). The circles
  in the main image mark six subhaloes that are shown enlarged in the
  surrounding panels, %, and in the bottom left panel, 
  as indicated by the labels.  Sub-subhaloes are clearly visible (corresponding to the
  configuration illustrated in the last step, $z_0$, in the left
  panel). The bottom row shows several generations of sub-subhaloes
  contained within subhalo f.}
\label{fig_subhaloes}
\vspace{-0.3\baselineskip}
\end{figure}

\addtocounter{footnote}{-2}
  \stepcounter{footnote}\footnotetext{\scriptsize{Reproduced from Carlo Giocoli et al. The substructure hierarchy in dark matter haloes . MNRAS (2010) 404 (1): 502–517, doi: 10.1111/j.1365-2966.2010.16311.x. By permission of Oxford University Press on behalf of the Royal Astronomical Society. For the original article, please visit the following \href{https://academic.oup.com/mnras/article/404/1/502/3101607}{link}. This figure is not included under the CC-BY license of this publication. For permissions, please email: journals.permissions@oup.com}}

 \stepcounter{footnote}\footnotetext{\scriptsize{Reproduced from Volker Springel et al. The Aquarius Project: the subhaloes of galactic haloes. MNRAS (2008) 391 (4): 1685–1711, doi: 10.1111/j.1365-2966.2008.14066.x. By permission of Oxford University Press on behalf of the Royal Astronomical Society. For the original article, please visit the following \href{https://academic.oup.com/mnras/article/391/4/1685/1747035}{link}. This figure is not included under the CC-BY license of this publication. For permissions, please email: journals.permissions@oup.com}}

\subsection{Halo mass assembly: smooth accretion vs mergers}\label{sec_smooth_mergers}

Haloes grow by accreting dark matter, either through mergers with
smaller haloes or by accretion of diffuse, smooth material. The
importance of each of these channels depends on the shape of the
primordial power spectrum and on the smallest mass halo that can be
formed, both of which, in turn, depend on the nature of dark matter
particles. For instance, in WIMP CDM models, the minimum halo mass is
in the range $(10^{-12}-10^{-6}$)~M$_\odot$
\cite{Green2005,Bringmann2009}, which is many orders of magnitude
below the resolution of current cosmological simulations. Thus, in
reality, the amount of smooth accretion measured in a simulation
consists of a combination of true unclustered dark matter and
unresolved dark matter haloes
(e.g. \cite{Genel2010,Wang2011}). However, estimates based, for
example, on the excursion set formalism can be used to extend the
results of simulations into the unresolved regime.  In the resolved
regime, the analytical expectations are in good agreement with
simulations \cite{Angulo2010,Wang2011}. These calculations show that the amount
of smooth accretion onto present-day CDM Milky Way-size haloes is in
the range $\sim (10-20)\%$ \cite{Angulo2010, Wang2011}.  Thus, in this
hierarchical structure formation model, haloes today are mainly
composed of the remnants of disrupted smaller haloes.  Of these, major
mergers (i.e., those with progenitor with mass ratios greater than 1:10)
contribute, on average, less than 20\% of the final mass
\cite{Wang2011}.

{\it When does a halo become a subhalo?} We mentioned earlier that the
boundary of a halo is not sharply defined, but rather chosen
approximately to separate the region within which the dark matter is
in dynamical equilibrium from an outer region where the dark matter is
still mostly infalling. In a similar way, the moment at which a halo
becomes a subhalo, i.e., when it crosses this transition region for
the first time, is somewhat arbitrary. For simplicity, it is common to
use the virial radius, e.g. $r_{200}$, as the boundary of the halo and
thus to define a subhalo as a halo that has crossed the virial radius
of a larger halo at some point in the past (see left panel of
Fig.~\ref{fig_subhaloes}). One could argue for a more physical
definition, based for instance in the relevance of the tidal forces
exerted by the dominant halo host in the local environment but, for
simplicity, and because of its common usage, we will use the former,
simple definition of a subhalo. We should remark that it is not
uncommon for subhaloes to leave the boundary of the main halo at some
point after first crossing \cite{Gill2005,Sales2007,Ludlow2009}, and
thus the subhalo population today extends to radii a few times the
current virial radius (\cite{Ludlow2009}; those systems beyond the
virial radius are commonly known as backsplash haloes
\cite{Gill2005}).

\subsection{Evolution of subhaloes: initial conditions}\label{evo}

{\it Halo merger trees and merger rates.--} $N$-body simulations have
been instrumental in determining the mass assembly history of
haloes. A particular powerful tool are halo merger trees (for an
overview of different algorithms to construct these trees see \cite{Srisawat2013}). A halo at
the redshift of interest is regarded as the trunk of the tree and the
merger tree structure consists of a catalogue of progenitors, which
constitute the secondary branches that eventually merge onto the main
branch of the tree (see left panel of Fig.~\ref{fig_subhaloes}). Thus,
a merger tree contains information about the accretion times and
masses of the haloes that eventually become subhaloes. Both of these
properties, together with the corresponding position and velocity
vectors, represent the initial conditions for the subsequent dynamical
evolution of the subhalo.

An interesting statistical property that can be extracted from a
merger tree is the {\it merger rate per halo}, $dN_m/d\xi/dz$
\cite{Fakhouri2008}, which gives the mean number of mergers, $dN_m$, per mass
ratio, $d\xi$ (relative to the main progenitor at the time of
accretion), per redshift interval, $dz$. This quantity has been found to have a
functional form that is nearly universal
\cite{Fakhouri2008,Fakhouri2010}:

\begin{equation}\label{merger_rate}
\dfrac{dN_m}{d\xi dz}(M_0,\xi,z)=A\left(\frac{M_0}{10^{12}~{\rm M}_\odot}\right)^\alpha\xi^\beta {\rm exp}\left[\left(\frac{\xi}{\tilde{\xi}}\right)^\gamma\right](1+z)^{\eta_z},
\end{equation}
where $M_0$ is the mass of the descendant and the fitting parameters
(for the Millennium simulation) are given in Table~1 of
\cite{Fakhouri2010} (see also \cite{Poole2017}). In fact, $\eta_z$ is
very small, which implies a very weak redshift dependence; $\alpha>0$
and $\beta>0$, implying that the merger rate is higher in more massive
haloes and for small mass ratios (the expected outcome in a
hierarchical model). We should remark that halo merger trees can also
be constructed from Monte Carlo realisations based on merger rates
computed using the extended Press-Schechter formalism,
e.g. \cite{Lacey1993}. This analytical approach, calibrated to
simulations \cite{Parkinson2008}, is also widely used to model the
assembly of dark matter haloes, particularly in the context of
semi-analytic models of galaxy formation (e.g. \cite{Cole2000}).

{\it Distribution of accretion times and orbital properties.--}
Eq.~\ref{merger_rate} can be used to compute the {\it average} number
of haloes that become subhaloes of a host at a given redshift in a
certain mass range. Halo merger trees can be used to compute other
statistics of the subhalo population that are directly linked to their
subsequent evolution, specifically: (i)~the distribution of accretion
(or infall) times, and (ii)~the distribution of orbital properties at
the time of accretion.  In a time-independent spherical potential only
two variables are needed to specify the orbit of a tracer particle (plus
the orientation of the orbit). Although subhaloes orbiting a host halo
are far from this idealised case, it is nevertheless useful to
describe the initial orbital parameters in this way since this
provides a point of comparison with the simple spherical
potential. These two parameters can be chosen to be the radial, $V_r$,
and tangential, $V_\theta$, velocities of the subhalo at the time of
infall \cite{Benson2005}, typically expressed in terms of the virial
velocity of the host halo, $V_{200}=\sqrt{GM_{200}/r_{200}}$. Another
common choice is to characterise the orbital properties at infall in
terms of a circular orbit of the same energy, $E$, and the same magnitude
of the angular momentum, $j$ \cite{Tormen1997}. The initial orbit is
then defined by the {\it circularity}, $\eta=j(E)/j_{\rm circ}(E)$,
and the infall radius, $r_{\rm 200}/r_{\rm circ}(E)$, at the time of
accretion\footnote{Another set of parameters that can be used to
  define the orbit are the apocentre and pericentre. Different
  parametrizations can be transformed into one another since they are
  all related to the potential, $\phi(r)$.}.

\begin{figure}[H]
\centering
\includegraphics[width=16 cm, height=9cm,trim=0.35cm 0.15cm 0.15cm 0.15cm, clip=true]{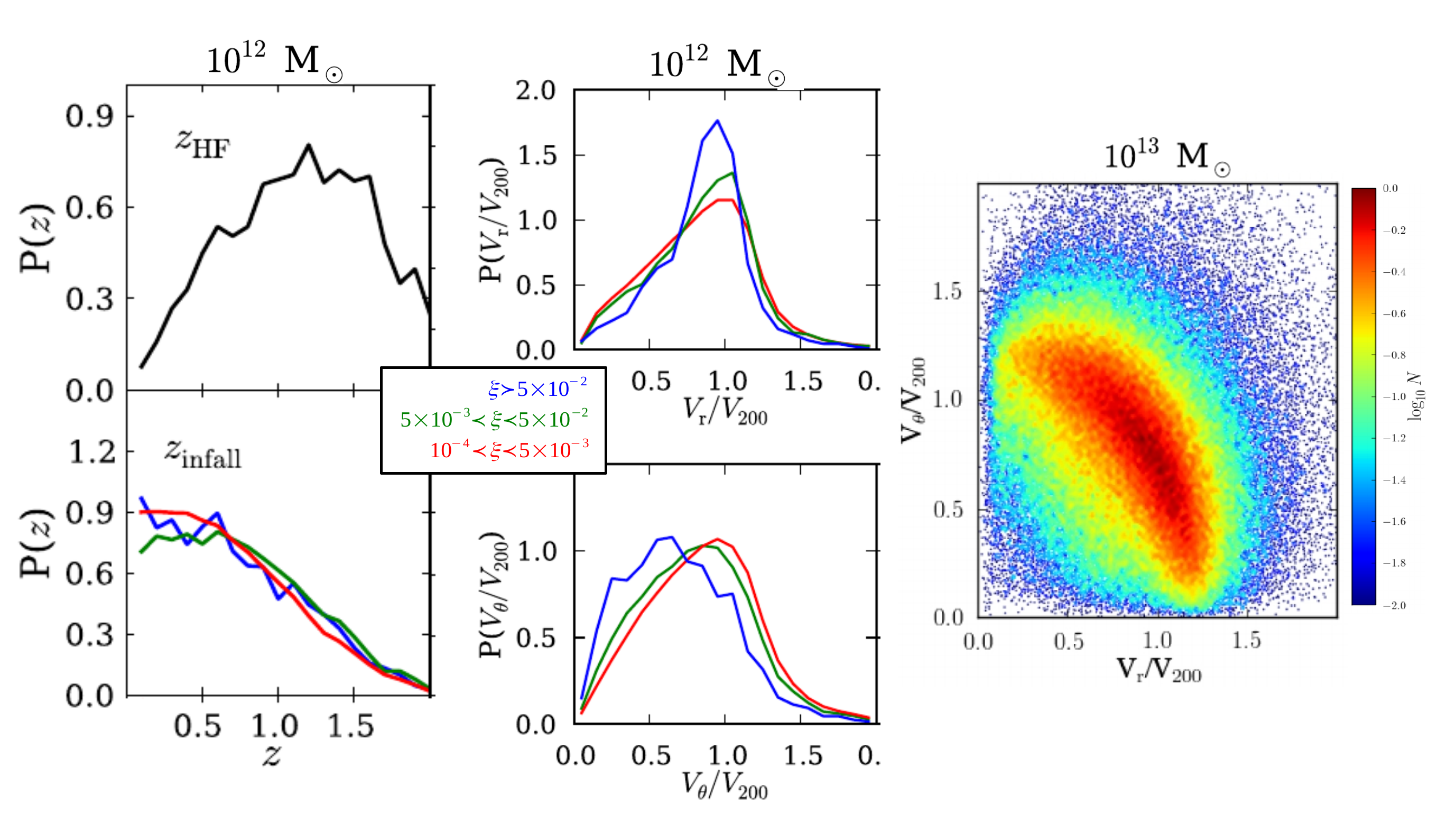}
\caption[Caption without FN]{Initial conditions for the orbits of subhaloes infalling into
  haloes of present-day mass $M_{200}=10^{12}$~M$_\odot$ (figures taken
  from \cite{Jiang2015}\footnotemark; see that paper for similar plots for other
  host masses). {\it Upper left:} the distribution of {\it formation
    redshifts} (defined as the redshift at which the mass of the main
  progenitor of the halo was half its present value). These and the
  other histograms in this figure are normalized such that the
  integral over the distribution is unity.  {\it Lower left}:
  distribution of infall (accretion) redshifts of subhaloes of
  different mass ratios, $\xi$ (relative to the host halo at the time
  of accretion; see legend). {\it Middle:} distributions of radial
  (upper panel) and tangential (lower panel) subhalo orbital
  velocities at infall, relative to the virial velocity of the host,
  for the same halo mass and subhalo-to-halo mass ratios as in the
  lower-left panel. %Although the distributions are broad, we can see
  %that, as expected, the median values are around the characteristic
  %velocity of the host halo, $V_{200}$. 
  {\it Right: } bivariate
  distribution of orbital parameters for infalling haloes into hosts
  of mass $M_{200}(z=0)=10^{13}$~M$_\odot$.}
\label{fig_subhaloes_ICs}
\end{figure}

Fig.~\ref{fig_subhaloes_ICs} shows a sample of the orbital parameters
of the subhalo population at the time of infall, calculated from
high-resolution $N$-body simulations \cite{Jiang2015}. The bottom left
panel shows the distribution of subhalo infall times for Milky
Way-size haloes, $M_{200}(z=0)=10^{12}$~M$_\odot$. This plot only
includes subhaloes accreted {\it after} the {\it formation redshift}
of the halo, $z_{\rm HF}$, defined as the time at which the main
progenitor had half the mass of the final halo. Since the merger rate
is higher for larger descendant masses (Eq.~\ref{merger_rate}), and
since the halo is growing from $z=z_{\rm HF}$ until today, we expect
the distribution of infall times to decrease with redshift down to a
minimum at $z_{\rm HF}$, independently of the mass ratio. Naturally,
recently accreted haloes will be found mostly near the virial radius
of the host, while haloes accreted long ago will be mostly found in
the central regions (we will return to this point below). For
subhaloes in bound orbits, we would expect orbital velocities at
infall to be close to the virial velocity of the host halo, $V_{200}$
\cite{Wetzel2011}.  The distributions of radial and tangential
velocities of infalling satellites for Milky Way-size haloes are shown
in the middle panels of Fig.~\ref{fig_subhaloes_ICs}. Although broad,
the distributions of orbital velocities do indeed have median values
around $V_{200}$. In fact, these distributions are not independent
since the total velocity of the subhalo orbit,
$(V_r^2+V_\theta^2)^{1/2}$, is determined by the potential of the host
halo. This is why the ridge line of the bivariate distribution,
($V_r, V_\theta$), shown in the right panel of
Fig.~\ref{fig_subhaloes_ICs}, is close to circular (see also
\cite{Wetzel2011}).

\footnotetext{\scriptsize{Reproduced from Lilian Jian et al. Orbital parameters of infalling satellite haloes in the hierarchical $\Lambda$CDM model. MNRAS (2015) 448 (2): 1674–1686, doi: 10.1093/mnras/stv053. By permission of Oxford University Press on behalf of the Royal Astronomical Society. For the original article, please visit the following \href{https://academic.oup.com/mnras/article/448/2/1674/1053529}{link}. This figure is not included under the CC-BY license of this publication. For permissions, please email: journals.permissions@oup.com}}

\subsection{Dynamics of subhaloes}\label{sec_dynamics}

The material content of a halo consists of: (i)~a smooth component
made up mostly of the debris of disrupted subhaloes but also of
material that was accreted in diffuse form; (ii)~gravitationally
self-bound substructure $-$ the subhaloes. As mentioned in
Section~\ref{sec_smooth_mergers}, the contribution of truly smooth
accretion to the total mass of a halo at $z=0$ is subdominant, and
even though most dark matter is accreted by (minor) mergers, only a
small fraction, $\sim10\%$, remains today in bound subhaloes, at least
in the CDM model \cite{Springel2008}\footnote{This estimate is based
  on an extrapolation over many orders of magnitude of the subhalo
  mass function determined in simulations down to the free-streaming
  mass of WIMP CDM particles. We discuss this in more detail
  below.}. Thus, most of the mass in a halo consists of the remnants
of the environmental processes responsible for stripping mass from
subhaloes after infall.

\begin{figure}[H]
\centering
\includegraphics[width=16 cm, height=9cm,trim=0.3cm 0.1cm 0.15cm 0.15cm, clip=true]{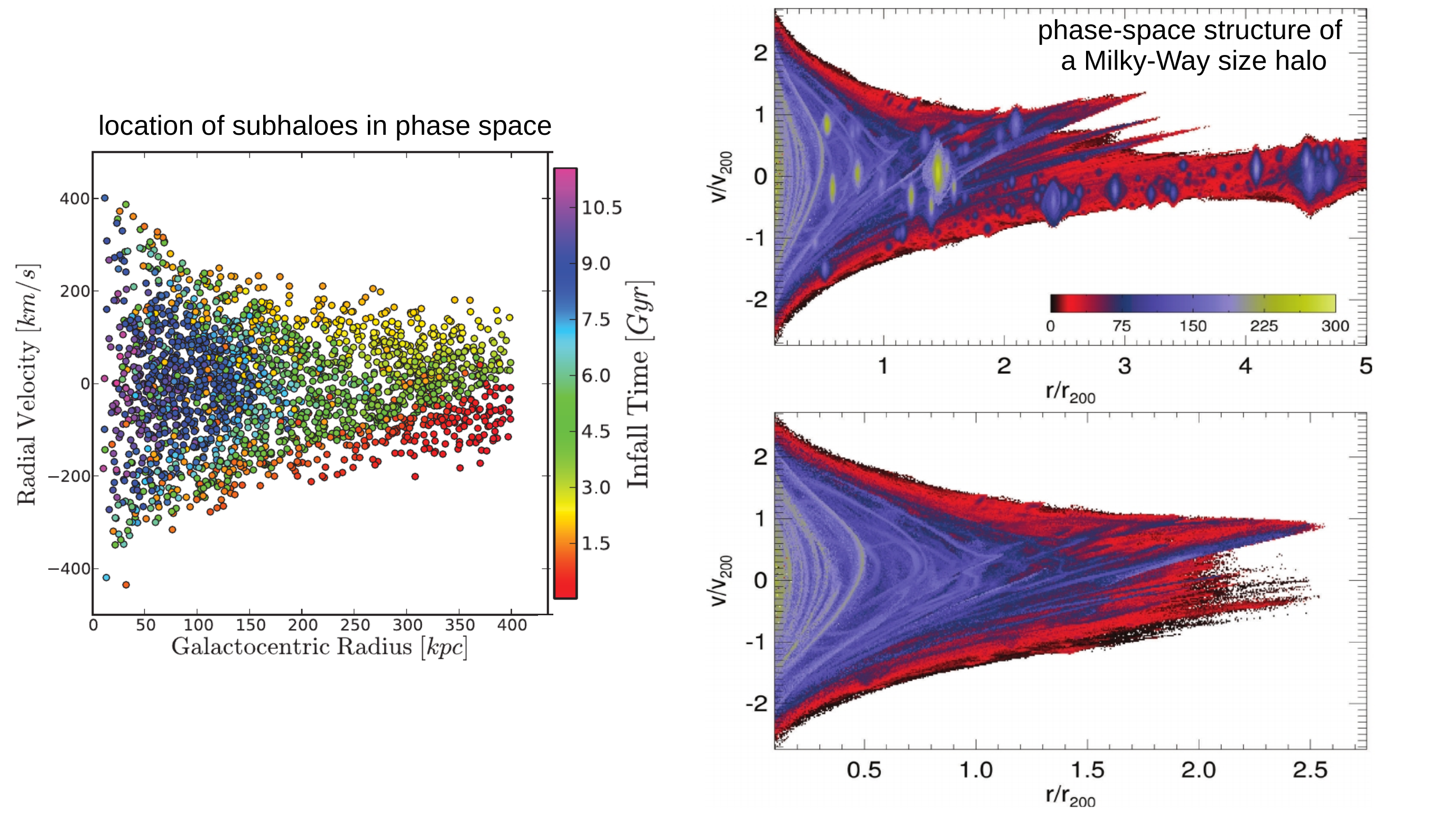}
\caption[Caption without FN]{{\it Left:} Distribution in the 2D radial phase-space plane of subhaloes
  identified in a Milky Way-size halo simulation (Via Lactea II
  \cite{Diemand2008}; figure adapted from \cite{Rocha2012}\protect\footnotemark).
  Subhaloes are colour-coded according to their infall time (measured
  from $z=0$). Subhaloes that are just being accreted are radially
  infalling, while those that were accreted earlier and have completed
  many orbits lose energy through dynamical friction and sink 
  towards the halo centre. {\it Right:} the 2D radial phase-space
  structure of simulation particles in a different Milky Way-size halo
  simulation (Aquarius \cite{Springel2008}; figure adapted from
  \cite{Vogelsberger2011}\protect\footnotemark). Each particle is color-coded according to
  the number of caustics it passes (roughly proportional to the number
  of orbits executed by a given particle). The top panel includes
  bound subhaloes, while the bottom one does not. In the latter, tidal
  streams from disrupted subhaloes are more clearly  visible.}
\label{fig_subhaloes_phase_space}
\end{figure}

Fig.~\ref{fig_subhaloes_phase_space} provides an illustration of the
richness of information contained in the phase-space structure (shown
here in its 2D radial projection) of haloes today that is relevant to
these environmental processes. In the left panel, the location in
phase space (at $z=0$) of the subhaloes of the Via Lactea II
simulation of a Milky Way-size halo \cite{Diemand2008} is shown
colour-coded according to the subhalo infall time. This figure reveals
the path of a subhalo as it orbits in the host halo: it first falls in
radially and 
\pagebreak
\addtocounter{footnote}{-2} 
 \stepcounter{footnote}\footnotetext{\scriptsize{Reproduced from Miguel Rocha et al. Infall times for Milky Way satellites from their present-day kinematics. MNRAS (2012) 425 (1): 231–244, doi: 10.1111/j.1365-2966.2012.21432.x. By permission of Oxford University Press on behalf of the Royal Astronomical Society. For the original article, please visit the following \href{https://academic.oup.com/mnras/article/425/1/231/998181}{link}. This figure is not included under the CC-BY license of this publication. For permissions, please email: journals.permissions@oup.com}}
 \stepcounter{footnote}\footnotetext{\scriptsize{Reproduced from Mark Vogelsberger and Simon D. M. White. Streams and caustics: the fine-grained structure of $\Lambda$ cold dark matter haloes. MNRAS (2011) 413 (2): 1419–1438, doi: 10.1111/j.1365-2966.2011.18224.x. By permission of Oxford University Press on behalf of the Royal Astronomical Society. For the original article, please visit the following \href{https://academic.oup.com/mnras/article/413/2/1419/1070092}{link}. This figure is not included under the CC-BY license of this publication. For permissions, please email: journals.permissions@oup.com}}
\noindent then loses energy as it is subjected to dynamical
friction and tidal forces in the host halo. The former causes the
subhalo to sink towards the centre while the latter gradually strip
mass from it, creating tidal streams. This picture can be appreciated
with clarity in the right-hand panels of the figure where the (2D)
phase-space structure of the dark matter particles is shown for a
different Milky Way-size halo simulation (Aquarius;
\cite{Springel2008}). The colour in this case encodes the number of
caustics that a given particle traverses\footnote{Caustics represent
  folds in the fine-grained phase-space distribution function, which
  in CDM evolves according to the collisionless Boltzmann equation
  (Eq.~\ref{CBE}). Before the formation of non-linear structures, CDM
  particles are distributed nearly uniformly in space with small
  density and velocity perturbations and very small thermal
  velocities. CDM particles thus occupy a thin, approximately three
  dimensional, sheet in 6D phase-space volume. Since CDM particles are
  collisionless and evolve according to Eq.~\ref{CBE}, the
  fine-grained phase-space density is conserved during gravitational
  evolution (this was discussed earlier in the context of the maximum
  phase-space density in WDM models in \ref{sec_dm_nature}), which
  implies that the original thin sheet can be stretched and folded but
  it cannot be broken. Caustics appear where folds occurs, and have
  very large spatial densities, limited only by primordial thermal
  motions
  (e.g. \cite{Natarajan2006,Vogelsberger2011,Vogelsberger2009b}).}. Since
caustics occur near orbital turning points, the number of caustics is
roughly proportional to the number of orbits each particle traverses.
The caustic count is thus an excellent way to highlight substructure
in the 2D phase-space structure seen in the right-hand panel of
Fig.~\ref{fig_subhaloes_phase_space} since particles that are part, or
were part, of a subhalo have undergone more particle orbits in their
earlier host. This plot thus shows the richness of the substructure
present in haloes today. As we mentioned earlier, most of the matter
in a halo today has been accreted through mergers and consists of
material that was stripped from subhaloes.

{\it Identifying substructure.--} Several algorithms are in common use
to identify subhaloes in $N$- body simulations and define their
boundaries and properties. These {\it subhalo finders} are based on
different techniques. Here we will list only the most popular ones;
for a comprehensive comparison study see \cite{Onions2012}.  A common
approach consists of finding local density maxima in the parent halo
density field and then associating adjacent particles with this peak
using a binding energy criterion; a subhalo is thus defined as the
collection of particles that is gravitationally self-bound, with the
density peak at its centre. Examples are: SUBFIND \cite{Springel2001};
Bound Density Maxima \cite{Klypin1999}; VOBOZ \cite{Neyrinck2005};
Amiga Halo Finder \cite{Knollmann2009}. A different approach is the ``time
domain subhalo finder'' which follows the time evolution of haloes and
tracks them when they become subhalos by identifying their bound
particles, as in the Hierachical Bound-Tracing or HBT
\cite{Han2012,Han2018}.

An improvement over this 3D spatial approach can
be made by including information on the particle velocities, which has
the advantage that subhaloes that are in close spatial proximity with
one another can be more easily disentangled. In this case a density
criterion is not enough, but the relative velocity between merging
subhaloes is a telltale sign of a merger. Modifications of these
algorithms can be used to identify the tidal streams that are the
remnants of the tidal stripping process (see below) and which are not
localised in real space, but have clear signatures in phase
space. Examples of phase-space finders are the Hierarchical Structure
Finder \cite{Macie2009} and ROCKSTAR \cite{Behroozi2013}.

Current subhalo finders are able to identify subhaloes down to
$20-100$ particles and different algorithms roughly agree with one
another on their location and main properties \cite{Onions2012}. Below
we review the main physical processes that affect the evolution and
inner structure of subhaloes along their orbits, as well as relevant
lessons learned from $N$-body simulations.

{\it Tidal stripping.--} Once a halo reaches the outer boundary of the
host halo into which it will merge, tidal forces begin to act,
suppressing the accretion of matter into the merging halo and
stripping mass from its outer layers in a process known as tidal
stripping. Since the enclosed overdensity of a subhalo depends on its
position within the host, the virial radius is no longer a meaningful
concept. A more relevant scale is the {\it tidal radius}, $r_t$,
defined as the radius at which the differential tidal force of the
host halo is equal to the gravitational force due to the mass of the
subhalo, or equivalently, as the radius within which the enclosed mean
mass density of the satellite is comparable to the mean mass density
of the main halo interior to the distance, $R$, to the satellite. The
expectation is that the matter beyond the tidal radius will be removed from
the subhalo, reducing its mass as it orbits around the host. For a circular
orbit and assuming that the subhalo mass, $m_{\rm sub}(<r_t)$, is much
smaller than the enclosed mass of the host, $M(<R)$, and that
$r_t\ll R$, the tidal radius is given by
\cite{Tormen1998}\footnote{This equation ignores the effects of the
  centrifugal force on the satellite as it orbits around the
  host. Including this effect (assuming circular orbits) modifies
  Eq.~\ref{eq_tidal_radius} slightly by changing the factor of 2 in
  the denominator to 3 \cite{Tollet2017}.}:
\begin{equation}\label{eq_tidal_radius}
r_t=R\left[\frac{m_{\rm sub}(<r_t)}{\left(2-{\rm dln}M/{\rm dln}r\right)M(<R)}\right]^{1/3}.
\end{equation}

\begin{figure}[H]
\centering
\includegraphics[width=16 cm, height=9cm,trim=0.4cm 0.1cm 0.15cm 0.15cm, clip=true]{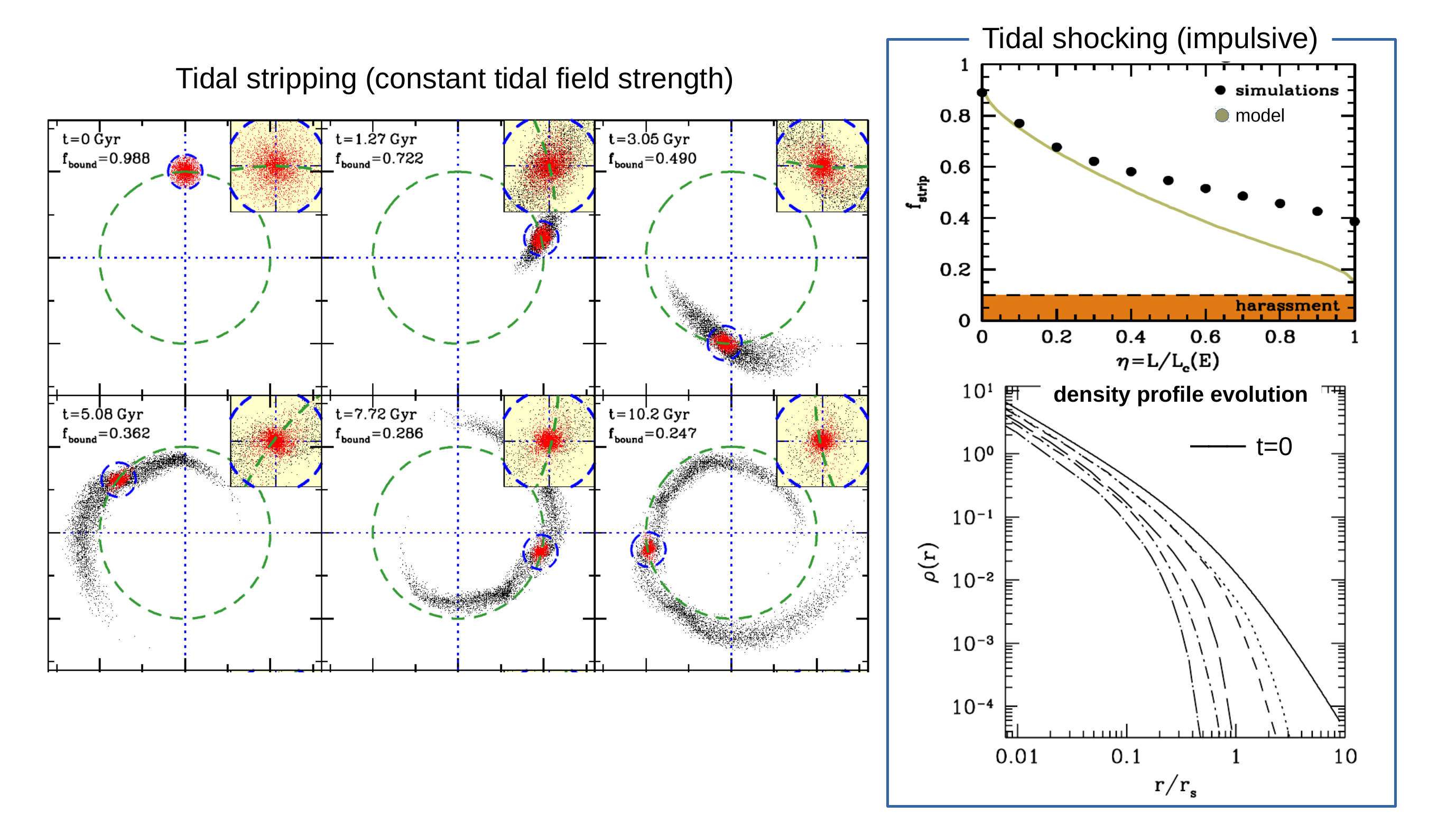}
\caption[Caption for LOF]{Tidal effects in subhaloes. {\it Left:} evolution of a
  subhalo in a circular orbit in a static host halo potential.  Since
  the tidal field strength is constant, the subhalo gradually loses
  mass (red particles are bound to the subhalo, black particles are
  unbound) as it orbits in the host halo creating characteristic tidal
  streams (figure adapted from \cite{vdBosch2018b}\footnotemark). {\it Right:} the
  effect of tidal shocks. For nearly radial orbits, the variations in
  the potential near pericentre are rapid (relative to the internal
  dynamical timescale of the subhalo) and this leads to an impulsive
  {\it tidal shock}, which causes a drastic removal of mass (upper
  right) and a change in the dark matter distribution (bottom
  right). In the upper panel the fraction of stripped mass as a
  function of circularity (see Section~\ref{evo}), given by the
  impulsive approximation, is compared with that in a controlled
  simulation (figure adapted from \cite{vdBosch2018}\footnotemark). The model works
  quite well for radial orbits but it fails for circular orbits (as in
  the left panel), for which an adiabatic model is more appropriate
  (Eq.~\ref{eq_mass_loss}). In the lower panel, tidal shocking is seen
  to reduce the mass in the central regions but preserves the
  asymptotic NFW behaviour, while the outer regions become
  considerably steeper than NFW (figure adapted from
  \cite{Kazantzidis2004}\footnotemark).}
\label{fig_tidal_modes}
\end{figure}

For non-circular orbits the situation is more complex (in fact in the
most general cases, the tidal radius can be ill-defined; see
e.g. \cite{Tollet2017}), but the principle behind the concept of tidal
radius remains valid: the relevant physical quantity to determine the
boundary of the subhalo is the relative strength of the gravitational
attraction of the subhalo and the tidal forces of the host. In this
way, the tidal radius is commonly used to model tidal stripping in a
variety of collisionless systems, not only subhaloes, but also, for
example, globular clusters. In particular, for a slowly varying
tidal field (i.e., in the adiabatic approximation), the mass loss due
to stripping may be modelled as:
\cite{Taylor2001b,Zentner2003,Zentner2005}:
\begin{equation}\label{eq_mass_loss}
dm_{\rm sub}=\alpha_t m_{\rm sub}(>r_t)\frac{dt}{t_{\rm orb}(R)},
\end{equation}
where $t_{\rm orb}(R)$ is the instantaneous orbital period at the
radius of the subhalo, and $\alpha_t$ is a tuning parameter, which
encapsulates departures from this simple approximation (e.g., the
details of subhalo structure); $\alpha_t$ is typically calibrated from
simulations but the values used vary significantly in the literature,
which is a major limitation. %of Eq.~\ref{eq_mass_loss}.
Eq.~\ref{eq_mass_loss} assumes that the relevant timescale for mass
loss is the orbital period of the subhalo, which is justified by
noting that the energy scale for the tidally stripped material is
given by the change in the potential of the host across the body of
the satellite $\sim r_td\phi_{\rm host}(R)/dR$
\cite{Johnston1998}. The left panel of Fig.~\ref{fig_tidal_modes}
shows an example of tidal stripping in this slow (adiabatic) mode for
a subhalo in a circular orbit around a static host potential (from
\cite{vdBosch2018b}). As pointed out by \cite{vdBosch2018}, a
combination of the ill-defined tidal radius and uncertainty in the
parameter $\alpha_t$, makes the modelling of this adiabatic case quite
complicated, with the end result that models based in
Eq.~\ref{eq_mass_loss} do not, in general, match simulation results
accurately. In any case, it is rare for subhalo orbits to be nearly
circular; for realistic orbits, most of the tidal mass loss
happens near pericentre.

{\it Tidal shock heating.--} While tidal stripping
(Eq.~\ref{eq_mass_loss}) refers to the gradual loss of loosely bound
material from a subhalo due to a slowly-varying external potential, a
rapid (impulsive) variation in the potential causes a transfer of the
satellite's orbital energy to the internal energy of its
particles. These tidal shocks are most important at pericentre where
the impulsive condition is best satisfied. The redistribution of
internal energy produced by the shock alters the inner structure of
the subhalo and can unbind some of its particles
\cite{Taylor2001b,Hayashi2003,vdBosch2018}.  This process is well
described by the ``impulsive approximation'' (see
\cite{Spitzer1958,Gnedin1999} for the case of globular clusters) in
which tidal forces are assumed to act during a much shorter time than
the dynamical timescale of the satellite (see
\cite{Aguilar1985,Aguilar1986}). The approximation gives the specific
energy change suffered by particles in the subhalo due to a tidal
shock as\footnote{It is possible to evaluate Eq.~\ref{eq_impulsive}
  for a given fixed potential and a given subhalo orbit (see e.g
  Eq. 20 of \cite{vdBosch2018} for an NFW halo).}:
\begin{equation}\label{eq_impulsive}
( \Delta E )_{\rm i,tid}=(\Delta v)^2_{\rm i,tid} \approx \left\vert\int_{\rm orbit} \vec{a}_{\rm i,tid}(t) dt\right\vert^2,
\end{equation}
where $\vec{a}_{\rm i,tid}$ is the acceleration experienced by a
particle in the subhalo and the integral is performed along the orbit
of the subhalo. If $( \Delta E )_{\rm i,tid}>E_{i,b}$, where $E_{i,b}$
is the binding energy of the particle, we may assume that the particle
will become unbound. The mass fraction
of particles that satisfies the inequality is then assumed to be
removed {\it instantaneously} from the subhalo. The impulsive
approximation accurately captures the results of simulations for
radial orbits. An example is shown in the right-hand panel of
Fig.~\ref{fig_tidal_modes} (taken from \cite{vdBosch2018}).

Although the energy injection in Eq.~\ref{eq_impulsive} from tidal
shocks might not be enough to unbind particles, it can still
affect the inner structure of a subhalo. As a result of the shock, the
orbits of dark matter particles in the centre expand, reducing the
inner density, although this process is not strong enough to create a
central core \cite{Hayashi2003,Kazantzidis2004}. The resulting density
profile is, in fact, still well described by an NFW profile in the
inner regions, albeit with a higher concentration, while the outer
regions are considerably steeper than NFW due to stripping. For
instance Ref.~\cite{Kazantzidis2004}, using idealised simulations in a
static external halo potential, found a profile of the form
$\rho\propto r^\gamma {\rm exp}(-r/r_b)$, with a central slope,
$\gamma\sim1$, and a cutoff radius due to tidal shocks, $r_b$ (see
bottom right panel of Fig.~\ref{fig_tidal_modes}).

For a general subhalo orbit, a combination of the adiabatic
(Eq.~\ref{eq_mass_loss}) and impulsive (Eq.~\ref{eq_impulsive})
approximations provides a good estimate of the amount of stripped
mass; the former is valid particularly near the apocentre of nearly
circular orbits while the latter is more appropriate near the
pericentre of eccentric orbits. The impulsive approximation reproduces
the results of simulations quite accurately for radial orbits, but the
adiabatic approximation is not very adequate for the reasons discussed
above (see also Section 4.3 of \cite{vdBosch2018}).

\addtocounter{footnote}{-3} 
 \stepcounter{footnote}\footnotetext{\scriptsize{Reproduced from Frank C van den Bosch and Go Ogiya. Dark matter substructure in numerical simulations: a tale of discreteness noise, runaway instabilities, and artificial disruption. MNRAS (2018) 475 (3): 4066–4087, doi: 10.1093/mnras/sty084. By permission of Oxford University Press on behalf of the Royal Astronomical Society. For the original article, please visit the following \href{https://academic.oup.com/mnras/article/475/3/4066/4797185}{link}. This figure is not included under the CC-BY license of this publication. For permissions, please email: journals.permissions@oup.com}}
 \stepcounter{footnote}\footnotetext{\scriptsize{Reproduced from Frank C van den Bosch et al. Disruption of dark matter substructure: fact or fiction? MNRAS (2018) 474 (3): 3043–3066, doi: 10.1093/mnras/stx2956. By permission of Oxford University Press on behalf of the Royal Astronomical Society. For the original article, please visit the following \href{https://academic.oup.com/mnras/article/474/3/3043/4638541}{link}. This figure is not included under the CC-BY license of this publication. For permissions, please email: journals.permissions@oup.com}}
 \stepcounter{footnote}\footnotetext{\scriptsize{\copyright AAS. Reproduced with permission. For the original article, please visit the following \href{https://iopscience.iop.org/article/10.1086/420840}{link}.}}

{\it Subhalo-subhalo encounters.--} Tidal heating can also be caused
by impulsive encounters with other subhaloes, which can add up to
produce a net effect on the subhalo inner structure and mass loss (a
process called {\it galaxy harassment} in the context of satellite
galaxies; \cite{Moore1996}). A similar impulsive approach to the one
above can be used to estimate the strength of this form of tidal
heating, but the calculation is more complicated because, among other
things, the distribution of subhaloes in the host and the encounter
rate need to be modeled. A recent study finds that tidal shocking
from encounters is subdominant (by a factor of several) compared to
shocking during pericentric passages \cite{vdBosch2018}.

\begin{figure}[H]
\centering
\vspace{-1.0\baselineskip}
\includegraphics[width=16 cm, height=9cm,trim=0.4cm 0.75cm 0.15cm 0.75cm, clip=true]{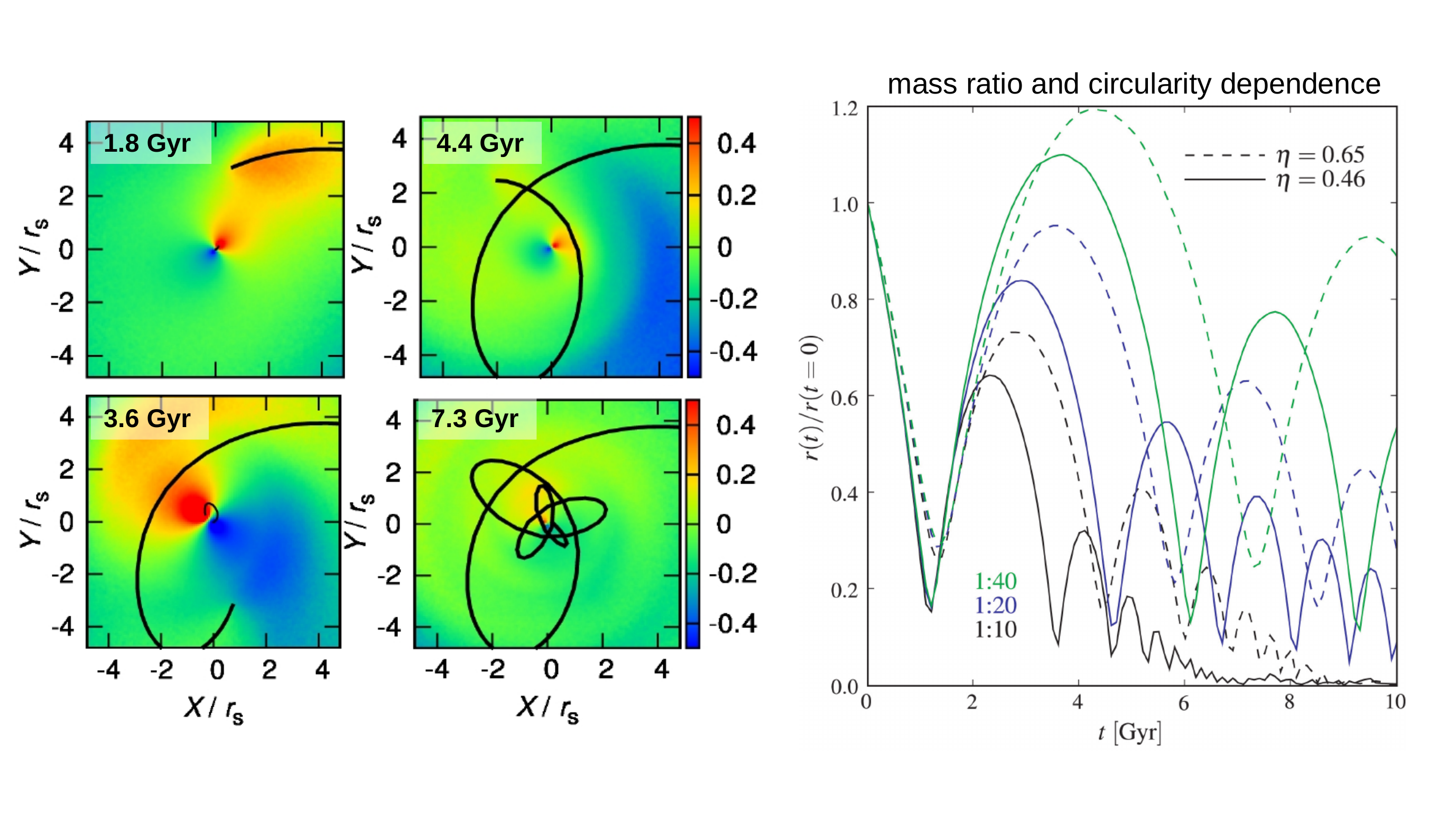}
\caption[Caption for LOF]{Dynamical friction experienced by subhaloes. {\it Left:}
  simulation of a subhalo orbiting a Milky Way-size halo; the initial
  mass ratio and circularity of the orbit are 0.1 and 0.5,
  respectively. The images show the projected over- (or under-)
  density relative to the initial value at $t=0$, at different times
  during the evolution. The thick solid line marks the subhalo orbit,
  which decays over time due to dynamical friction. This gravitational
  process induces a wake in the host halo trailing behind the
  satellite (most clearly visible in the top left panel). The dipole
  feature at the centre of the host is caused by the tidal effect of
  the subhalo, which perturbs the position of the halo potential
  minimum. This effect diminishes with time as the satellite is
  stripped of mass (figure adapted from \cite{Ogiya2016}\footnotemark). {\it
    Right:} evolution of the radial distance of a simulated subhalo
  orbiting a Milky Way-size halo (figure taken from \cite{BK2008}\footnotemark). The orbit decays by
  dynamical friction on a timescale that strongly depends on the
  initial mass ratio (different colours) and circularity of the orbit
  (dashed and solid lines). The timescales are well approximated by
  the fitting formula (Eq.~\ref{eq_tdf_imp}), which is an improvement
  over the classical Chandrasekhar formula (Eq.~\ref{eq_tdf}).}
\label{fig_df}
\vspace{-1.0\baselineskip}
\end{figure}

{\it Dynamical friction.--} When an object of mass, $M_s$, moves
through an ambient medium of collisionless particles of mass
$m\ll M_s$, the object experiences a drag force known as dynamical
friction. This force may be thought of as the gravitational pull
exerted by a local enhancement in the ambient density formed behind
the moving object (a trailing wake) as the object gravitationally
focuses the surrounding particles (see left panels of
Fig.~\ref{fig_df}).

\addtocounter{footnote}{-2}
 \stepcounter{footnote}\footnotetext{\scriptsize{Reproduced from Go Ogiya and Andreas Burkert. Dynamical friction and scratches of orbiting satellite galaxies on host systems. MNRAS (2016) 457 (2): 2164–2172, doi: 10.1093/mnras/stw091. By permission of Oxford University Press on behalf of the Royal Astronomical Society. For the original article, please visit the following \href{https://academic.oup.com/mnras/article/457/2/2164/970692}{link}. This figure is not included under the CC-BY license of this publication. For permissions, please email: journals.permissions@oup.com}}

 \stepcounter{footnote}\footnotetext{\scriptsize{Reproduced from Michael Boylan-Kolchin et al. Dynamical friction and galaxy merging time-scales . MNRAS (2008) 383 (1): 93–101, doi: 10.1111/j.1365-2966.2007.12530.x. By permission of Oxford University Press on behalf of the Royal Astronomical Society. For the original article, please visit the following \href{https://academic.oup.com/mnras/article/383/1/93/1067887}{link}. This figure is not included under the CC-BY license of this publication. For permissions, please email: journals.permissions@oup.com}}

The net result of
dynamical friction is a transfer of orbital angular momentum and
energy from the moving object into the surrounding medium. The process
can be analysed as a series of uncorrelated sequential encounters
between the object of mass, $M_s$, and velocity, $\vec{v}_s$, and
particles randomly taken from the ambient medium with velocity
distribution, $f_m(\vec{u})$. These interactions occur on a timescale
much shorter than the variations in the velocity, $\vec{v}_s$, of the
object. If the ambient medium is assumed to have a homogeneous
density, $\rho_m$, then the changes to the velocity of the object
perpendicular to its motion average to zero by symmetry, while the
velocity changes parallel to the direction of motion are given by the
well-known Chandrasekhar dynamical friction formula
\cite{Chandra1943}, which, for the drag force, $\vec{F}_{\rm df}$,
takes the form: %\footnote{This particular form of Chandrasekhar's formula was taken from \cite{Mo2010}.}:
\begin{eqnarray}\label{eq_chandra}
\vec{F}_{\rm df}=M_s\dfrac{d\vec{v}_s}{dt}&=&-16\pi^2M_s^2m{\rm ln} \Lambda\left[\int_0^{\vert \vec{v}_s\vert} f_m(\vert \vec{u}\vert)u^2d\vert \vec{u}\vert\right]\frac{\vec{v}_s}{\vert \vec{v}_s\vert}\\\nonumber
&=&-4\pi\left(\frac{GM_s}{v_s}\right)^2{\rm ln} \Lambda~\rho(<\vert \vec{v}_s\vert)\frac{\vec{v}_s}{\vert \vec{v}_s\vert},
\end{eqnarray}
where $\rho(<\vert\vec{v}_s\vert)$ is the density of ambient particles
with speed less than $\vert \vec{v}_s\vert$ and
${\rm ln}\Lambda\equiv{\rm ln}\left[1+(b_{\rm max}/b_{90})^2\right]$
is the Coulomb logarithm, with $b_{90}=G(M_s+m)/v_\infty^2$, 
$v_\infty$ the initial relative velocity of an individual encounter,
and $b_{\rm max}$ the maximum impact parameter
($b_{\rm max}\gg b_{90}$). As a consequence of dynamical friction, the
orbit of the object decays in time sinking towards the centre of the
host halo. For circular orbits in a spherical singular isothermal host
halo (implying a Maxwellian velocity distribution, $f_m$) of mass,
$M_h$, the timescale for the orbit to decay to zero (i.e., the
dynamical friction time) is approximately given by \cite[e.g.,][]{Mo2010}: 
\begin{equation}\label{eq_tdf}
\frac{t_{\rm df}}{t_{\rm dyn}}\approx \frac{1.17}{{\rm
    ln}(M_h/M_s)}\left(\frac{M_h}{M_s}\right), 
\end{equation}
where $t_{\rm dyn}=r_{\rm vir}/V_c$ is the dynamical timescale at the
virial radius of the halo, $r_h$, with $V_c$ the circular velocity of
the host halo, which is independent of radius for a singular
isothermal sphere.  For more general orbits, Eq.~\ref{eq_tdf} requires
a correction that scales with the circularity, $\eta$, as
$t_{\rm df}\propto\eta^{\gamma_\eta}$, where $\gamma_\eta\sim0.53$
\cite{vdBosch1999}, implying that more eccentric orbits decay more
rapidly (see right panel of Fig.~\ref{fig_df}).

Eq.~\ref{eq_chandra} is derived under the following assumptions:
(i)~both the satellite and the particles that make up the ambient
medium can be treated as point masses; (ii)~the self-gravity of the
ambient medium can be ignored, and (iii)~the distribution of ambient
medium particles is infinite, homogeneous and isotropic. None of these
assumptions is strictly valid in realistic situations. Nevertheless,
Chandrasekhar's formula provides a reasonable description of dynamical
friction, particularly when modifications are included to account for
the density profile of the subhaloes and their orbits; in practice,
this can be done by regarding the Coulomb logarithm as a free
parameter that depends on these properties. For example, \cite{BK2008}
carried out a series of idealised $N$-body simulations of a subhalo
infalling into a host, both described by a Hernquist density
profile,\footnote{The Hernquist halo profile \cite{Hernquist1990} has
  the same asymptotic behaviour at the centre as the NFW halo and has
  the advantage that the velocity distribution function in the
  isotropic case has an analytic form (see Eq.~\ref{Eddington}), which
  makes it particularly simple to set up initial conditions for
  simulating haloes in dynamical equilibrium.} and found the following fitting function 
  to the dynamical friction timescale
  (a few examples of the orbital evolution in this study are shown in the right
panel of Fig.~\ref{fig_df}):
\begin{equation}\label{eq_tdf_imp}
\frac{t_{\rm df}}{t_{\rm dyn}}=A\frac{(M_h/M_s)^b}{{\rm
    ln}(1+M_h/M_s)}{\rm exp}[c\eta(E)]\left[\frac{r_c(E)}{r_{200}}\right]^d,
\end{equation}
where $E$ is the initial orbital energy of the satellite (we recall
that $r_c(E)$ is the radius of a circular orbit of the same energy,
$E$), and the fitting parameters are of order one\footnote{The values
  for these parameters reported in \cite{BK2008} are: $A=0.216$,
  $b=1.3$, $c=1.9$, and $d=1.0$, but we point out that in this study
  both the halo and the subhalo were modelled as Hernquist
  profiles.}. Eq.~\ref{eq_tdf_imp} was found to be valid over a wide
range of orbital parameters; the most relevant restriction is
$0.025\leq M_s/M_h\leq0.3$. For smaller mass ratios, the dynamical
friction timescale becomes much larger than the age of the Universe,
while for larger mass ratios, the relevant timescale is just the
dynamical or free-fall time\footnote{Eq.~\ref{eq_tdf_imp} was only
  explored for values of the circularity in the range
  $0.3\leq\eta\leq1.0$ and for $0.65\leq r_c/r_{200}\leq1.0$; the
  lower limits were imposed in order to avoid radial orbits that would
  take the subhalo so close to the centre of the halo in the first
  orbit that the tidal effects of the galaxy cannot be ignored.  So
  far we have not discussed baryonic effects, but it is worth
  mentioning them here since Eq.~\ref{eq_tdf_imp} was not investigated
  outside this range and might not be valid there even in the absence
  of a central galaxy.}.

\subsection{The abundance, spatial distribution and internal structure
  of dark matter subhaloes}

The abundance, spatial distribution within the host halo and 
internal structure of subhaloes are determined by the combined effects
of the initial conditions at the time of accretion, which depend on
cosmology, and the dynamical processes described in the previous
section. These properties are best derived in full generality using
cosmological $N$-body simulations but analytic models can provide
valuable physical insights \cite{Han2016}.  In this section we present
some of the key structural properties of the subhalo population, as 
revealed by simulations. Naturally, these properties are closely
related to those of {\it isolated} haloes (discussed in
Section~\ref{sec_dm_nature}) with a few relevant modifications.

\begin{figure}[H]
\centering
\vspace{-0.75\baselineskip}
\includegraphics[width=16 cm, height=7cm,trim=0.4cm 1.5cm 0.15cm 1.75cm, clip=true]{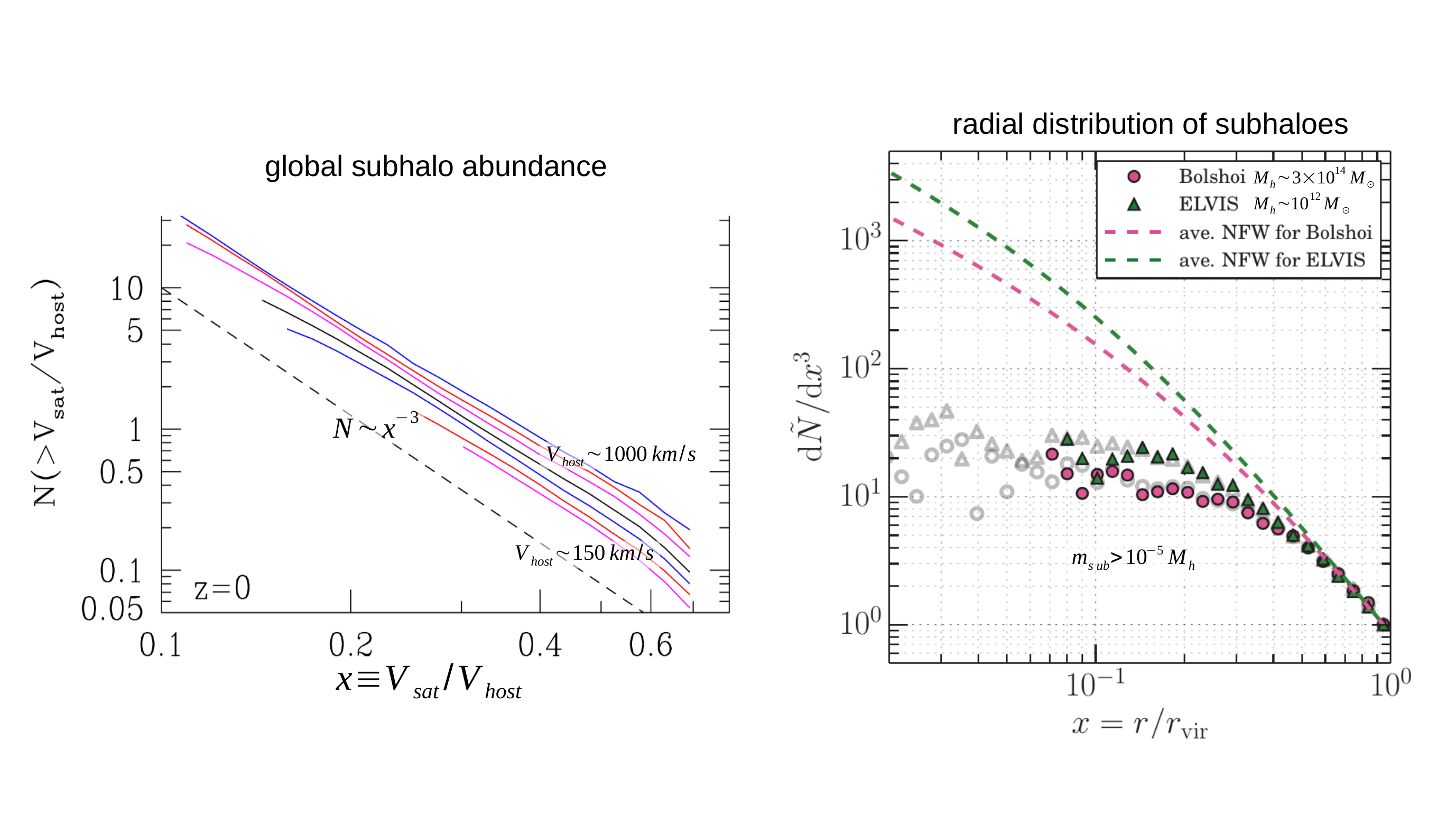}
\caption[Caption without FN]{Subhalo abundance. {\it Left:} the subhalo velocity
  function at $z=0$ for haloes of different maximum circular velocity, 
  from $\sim$150 km/s to $\sim$1000 km/s (bottom to top). In terms
  of the velocity ratio, $x=V_{\rm sub}/V_h$, the velocity function is nearly
  universal, scaling as $x^{-3}$ (dashed line) with a scale-dependent
  normalization (see Eq.~\ref{eq_kp}; figure adapted from
  \cite{Klypin2011}\footnotemark). {\it Right:} the number density of
  subhaloes as a function of halocentric distance in units of the
  virial radius for Miky Way-size haloes (triangles) and cluster-size
  haloes (circles). All subhaloes with $m_{\rm sub}/M_h>10^{-5}$ have
  been included. The dashed lines are the average NFW fits to the
  density profiles of the hosts. These functions have been normalised
  to unity at the virial radius. Figure adapted from \cite{Jiang2017}.} 
\label{fig_structure_1}
\vspace{-0.3\baselineskip}
\end{figure}

\footnotetext{\scriptsize{\copyright AAS. Reproduced with permission. For the original article, please visit the following \href{https://iopscience.iop.org/article/10.1088/0004-637X/740/2/102}{link}.}}

{\it The subhalo mass function.--} As in the case of isolated haloes,
the total CDM subhalo mass function (measured within the virial
radius) is remarkably close to universal and, in fact, has a similar
low-mass slope as the halo mass function
\cite{Gao2004,Springel2008,Diemand2007,Gao2012,GK2014,Griffen2016}:
$dn_{\rm sub}/dm_{\rm sub}\propto m_{\rm sub}^\alpha$, where
$\alpha~\sim -1.9$ (see Eq.~\ref{eq_halo_mf}). This similarity to the
halo mass function is partly due to the fact that most subhaloes
identified at a given time were accreted relatively recently and thus
tidal effects have not had time to act; see
Fig.~\ref{fig_subhaloes_ICs}. The normalisation of the subhalo mass
function depends on the mass of the host halo, with more massive
haloes having, on average, larger subhalo populations
\cite{Gao2011,Gao2012,Wang2011}. This reflects the earlier assembly of
low-mass haloes, which allows tidal effects more time to act and
disrupt subhaloes. For similar reasons other properties of the host
halo can have second-order effects on the amplitude of the subhalo
mass function, e.g., at fixed mass, more concentrated haloes (which
assemble earlier) have fewer subhaloes.

When the subhalo mass function is scaled to the host halo
mass it becomes nearly universal across halo masses, with a functional
form that is well fitted by:
\begin{equation}\label{eq_gao}
N(>\mu\equiv m_{\rm sub}/M_h)=\left(\frac{\mu}{{\widetilde \mu_1}}\right)^{1+\alpha}{\rm exp}\left[-\left(\frac{\mu}{\mu_{\rm cut}}\right)^b\right],
\end{equation}
\cite{Angulo2009,BK2010,Gao2011,RP2016} where the exponential cutoff
accounts for the increasing rarity of subhaloes of mass close to that
of the host halo mass\footnote{The fitting parameters in
  Eq.~\ref{eq_gao} in the case of the Millennium simulations may be
  found in \cite{Gao2011}, where a redshift dependence is also
  provided.}. The parameter, ${\widetilde \mu_1}$, is the typical mass
fraction of the most massive subhalo (relative to the host halo mass),
which, for a Milky Way-size halo, is of order 0.01, but with a large
spread \cite{BK2010}. The universality of the subhalo mass function
is, however, not perfect; the remaining dependence on host halo mass
can be captured by allowing a relatively weak scaling of the
normalisation parameter, ${\widetilde \mu_1}$. This dependence is
amplified, and perhaps better expressed, if the subhalo {\it velocity
  function} is used instead, that is, if the abundance is given in
terms of the maximum circular velocity instead of the mass (see left
panel of Fig.~\ref{fig_structure_1}). In this case, an accurate
approximation is given by \cite{Klypin2011}:
\begin{equation}\label{eq_kp}
N(>x\equiv V_{\rm sub}/V_h)\propto V_h^{1/2}x^{-3}, \ \ \ \ \ x<0.7,
\end{equation}
which implies, for instance, that a cluster-sized halo
($V_h\sim1000$~km/s) has $\sim2.2$ times more substructure of a given
velocity ratio than a Milky Way-size halo ($V_h\sim200$~km/s). Notice
that this difference is considerably weaker if the mass ratio is used
since, in this case, the abundance scales as a power law of exponent
$\sim-0.9$ rather than $\sim-3$.

The fact that the power-law exponent of the subhalo mass function at
low masses, $\alpha$, is greater than $-2$ is important; a steeper
slope would imply that the total mass in substructures diverges when
extrapolated to arbitrarily low masses. For a given particle dark
matter model we know, of course, that this extrapolation cannot be
continued beyond the truncation mass below which the properties of the
dark matter particle prevents the formation of smaller structures (due
to the suppression mechanisms mentioned in
Section~\ref{subsec_linear}). In the case of CDM WIMPs, the
extrapolation of the subhalo mass function down to the Earth's mass
($10^{-6}$~M$_\odot$) implies that the fraction of mass contained in
{\it unresolved} subhaloes is $\sim4.5\%$, in contrast to the
$\sim13\%$ mass fraction found in the highest resolution simulation
(with a particle mass of $2\times 10^4$~M$_\odot$) to date of a Milky
Way-size halo \cite{Springel2008}. As mentioned earlier, most of the
mass in haloes is not, in fact, in the form of self-bound subhaloes,
but in the remnants of the tidal stripping process accumulated over
the entire history of the halo.

{\it The radial distribution of subhaloes.--} The spatial distribution
of a subhalo population reflects the balance between accretion of new
subhaloes and tidal disruption of older ones. This distribution has
been studied extensively in $N$-body simulations
\cite{Ghigna2000,Diemand2004,Gao2004,Nagai2005,Diemand2007b,Springel2008,Angulo2009}
and the picture that emerges is that the radial distribution of
subhaloes is significantly less centrally concentrated than the dark
matter distribution (i.e., the smooth halo), and is relatively
independent of the host halo mass (see right panel of
Fig.~\ref{fig_structure_1}). Most remarkably, when subhaloes are
selected according to mass (rather than maximum circular velocity) and
the distribution is normalized to the mean number density of subhaloes
of a given mass within the virial radius, there appears to be no trend
in the shape of the number density profile with subhalo mass
\cite{Springel2008,Angulo2009,Ludlow2009}.  A recent analysis using
the HBT finder, however, has shown that the most massive subhaloes are
actually more concentrated in the central regions than lower mass
subhaloes \cite{Han2018}, which seemingly reflects the resilience of
very massive subhaloes to tidal stripping despite suffering from
substantial dynamical friction. The near universality of the radial
distribution of subhaloes is then the result of a convolution of the
distribution of subhaloes before infall (sometimes called the {\it
  unevolved} radial distribution of subhaloes), which is nearly scale
free, with a tidal stripping process that is also nearly scale free,
except at the massive end (see \cite{Han2016} for an analytical model
of the subhalo distribution).
%This trend is not fully
%understood although it is clearly related to the dynamical processes
%described in the previous section: dynamical friction acts more
%efficiently on more massive haloes, causing them to sink more quickly
%towards the centre where tidal effects disrupt them, in contrast to
%smaller subhaloes, which sink less rapidly and are more resilient to
%tidal disruption given their higher concentrations 
%(see \cite{Han2016} for an analytical model of the subhalo distribution). 
It is interesting that the radial distribution of subhaloes with
maximum circular velociy $>V_{\rm sub}$ is steeper than that of
subhaloes with mass $>m_{\rm sub}$ (see e.g. \cite{Gao2004b}), since
the latter is more heavily influenced by tidal stripping.

%We will comment
%below, however, that numerical artefacts might still be present even
%in high resolution simulations, which might affect these results to
%some extent.

The ratio between the average, mass-selected subhalo radial
distribution and the average NFW mass density profile of their host
haloes (both normalised to the virial radius as defined in
\cite{Jiang2017} and shown in the right-hand panel of
Fig.~\ref{fig_structure_1}), is approximated quite accurately by the
following functional form \cite{Jiang2017}:
\begin{equation}\label{radial_bias}
\phi(x\equiv r/r_{\rm vir})=\frac{d{\widetilde N}/dx^3\vert_{\rm sub}}{d{\widetilde N}/dx^3\vert_{\rm NFW}}=4\frac{x^4}{(1+x)^2}
\end{equation}

\begin{figure}[H]
\centering
\includegraphics[width=16 cm, height=9cm,trim=0.4cm 0.15cm 0.15cm 0.15cm, clip=true]{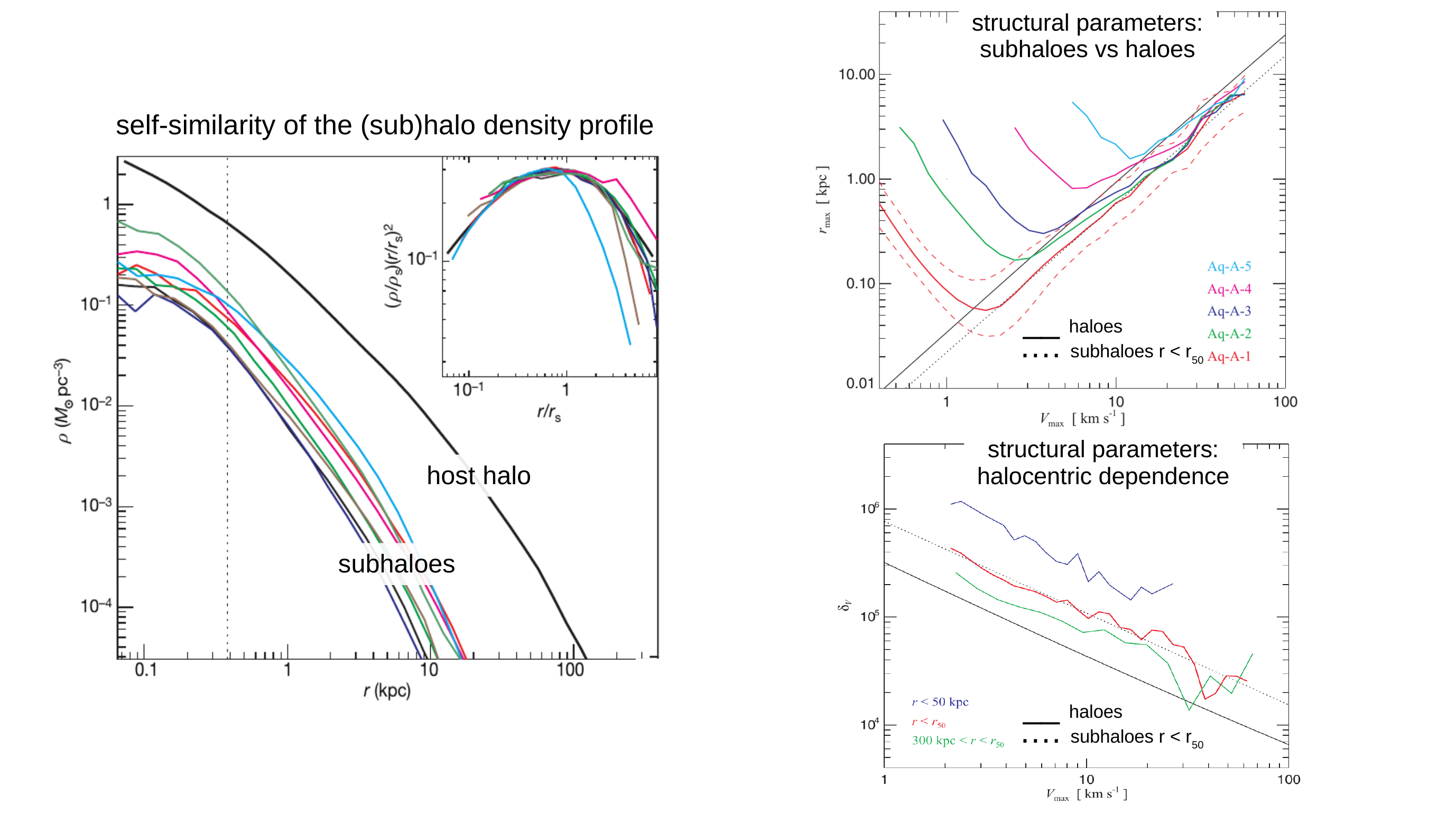}
\caption[Caption without FN]{The inner structure of subhaloes. {\it Left:} spherically
  averaged density profile of subhaloes (which is remarkably similar
  to that of isolated haloes). The plot shows the density profile of a
  Milky Way-size halo (solid black line) and eight of its largest
  subhaloes (colour lines). The vertical dotted line marks the radius
  beyond which the simulation results are converged. The
  self-similarity in the central region is better appreciated in the
  inset where the density and radius are scaled to their values at the
  scale radius, $r_s$. The figure is for the Via Lactea II simulation
  and is adapted from \cite{Diemand2008}.  {\it Upper right:} mean
  relation between the maximum circular velocity, $V_{\rm max}$, and
  the radius at which it is achieved, $r_{\rm max}$, for subhaloes
  within $r_{50}$ (the radius within which the mean enclosed density
  is 50 times the critical density) of one the Milky Way-size halo
  simulations in the Aquarius project, at different resolution
  levels (colour lines). The red dashed lines show the scatter ($68\%$
  of the distribution) for the highest resolution level. The dotted
  line is a fit to the mean relation for subhaloes and lies
  systematically below the equivalent line for isolated haloes (solid
  line).  {\it Lower right}: a measure of concentration for subhaloes (see Eq.~\ref{delta_V_2})
  within different radial ranges, as given in the legend. The solid
  line corresponds to isolated haloes. Figures adapted from
  \cite{Springel2008}\footnotemark.}
\label{fig_structure_2}
\end{figure}

{\it The inner structure of dark matter subhaloes.--} Cosmological
$N$-body simulations have shown that the density profiles of subhaloes
retain the near universal properties of isolated field haloes but with
modifications that reflect the tidal effects discussed in
Section~\ref{sec_dynamics}. These modifications are consistent with
expectations of analytical estimates and controlled simulations.  In
particular, for CDM, the subhalo radial density profile exhibits the
same central cusp as an isolated halo in equilibrium (left panel of
Fig.~\ref{fig_structure_2}), while the outer regions show a steep
truncation at a radius approximately equal to the tidal radius given
in Eq.~\ref{eq_tidal_radius} (see Fig. 15 of \cite{Springel2008}). We
should remark that, as has been found for field haloes
\cite{Navarro2010}, a better fit to the density profile of subhaloes
is given by the 3-parameter Einasto profile
\cite{Navarro2004}\footnote{Although the introduction of a third
  parameter will obviously improve the quality of the fit, the Einasto
  profile is, in fact, a slightly better fit to simulations than the
  2-parameter NFW profile even after one of the parameters
  ($\alpha_E$) is fixed to an appropriate value. For instance, fixing
  $\alpha_E\sim0.16$ gives a better fit than NFW to haloes across a
  range of halo masses \cite{Gao2008}.}:
\begin{equation}\label{rho_Ein}
\rho_E(x_E\equiv
r/r_{-2})=\rho_{-2}e^{-2(x_E^{\alpha_E}-1)/\alpha_E},%{\rm
                                %exp}\left(\frac{-2}{\alpha}\right), 
\end{equation}
where $\alpha_E$ is a shape parameter and $\rho_{-2}$ and $r_{-2}$ are
the density and radius at which the logarithmic slope of the density
profile is equal to $-2$.  The Einasto and NFW profiles are quite
similar, and both give good fits to the subhalo profiles in the range
$0.01<x_E<100$ if $\alpha_E\sim0.22$
\cite{VeraCiro2013,Dutton2014}. Although for isolated haloes the
parameters $\alpha_E$ and $r_{-2}$ can be related to the virial mass
of the halo, $M_{200}$ \cite{Gao2008,Klypin2016}, in a similar way as the
halo (NFW) concentration is connected to the virial mass, the
situation is less clear for subhaloes \cite{VeraCiro2013}, and the
spread of the parameters across subhalo masses is large. Thus, for its
simplicity, the NFW profile remains a reasonable approximation to the
structure of both haloes and subhaloes.

Since for subhaloes the virial radius no longer has a proper meaning
as the ``boundary'' of the object, the concentration parameter,
defined as $c=r_{200}/r_s$ commonly used to characterize NFW haloes,
is no longer appropriate. Instead, it is convenient to define the
concentration of a subhalo in a way that is independent of its size.
One such measure of concentration is the characteristic overdensity,
$\delta_V$, defined as the mean density within the radius,
$r_{\rm max}$, where the circular velocity peaks, at a value of
$V_{\rm max}$, relative to the critical density
\cite{Diemand2007,Springel2008}:
\begin{equation}\label{delta_V}
\delta_V=\frac{{\overline \rho(<r_{\rm max})}}{\rho_{\rm crit}}=2\left(\frac{V_{\rm max}}{Hr_{\rm max}}\right)^2,
\end{equation}
where $H$ is the Hubble constant. Eq.~\ref{delta_V} can be related to
the standard scale density of the NFW profile ($\delta_c$ in
Eq.~\ref{eq_delta_c}), and thus to the NFW concentration, in a
straightforward way \cite{Diemand2007}:
\begin{equation}\label{delta_V_2}
\delta_V=\left(\frac{c}{2.163}\right)^3\frac{K_c(2.163)}{K_c(c)}\Delta, 
\end{equation}
where $K_c$ was defined just after Eq.~\ref{eq_delta_c}. We note that
for the NFW profile, $r_{\rm max}/r_s=2.163$.

 \footnotetext{\scriptsize{Reproduced from Volker Springel et al. The Aquarius Project: the subhaloes of galactic haloes. MNRAS (2008) 391 (4): 1685–1711, doi: 10.1111/j.1365-2966.2008.14066.x. By permission of Oxford University Press on behalf of the Royal Astronomical Society. For the original article, please visit the following \href{https://academic.oup.com/mnras/article/391/4/1685/1747035}{link}. This figure is not included under the CC-BY license of this publication. For permissions, please email: journals.permissions@oup.com}}

Since the concentration of haloes (and subhaloes) is tightly
correlated with their mass (see Section~\ref{sec_dm_nature}),
Eq.~\ref{delta_V} implies a tight correlation between $V_{\rm max}$
and $r_{\rm max}$, which indeed has been found and characterised in
simulations (see right panel of Fig.~\ref{fig_structure_2}).  For
instance in the case of the Aquarius-A Milky Way-size halo, the
following fitting functions (to the mean relations in subhaloes)
provide a direct connection in terms of the subhalo mass\footnote{We
  note that there is a typo in the caption of Fig.~28 in
  \cite{Springel2008}, which gives the fitting function for $\delta_V$
  and $m_{\rm sub}$
  ($5.8\times10^8$~M$_\odot\rightarrow5.8\times10^4$~M$_\odot$).}:
\begin{eqnarray}\label{rmax_vmax}
V_{\rm max}&=&10~{\rm km/s}\left(\frac{m_{\rm sub}}{3.37\times10^7{\rm M}_\odot}\right)^{0.29}\nonumber\\
\frac{\delta_V(z=0)}{2}=\left(\frac{V_{\rm max}}{H_0r_{\rm max}}\right)^2&=&2.9\times10^4\left(\frac{m_{\rm sub}}{10^8{\rm M}_\odot}\right)^{-0.18}.
\end{eqnarray}
The fitting function for the $r_{\rm max} - V_{\rm max}$ relation
implied by Eq.~\ref{rmax_vmax} is shown as a dotted line in the upper
right panel of Fig.~\ref{fig_structure_2}, while the corresponding
relation for isolated haloes in the Millennium I simulation is shown
as a solid line \cite{Neto2007}\footnote{Obtained by taking the
  power-law concentration-mass relation in \cite{Neto2007} (their
  Eq.~4), and using Eqs.~\ref{delta_V} and \ref{delta_V_2}.}. The most
relevant result when comparing haloes and subhaloes in the
$r_{\rm max}-V_{\rm max}$ plane is that both share the same relation,
but subhaloes have systematically higher concentrations at a given
$V_{\rm max}$ \cite{Bullock2001}: in Fig.~\ref{fig_structure_2} the
dotted line is a factor of $0.62$ lower than the solid line,
i.e., subhaloes have on average $r_{\rm max}$ values that are smaller
than haloes of the same $V_{\rm max}$ by this factor. Equivalently,
the characteristic overdensity, $\delta_V$, in subhaloes is roughly a
factor of $(1/0.62)^2\sim2.6$ larger for subhaloes than for haloes of
the same $V_{\rm max}$ (lower right panel of
Fig.~\ref{fig_structure_2}), which roughly corresponds to a $30\%$
increase in the NFW concentration. This relative increase in
concentration is larger for subhaloes nearer the centre of the host,
as expected from the nature of the tidal forces experienced by the
subhaloes as described in Section~\ref{sec_dynamics}: while tidal
stripping naturally reduces $V_{\rm max}$, it reduces $r_{\rm max}$
even further \cite{Hayashi2003}\footnote{For a clear illustration of
  the evolutionary track of suhaloes in the $r_{\rm max}-V_{\rm max}$
  plane due to tidal stripping, see Fig. 8 of \cite{Penarrubia2008}.},
making the subhalo effectively more concentrated; the stronger the
mass loss, the stronger the effect, and hence the trend with
halocentric distance.

It is thus possible to model the inner density profile of the subhalo
population by assuming a model for the concentration-mass relation of
field haloes and making a simple correction to the subhalo
concentration depending on the location of the subhalo.  More
exhaustive studies of subhalo concentration exist that provide fitting
functions across a wide range of subhalo masses, host halo masses, and
distance to the halo centre (e.g. \cite{Moline2017} for the case of
Milky Way-size haloes).

{\it The shapes and internal kinematics of subhaloes.--} The impact of
tidal forces in the structure of subhaloes is reflected also in their
shapes. Although tides tend to elongate objects, these distortions are
short-lived features accentuated during pericentric passages. Once the
tidal streams cease to be bound to the subhalo, simulations have shown
that the bound material remains in an equilibirum configuration that
is, in fact, more spherical than it was at the time of infall; the
stronger the tidal effects, the more spherical the subhalo becomes
\cite{Barber2015}. Although these differences are significant for the
fraction of the subhalo population whose orbits are strongly
influenced by the tides of the host, the subhalo population as a whole
is only slightly affected and exhibits a small systematic shift
towards less aspherical shapes compared to field haloes
\cite{VeraCiro2014}. This is because the global subhalo population is
dominated in number by subhaloes near the virial radius of the host,
which have only recently fallen in.

When tidal effects are strong, the internal kinematics of subhaloes
are also substantially altered. In particular, the velocity anisotropy
of the dark matter particles becomes increasingly tangential
($\beta<0$) from the subhalo centre outwards \cite{VeraCiro2014}, in
contrast to field haloes that are radially anisotropic at larger
radii. This is the result of the preferential stripping by tides (when
the subhalo is near pericentre) of subhalo particles with radial
orbits. On the other hand, the pseudo-phase space density,
$Q_{\rm sub}$, of subhaloes in equilibrium retains the universal
power-law behaviour of CDM field haloes, but with a slightly shallower
slope, $Q_{\rm sub}\sim r^{-1.6}$ \cite{VeraCiro2014}, compared to
$\sim r^{-1.9}$ for field haloes.

\subsection{The impact of the nature of the dark matter}

{\it Subhalo abundance.--} By far the main difference in the subhalo
populations predicted in models with different kinds of dark matter is
the abundance of low-mass subhaloes. %The abundance of haloes of galactic
%mass an higher is strongly constrained to be close to the CDM
%prediction by a variety of observational data such as the galaxy
%luminosity function and galaxy clustering. 
In particular, as we
discussed in Section \ref{sec_dm_nature}, models in which the
primordial power spectrum of density perturbations has a cutoff at
relatively low $k$ (such as WDM and interacting dark matter) have a corresponding cutoff in the mass
function of haloes and subhaloes. These models predict far fewer haloes
and subhaloes than CDM, and this offers the best prospect for
distinguishing between them and perhaps constraining the properties of
the particles themselves (such as the WDM particle mass).

A cutoff in the mass function breaks the universal behaviour of the
halo and subhalo mass functions at low masses in a way that also
depends on the nature of the dark matter particle. For example, the
self-similarity of the abundance of CDM subhaloes as a function of
relative mass, exhibited in Eqs.~\ref{eq_gao} and \ref{eq_kp}, is
broken \cite{Bose2017} because the cutoff scale expressed in terms of
the ratio, $\mu=m_{\rm sub}/M_h$, occurs at larger values of $\mu$ for
smaller values of $M_h$. The radial distribution of subhaloes in WDM
models is quite similar to that in the CDM case, with only minor
differences explained by the enhanced tidal stripping of low-mass WDM
haloes resulting from their lower concentrations
\cite{Han2016,Bose2017}.

In many SIDM models the subhalo mass function remains largely
unchanged as long as the interaction cross-section,
$\sigma_T/m_\chi<10$~cm$^2/$g
\cite{Vogelsberger2012,Rocha2013,Dooley2016}. For higher values,
collisions between dark matter particles within subhaloes and in the
host are frequent enough to unbind material from the halo. This form
of {\it subhalo evaporation} is energetically efficient because the
energy transfer is determined by the relative velocity of the
colliding particles, which is of the order of the orbital velocity.
In this case, the mass loss in subhaloes is enhanced and the subhalo
abundance is depleted relative to the CDM case, particularly in the
central regions \cite{Vogelsberger2012}.

{\it Inner structure of subhaloes.--} The inner structure of WDM and
SIDM subhaloes is rather similar to that of field haloes (see
Fig.~\ref{fig_structure_nonCDM}), and the outer structure is altered
by tidal effects in a very similar way as in CDM. The main difference
is an enhancement in the concentration of subhaloes relative to their
field counterparts in WDM models \cite{Bose2017} due to the increased
efficiency of tidal stripping of WDM subhaloes, which are less
concentrated than in CDM at the time of infall.  Tidal stripping also
plays a greater role in SIDM subhaloes \cite{Dooley2016}. Most
importantly, it can trigger a gravothermal catastrophe and this can
give rise to segregation according to particular orbits, with cuspy
profiles for subhaloes which have experienced substantial tidal mass
loss and central cores for those where tidal effects have been minimal
\cite{Nishikawa2019}.

\section{Outlook}

It is fair to say that the evolution of the phase-space distribution
of classical, non-relativistic, collisionless dark matter (CDM) down
to galactic-scale haloes and subhaloes is now essentially a solved problem,
largely through the application of $N-$body simulations over the past
40 years\footnote{By galactic-scale haloes and subhaloes, we mean
  self-bound dark matter structures that can potentially host a
  galaxy, that is haloes of mass above $\sim10^8$~M$_\odot$, in which
  gas can cool by atomic processes
  (e.g. \cite{Mo2010,Sawala2016}).}. This strong statement carries
a couple of major caveats, which define today's frontier in $N-$body
simulations of cosmological structure formation.

Firstly, the statement above can still hold if, instead of CDM, most
of the dark matter consists of other types of particles, such as WDM
and SIDM\footnote{This is true only for elastic SIDM, and for cross
  sections that do not exceed the gravothermal collapse threshold,
  $\sigma_T/m_\chi\sim10$~cm$^2/$g, for dwarf-size haloes (see the
  last paragraphs of Section \ref{sec_dm_nature}). Although the regime
  of gravothermal collapse has been known for a couple of decades
  \cite{Balberg2002,Colin2002}, a comprehensive analysis of this
  regime has yet to be carried out (see
  \cite{Nishikawa2019,Zavala2019,Sameie2019,Kahl2019} for recent
  developments in this interesting regime).}, for which $N-$body
simulations with appropriate modifications have been applied at a
similar level of detail as in CDM; in this review we have discussed
the most important changes in the dark matter phase-space structure
that occur in these alternative models. Nevertheless, there are still
dark matter models that remain unexplored, or only partially explored
with $N-$body simulations, e.g. hidden dark sector models with DAOs
\cite{CyrRacine2016,Vogelsberger2016} and inelastic
SIDM\footnote{There is a class of inelastic SIDM models in which the
  dark matter can have ground and excited states
  (e.g. \cite{Arkani-Hamed2009}), and in which scattering between the
  excited and ground states can result in energy injection at the
  centre of dark matter haloes thus altering their structure. Only
  until very recently have these models began to be explored with
  simulations \cite{Vogelsberger2019,Todoroki2019}.}.  Secondly, and
crucially, the statement above does not take account of the interplay
between baryons and dark matter, which are dynamically coupled through
gravity. Several mechanisms that can radically modify the
dark-matter-only predictions of $N-$body simulations and which are, of
course, crucial for a complete theory of structure formation and its
connection to reality, have been studied extensively for several
decades. We briefly summarise these in Subsection~\ref{sec_baryons}
below.

Finally, we make two remarks concerning the limited
resolution of current $N$-simulations: (i)~there have been recent
claims that subhaloes can be artificially disrupted in cosmological
simulations due to discreteness effects and inadequate force
resolution \cite{vdBosch2018,vdBosch2018b}; if correct, these effects
could alter some of the current results on the abundance and structure
of subhaloes, particularly at low masses; (ii)~as we have seen, the
best current $N$-body simulations only resolve haloes of mass greater
that $\sim 10^5$~M$_\odot$, many orders of magnitude larger than the
cutoff mass in the linear density power spectrum for CDM. Yet, if the
dark matter is made of Majorana particles, these, so far unresolved, 
haloes could be crucial for predicting the properties of their
annihilation radiation and thus for elucidating the nature of dark
matter. The first attempts at understanding the properties of haloes
down to the cutoff in the CDM primordial power spectrum have been made
\cite{Anderhalden2013,Ishiyama2014,Angulo2017} but new techniques will
be required to tackle this problem in full generality.

\subsection{The impact of baryonic physics on dark matter structure}\label{sec_baryons}

In the linear regime, the (gravitational) impact of baryons (and
electrons and photons) in the dark matter distribution, of which
baryonic acoustic oscillations is perhaps the best known outcome
\cite[e.g.][]{Eisenstein1998}, is fairly well understood. In the
non-linear regime, on the other hand, the complexity of baryonic
physics is much greater and the list of relevant processes is
extensive: gasdynamics, radiative processes, star formation and
evolution, supermassive black hole formation and evolution, etc. Here
we focus on some of the most important mechanisms that modify the
predictions for the abundance and structure of CDM haloes from
$N-$body simulations.

{\it Condensation of baryons into haloes: adiabatic gas cooling and
  mergers.-} In the classical theory of galaxy formation, gas
initially follows dark matter; as haloes collapse and virialize, the
associated gas heats up by shocks and adiabatic compression to the
virial temperature of the halo
\cite{White1978,White1991}\footnote{Large relative velocities between
  gas and dark matter inherited from the photon-baryon coupling
  before recombination can impede the growth of gravitational
  perturbations and also stop gas from accreting into the first haloes
  \cite{Tseliakhovich2010}. This process, however, is only thought to
  be relevant for the formation of the first stars.}.  Subsequently,
the gas can radiatively cool and condense towards the centre of the
halo if the cooling time is shorter than the free-fall time. The halo
mass threshold for effective cooling depends on the density,
temperature and metallicity of the gas; cooling is quite efficient in
low-mass haloes down to the atomic cooling limit (virial temperatures
$\sim10^4$~K, corresponding to halo masses $\sim10^8$~M$_\odot$) below
which cooling becomes highly inefficient. At higher masses
($\sim10^{13}$~M$_\odot$ for gas with solar metallicity,
e.g. \cite{White1991}), cooling is also suppressed because the cooling
time exceeds the free-fall time, limiting the condensation of baryons,
a process that can be exacerbated when the gas is heated by energy
input from Active Galactic Nuclei (AGN) \cite{Bower2006,Croton2006}. A
hot, quasi-hydrostatic corona forms from which gas can subsequently
cool at the centre. Additional gas may be brought in by galaxy
mergers.  Regardless of the condensation mode, the assembly of the
central galaxy ultimately results in an enhancement of the central
gravitational potential, compared to the situation where the galaxy is
absent. The dark matter distribution reacts dynamically, becoming more
concentrated, a process first modelled assuming an adiabatic response
leading to the contraction of the halo
\cite{Blumenthal1986,Mo1998}. Even though the assembly of baryonic
matter by mergers is not, in general, adiabatic, the simple adiabatic
model remains a reasonable approximation \cite{Gnedin2004}. In the
absence of heating processes, the general expectation is thus that
haloes should be cuspier than the NFW profile in the central regions,
as indeed is seen in cosmological hydrodynamics simulations
\cite[e.g.][]{Schaller2015,Lovell2018}.

{\it Energy injection into haloes: UV background photoheating.-} The
hydrogen emerging from recombination is, of course, neutral. However,
the UV radiation produced by stars in the first generations of
galaxies reionises this gas and heats it up, suppressing gas cooling
into low-mass haloes and subsequent star formation
\cite{Efstathiou1992,Babul1992}. This heating mechanism moves the
minimum scale for galaxy formation from the atomic cooling limit to
larger halo masses of order $10^9$~M$_\odot$ today\footnote{This mass
  threshold is smaller at higher redshifts, see e.g. Fig. 3 of
  \cite{Okamoto2008}.}
\cite{Thoul1996,Barkana1999,Bullock2000,Gnedin2000,Benson2002,Sommerville2002,Hoeft2006,Sawala2016,Ockvirk2016}.
This baryonic process is important also because, in conjunction with
the expulsion of gas from haloes by supernova feedback (see below) at
high redshift, it reduces the overall baryonic content, and thus, the
total mass content of low-mass haloes; this reduces the growth rate and
final masses of these haloes compared to their counterparts in
simulations without baryons \cite{Sawala2013,Sawala2016}.

{\it Energy injection into haloes: supernova and AGN feedback.-} When
massive stars explode as supernovae in the final stages of their
evolution they release vast amounts of energy, a fraction of which
may couple effectively to the surrounding interstellar medium (ISM),
heating it and pushing it in a violent blowout. The combined impulsive
removal of baryonic outflows from several supernovae creates a
collective effect in the host galaxy known as supernova feedback,
which has a fundamental role in regulating star formation
\cite{White1991}. Supernova feedback affects the evolution of the
galaxy population at all galactic masses, but is particularly
important in low-mass haloes which have shallow potential wells;
supernova-driven galactic winds affect both the abundance
\cite{Larson1974,White1978,Dekel1986} and inner structure of
low-luminosity galaxies.  Acting in conjunction with reionisation,
such winds strongly suppress galaxy formation in small haloes, reducing
the abundance of luminous low-mass galaxies
\cite{Bullock2000,Benson2002,Sommerville2002}.

Energy injection from supernova can potentially alter the inner
structure of dark matter haloes: if gas becomes gravitationally
dominant in the centre and most of it is removed suddenly, as could
happen in a starburst, energy can be transferred from the gas to the
dark matter and this can cause the centre to expand, turning the
original NFW cusp into a core. This mechanism, first proposed in the
1990s \cite{Navarro1996}, became fashionable again several years later
\cite{Read2005,Gnedin2002,Governato2010,Pontzen2012,DiCintio2014,Chan2015,Tollet2016,Read2016}
when tentative observational evidence for cores, particularly in dwarf
galaxies, began to emerge \cite{Moore1994}. This evidence, however, is
controversial \cite{Oman2015,Oman2019}. While the proof of concept in
\cite{Navarro1996} was based on a single explosive event, recent
simulations have shown that repeated outflows can create rapid
fluctuations in the gravitational potential which efficiently transfer
energy to the dark matter \cite{Pontzen2012}. This core-formation
mechanism depends on the details of the baryon physics implemented in
the simulation \cite{Benitez-Llambay2018} and not all cosmological
simulations produce cores in dwarf galaxies \cite{Bose2019}.  On
scales larger than dwarfs, energy injection by AGN has been invoked as
a mechanism for core formation; however, the conditions required to
alter the deep potential wells of massive galaxies appear quite
extreme
\cite{Peirani2008,Duffy2010,Teyssier2011,Martizzi2013,Peirani2017}.

{\it Energy injection into subhaloes: tidal effects from baryonic
  structures.-} In Section~\ref{sec_dynamics} we described the tidal
effects that the host halo induces on the dynamics and structure of
subhaloes. The presence of a central galaxy enhances these effects
both in subhaloes and in the satellite galaxies within them,
particularly when their orbits cross the region where the central
galaxy dominates the tidal field. Tidal shocking by a galactic disc
can result in the total disruption of subhaloes around the central
regions of the host \cite{Donghia2010,Kazantzidis2011} and other
structural changes. Current hydrodynamical simulations of
Milky-Way-like galaxies and their environment seem to agree that the
overall effect is a substantial reduction in the number of subhaloes
near the centre \cite{Zolotov2012,Sawala2017,GK2018}.

There has been great progress in the past decade in incorporating
baryonic physics into full cosmological simulations; today galaxy
formation and evolution can be modelled in unprecedented detail
\cite{Dubois2014,Vogelsberger2014,Schaye2015,Khandai2015,Pillepich2018}. In
this way the effect on the dark matter phase-space distribution of the
complex interplay between the cooling and heating mechanisms of
baryons described above can be studied in their full cosmological
setting. In spite of this undeniable progress, many aspects of
baryonic physics remain poorly understood and, when they involve
processes on scales below the resolution of the simulation, they need to be
included as a {\it subgrid model}.  There are different approaches to
this problem which are often difficult to validate and this translates
into substantial uncertainties in some of the predictions of the
simulations (see \cite{Schaye2015} for a discussion of the limitations of
gasdynamic simulations).

%\subsection{The unresolved regime of structure formation}\label{sec_unresolved}

\subsection{Astrophysical tests of the nature of the dark matter}\label{sec_astrotests}

Laboratory searches for dark matter have so far proved
unsuccessful. This, and the failure to find evidence for {\sc
  Supersymmetry}, has generated gloom amongst proponents of the
lightest stable supersymmetric particle as the dark matter (even
though the mass of the Higgs boson suggests that the supersymmetry
scale is likely to be larger than a few~{\sc t}e{\sc v}, beyond the
reach of the {\sc lhc}).  There have been, however, claims that both
CDM-WIMPs, and WDM particles in the form of sterile neutrinos of
mass 7~ke{\sc v}, have been discovered, the former through $\gamma$-ray
annihilation radiation from the Galactic Centre \cite{Goodenough2009},
the latter through a 3.5~ke{\sc v} decay line in the {\sc x}-ray
spectra of galaxies and clusters
\cite{Bulbul2014,Boyarsky2014}\footnote{Contrary to some claims, {\sc
    xmm} data for Draco, and Hitomi data for Perseus, are consistent
  with a 7~ke{\sc v} neutrino \cite{Ruchayskiy2016,Hitomi2017}.}.
These claims are highly controversial but, since cosmogonic models
based on such particles have strong predictive power, they are
disprovable with appropriate astrophysical observations.

The standard CDM model has naturally come under the closest
scrutiny. Perhaps the two most important predictions of this model
(derived from $N$-body simulations) are: (i)~the existence of a vast
population of haloes and subhaloes which, below a mass of order
$10^9$~M$_\odot$, are dark; (ii)~the presence (in the absence of the
baryon effects discussed in the preceding section) of a steep cusp
($\rho\propto r^{-1}$) in the density profile of dark matter haloes of
all masses. These two predictions are related to three of the much
publicized four problems of the CDM model on subgalactic
scales (often referred to as the ``small-scale crisis'' of
CDM): the {\em (i)} missing satellites; {\em (ii)}
too-big-to-fail and {\em (iii)} core-cusp problems. The fourth is the
so-called {\em (iv)} planes of satellites problem. Indeed some of
alternative dark matter particle models, such as SIDM, have been
proposed specifically to solve some or all of these perceived
astrophysical problems.

The missing satellites problem is the discrepancy between the
relatively small number of satellites observed around the Milky Way
and M31 and the many orders of magnitude larger number of halo
substructures predicted by CDM $N$-body simulations
\cite{Klypin1999ms,Moore1999}. The ``too-big-to-fail'' problem is the
existence in CDM $N$-body simulations of massive, dense galactic
subhaloes (maximum circular velocities, $V_{\rm max} > 30$ km/s) whose
kinematics appear inconsistent with those of the brightest Milky Way
satellites \cite{Boylan-Kolchin2011}.  The core-cusp problem is the
discrepancy between the cuspy universal NFW density profiles predicted
for pure CDM/WDM haloes and the inference of central cores in some
galaxies, particularly dwarfs \cite[e.g.][]{Walker2011}.  The ``planes
of satellites'' problem is the arrangement of the bright satellites of
the Milky Way, M31 (and a few others) on a thin plane in which the
satellites seem to be coherently rotating and which have been claimed
to be incompatible with CDM \cite{Kroupa2005,Ibata2014,Pawlowski2014}.

The first three of the four perceived problems can be solved once the
effects of baryons discussed in Section~\ref{sec_baryons} are taken
into account. Perhaps paradoxically, the solution to what later became
known as the ``missing satellites'' problem was understood long before
it came to be regarded as a problem for CDM.  The strong suppression
of galaxy formation in haloes below a mass of $\sim 10^{10}$M$_\odot$
was originally calculated using semi-analytic techniques
\cite{Thoul1996,Efstathiou1992,Babul1992}, as were the implications
for the abundance of galactic satellites in the CDM model
\cite{Bullock2000,Benson2002,Sommerville2002}. This solution has been
repeatedly confirmed by modern gasdynamic simulations
\cite[e.g.][]{Okamoto2008,Wadepuhl2011,Sawala2016,Simpson2018}. Similarly,
the ``too-big-to-fail'' problem disappears when baryons are taken into
account, in this case through the more subtle effect of the reduced
growth of subhaloes arising from the early loss of baryons mentioned
above \cite{Sawala2016}. The ``core-cusp problem'', if it exists at
all, can also be solved by the type of explosive baryonic effects
discussed in Section~\ref{sec_baryons}, which can transform NFW cusps
into cores\footnote{Other baryon effects that can transform cusps into
  cores have been proposed (e.g. \cite{Mashchenko2006,Weinberg2002})
  but have been less studied.}.  The existence of ``planes of
satellites'' in the Milky Way and M31 turns out not to be as unlikely
as has been claimed \cite[e.g.][]{Ibata2014,Muller2016}, once the
statistics are calculated rigorously, taking into account the ``look
elsewhere effect'' \cite{Cautun2015}\footnote{See \cite{Muller2018}
  for an opposed view.}. The origin of these planes is almost
certainly the anisotropic nature of the accretion of satellites along
filaments of the cosmic web \cite{Libeskind2005,Shao2019} although the
exact mechanism is still unclear as is the expected frequency of these
structures.

While it is now generally agreed amongst practitioners of the field
that CDM is not afflicted by a ``missing satellite'' or a
``too-big-to-fail'' problem, the data on the satellites of the Milky
Way can be used to constrain alternative dark matter models,
particularly those with a cutoff in the primordial power spectrum. In
WDM, the cutoff lengthscale varies roughly inversely with the mass of
particle. Thus, if this mass is too small, then too few small-mass
haloes would form and their abundance could be too low to account for
the observed number of satellites of the Milky Way. The expected
subhalo abundance increases roughly in proportion to the mass of the
parent halo \cite{Wang2012} so, in reality, the observed abundance of
satellites constrains both the particle mass and the host halo mass
simultaneously.  For instance, using a semi-analytic model of galaxy
formation, the thermal WDM model was found to be in conflict with the
data if the Milky Way halo mass is smaller than
$1.1 \times 10^{12}$~M$_\odot$ \cite{Kennedy2014}. Using a similar
approach, \cite{Lovell2016} have ruled out a significant fraction of
the parameter space of sterile neutrinos and conclude that the models
that are in best agreement with the observed 3.5 keV line require 
the Milky Way halo to have a mass no smaller than
$1.5\times 10^{12}$~M$_\odot$, a value that may already be in conflict
with the most recent determinations of the Milky Way halo mass
\cite{Callingham2019}. We should note that since the central densities
of WDM haloes are lower than those of their CDM counterparts, the
``too-big-to-fail problem'' is easily avoided in WDM
\cite{Lovell2012,Lovell2017}.

Although the strongest constraints on the SIDM cross section come from
the shapes and dynamics of massive haloes (particularly of galaxy
clusters, see e.g. Table~1 of \cite{Tulin2018}), the Milky Way
satellites are perhaps the best testbed for SIDM, since it is in these
systems that the model shows its greatest promise as an alternative to
CDM.  A few years ago it was suggested that the interesting range of
cross sections for the SIDM model to alleviate the ``too-big-to-fail''
problem (without taking into account the baryonic processes just
mentioned) is $0.1\lesssim\sigma_T/m_\chi\lesssim10~$cm$^2$/g
\cite{Zavala2013}.  Since then, several studies have taken a closer
look at the properties of the Milky Way satellites within the context
of SIDM and the picture that is emerging points to promising tests for
the near future which will either strengthen SIDM as an alternative to
CDM or narrow the range of allowed cross sections. For instance, the
diversity of dark matter densities on subkiloparsec scales in the
Milky Way satellites is difficult to accommodate for SIDM cross
sections $\sigma/m_\chi\sim1~$cm$^2/$g \cite{Zavala2019}. The
inferred high dark matter densities in the ultra-faint satellites
(albeit uncertain due to possible systematic effects) are at first
sight difficult to explain within SIDM, which naturally predicts
cores, particularly in low-mass dark matter-dominated haloes. However,
a gravothermal collapse phase in SIDM haloes has recently been
proposed \cite{Nishikawa2019,Zavala2019,Sameie2019,Kahlhoefer2019} as
a mechanism to create a diverse population of dwarf-size haloes, some
of which would be cuspy (those that collapse), and others that would
have cores. If cores are indeed shown to be present in (some) dwarf
galaxies, then dark matter self-interactions and the explosive baryon
effects in CDM mentioned above provide alternative explanations that
need to be contrasted. A promising way to achieve this, recently put
forward \cite{Burger2019}, is to search for distinct signatures in the
detailed kinematics of the stellar population as they respond
differently to these two core formation mechanisms, one impulsive
(supernova feedback) and the other adiabatic (SIDM).

Since, as we have seen, the simplicity of the predictions of $N$-body
simulations can be easily obscured by the complexity of baryon
effects, testing dark matter models with astronomical observations
might, at first sight, seem a hopeless task. In fact, this is not the
case: the vast majority of haloes in CDM (and in many alternative dark
matter models) are dark, that is, unaffected or almost unaffected by
baryons. It is the existence of a vast population of such small-mass
haloes ($m \lesssim 5\times 10^9$~M$_\odot$) that is the hallmark of
the CDM model that distinguishes it from, for example, the WDM model.
Fortunately, nature has provided us with several tools to detect dark
objects in the Universe. One of these takes advantage of a side effect
of cosmic reionization which allows haloes in a small mass window
($10^8 \lesssim 5\times 10^9$)~M$_\odot$ to retain neutral hydrogen in
hydrostatic equilibrium with the dark matter potential and in thermal
equilibrium with the ionizing UV background, gas which is, however,
too diffuse to make stars \cite{Rees1986}. These objects called
RELHICs (REionization-Limited HI Clouds, \cite{Benitez-Llambay2017})
may be detectable in forthcoming blind HI surveys and provide, in
principle, a critical test of CDM and related models in a regime that
has not been proved before.

An interesting idea that has been proposed to infer the existence of
small dark subhalos orbiting in the Milky Way halo is the disturbance
they cause when they cross a tidal stellar stream \cite{Carlberg2012}.
When a subhalo crosses a stream it induces velocity changes along and
across the stream that can give rise to a visible gap, particularly in
cold streams such as those stripped from globular clusters. The
cross-section for gap creation is dominated by the smallest subhalos
so gaps can, in principle, constrain the identity of the dark matter.
The creation of gaps has been investigated with analytical treatments
or idealized numerical studies and it has been suggested that
perturbers of mass $\sim 10^7$~M$_\odot$ could be detected in the {\sc
  gd-1} and Pal~5 globular cluster stellar streams \cite{Erkal2016}. A
complication of this method is that perturbations on the streams can
be induced not only by dark subhaloes but also by giant molecular
clouds and the bar at the centre of the Milky Way
\cite{Amorisco2016}. Recent deep imaging around the Pal 5 stellar
stream does indeed reveal significant disturbances, in particular two
gaps which have been attributed to the impact of subhalos of mass in
the range $10^6-10^7$~M$_\odot$ and $10^7-10^8$~M$_\odot$ respectively
(although the smaller gap could also be due to the impact of a giant
molecular cloud) \cite{Erkal2017,Bovy2017}.

But perhaps the most direct method to search for the ubiquitous
small-mass dark haloes is gravitational lensing.  There are two
specific instances where strong gravitational lensing could provide
the means to do this.  The first are the ``flux-ratio anomalies'' seen
in some multiply-lensed quasars; the second are small distortions of
Einstein rings and large arcs.

In a multiply-lensed image, the magnifications are determined by high
order derivatives of the lensing potential and are therefore
particularly sensitive to small changes in the potential such as those
produced by intervening small-mass structures. If the mass
distribution of the lens is smooth, the ratios of the fluxes of close
images (formed when the sources are close to a fold or a cusp of the
caustic) follow a certain asymptotic relation
\cite{Mao1998,Schneider1992}. These smooth-lens relations are violated
if there are intervening structures or substructures in the lens
giving rise to flux-ratio anomalies, which probe the total amount of
mass in structures along the line of sight to the lens
\cite{Mao1998,Metcalf2001, Dalal2002}.  Flux-ratio anomalies have been
observed in several quadruply-lensed quasars but dark substructures
alone are insufficient to explain the observed anomalies
\cite{Xu2015}, implying that other effects such as inadequate lens
modelling may be at work. With better modelling of the lens (including
stellar discs and luminous satellites), it has been possible to set a
lower limit to the mass of a thermal WDM particle (see
\cite{Hsueh2019} and Harvey et al., in preparation), similar to the
limits from satellite counts discussed above and also to those derived
from the observed inhomogeneity of the gas distribution at high
redshift probed by the Lyman-$\alpha$ forest \cite{Irsic2017}.

%(Mao & Schneider 1998; Metcalf & Madau 2001; Metcalf & Zhao 2002; Dalal & Kochanek 2002; Chiba 2002; Kochanek & Dalal 2004)

A more direct strategy for detecting dark structures and substructures
is to search for distortions in strongly lensed images. When the
source (a background galaxy), the lens (a massive halo) and the
observer are perfectly aligned, a circular feature near the centre of
the lens, an Einstein ring, is formed; if the alignment is not
perfect, then giant arcs are formed.  If the lens is a halo of mass
larger than $\sim 10^{13}$~M$_\odot$, the radius of the Einstein ring
is generally larger than the image of the central galaxy and can thus
be studied in detail. If a halo or subhalo happens to be projected
onto the Einstein ring, it too will gravitationally lens the light
from the source producing a small distortion in the image of the
Einstein ring or giant arc \cite{Vegetti2009a}.  This strategy has
already yielded a halo of $\sim 10^8$~M$_\odot$  \cite{Vegetti2012}
\footnote{This halo mass was estimated assuming a truncated
  pseudo-Jaffe profile (see e.g. eq.~42 in \cite{Keeton2001}). The
  inferred mass is likely to be larger if an NFW profile is assumed
  instead. For instance, a similar  dark matter substructure detected
  with lensing was reported by \cite{Vegetti2010} with a mass of
  $\sim3.5\times10^9$~M$_\odot$ assuming a truncated pseudo-Jaffe
  profile, while assuming an NFW profile this substructure is estimated to
  have a mass of $\sim10^{10}$~M$_\odot$  \cite{Vegetti2018}.} and 
could detect haloes as small as $\sim 10^7$~M$_\odot$
\cite{Vegetti2014,Hezaveh2016}.

Detecting the small signal generated by individual projected haloes or
subhaloes requires accurate modelling of the source and the lens (the
``macro'' model; \cite[e.g.][]{Vegetti2009a,Nightingale2017}) and sophisticated
statistical techniques to analyse the image residuals. Dark haloes
imprint other observable features onto strong arcs. For example,
distortions to the lensing potential caused by the cumulative
contribution of many hundreds of projected structures produce unique
correlated residuals in the lensed image, the nature of which is
dependent on the abundance and mass distribution of the halo
population and, therefore, on the nature of the dark
matter \cite{Brewer2016,Diaz2017}.  The mass function of dark haloes may
also be detectable through the {\sc n}-point functions of the
projected density field or the substructure convergence power
spectrum \cite{Diaz2017}.

A very attractive feature of strong lensing as a means to detect
small-mass objects is that, for lens configurations of interest, the
dominant source of strong arc distortions are field haloes along the
line of sight, rather than subhaloes resident in the lens
\cite{Li2017,Despali2018}\footnote{This has been demonstrated
  explicitly for the case of Einstein ring distortions but it may hold
  true for other tests as well.}.  This makes this test uniquely
powerful because, as we have seen, the haloes of interest, of mass less
than $\sim 10^8$~M$_\odot$, are completely dark: they have never been
modified in any way by bayrons. Thus, the test depends mostly on the
abundance of pristine ``field'' dark matter haloes which we know very
well how to calculate rigorously and precisely with N-body simulations
for cosmological models of interest.

Approximately a few hundred high quality strong lens systems would
suffice to rule out either the 7~keV sterile neutrino model or
CDM itself \cite{Li2016}. Very high resolution imaging is the
primary requirement, either in the optical or UV, or using
interferometry at submillimeter and longer wavelengths \cite{Hezaveh2016}.  At least
several tens of systems with high quality data are already available
and future imaging facilities such as LSST and Euclid will increase
the number of suitable strong lenses by orders of magnitude.  By
bypassing the complications introduced by baryons, which have spoiled
all previous efforts to test the CDM model unambiguously and
distinguish it from alternative models, be they on small or large
scales, gravitational lensing offers a unique opportunity for a
breakthrough in this quest from astrophysics evidence alone.

\acknowledgments{We are very grateful to Alejandro Benitez-Llambay,
  Sownak Bose, Jiaxin Han,Mark Lovell and Simon White for their
  valuable comments.  Special thanks to Simon for his careful reading
  of the manuscript, insightful comments and suggestions. All of these
  have significantly improved our paper. We also thank Sebastian Bohr for
  his help in creating the right panel of Fig. 2.
  JZ acknowledges support by a
  Grant of Excellence from the Icelandic Research fund (grant number
  173929$-$052). CSF acknowledges support from the European Research
  Council (ERC) Advanced Investigator grant DMIDAS (GA 786910) and the
  Science and Technology Facilities Council (STFC) [grant number
  ST/F001166/1, ST/I00162X/1, ST/P000541/1]. Some of the simulations
  reported in this paper were carried out on the DiRAC Data Centric
  system at Durham University, operated by the ICC on behalf of the
  STFC DiRAC HPC Facility (www.dirac.ac.uk). This equipment was funded
  by BIS National E-infrastructure capital grant ST/K00042X/1, STFC
  capital grant ST/H008519/1, and STFC DiRAC Operations grant
  ST/K003267/1 and Durham University. DiRAC is part of the National
  E-Infrastructure.}

\newpage
\externalbibliography{yes}
\bibliography{subhalo_review}

%%%%%%%%%%%%%%%%%%%%%%%%%%%%%%%%%%%%%%%%%%
\end{document}